\documentclass[prb,aps,amsmath,amssymb,amsfonts,floatfix,superscriptaddress,
showpacs,twocolumn]{revtex4}
\usepackage{graphicx}
\usepackage[dvips]{color}
\usepackage{dcolumn}
\newcommand{\bk}{\mathbf{k}}
\newcommand{\bq}{\mathbf{q}}
\newcommand{\br}{\mathbf{r}}
\newcommand{\bR}{\mathbf{R}}
\newcommand{\bT}{\mathbf{T}}
\newcommand{\bG}{\mathbf{G}}
\newcommand{\bH}{\mathbf{H}}
\newcommand{\bD}{\mathbf{D}}

\newcommand{\bM}{\mathbf{M}}
\newcommand{\bU}{\mathbf{U}}

\newcommand{\bsig}{\mathbf\Sigma}
\newcommand{\tr}{\mbox{tr}\,}
\newcommand{\tg}{t_{2g}}

\newcommand{\hks}{\mathbf{H}_{\rm{KS}}}

\newcommand{\brT}{\mbox{$\br$$-$$\bT$}}
\newcommand{\brR}{\mbox{$\br$$-$$\bR$}}
\newcommand{\brrR}{\mbox{$\br'$$-$$\bR$}}
\newcommand{\brRT}{\mbox{$\br$$-$$\bR$$-$$\bT$}}
\newcommand{\brrRT}{\mbox{$\br'$$-$$\bR$$-$$\bT$}}
\def\bavs3{BaVS$_3$}
\def\srvo3{SrVO$_3$}
\def\t2g{$t_{2g}$}
\def\eg{$e_g$}
\def\eeg{$E_g$ }
\def\A1g{$A_{1g}$}
\def\fermi{$\varepsilon_{\rm{F}}^{\hfill}$ }

\newcommand{\hP}{\hat{P}}

\newcommand{\iomn}{i\omega_n}
\newcommand{\Gloc}{\hat{G}_{\rm{loc}}}
\newcommand{\Gimp}{\hat{G}_{\rm{imp}}}
\newcommand{\Bka}{B_{{\bf k}\alpha}}
\newcommand{\Bkap}{B_{{\bf k}\alpha'}}
\newcommand{\HKS}{\hat{H}_{\rm{KS}}}
\newcommand{\DSig}{\Delta\hat{\Sigma}}
\newcommand{\psiknu}{\psi^{\hfill}_{{\mathbf k}\nu}}
\newcommand{\psiknut}{\tilde{\psi}^{\hfill}_{{\mathbf k}\nu}}

\begin{document}

\title{Dynamical mean-field theory using Wannier functions:\\
a flexible route to electronic structure calculations of strongly correlated 
materials}
\author{F. Lechermann}
\email{Frank.Lechermann@cpht.polytechnique.fr}
\affiliation{Centre de Physique Th\'eorique, \'Ecole Polytechnique,
91128 Palaiseau Cedex, France}
\affiliation{LPT-ENS, 24 Rue Lhomond, 75231 Paris Cedex 05, France}
\author{A. Georges}
\affiliation{Centre de Physique Th\'eorique, \'Ecole Polytechnique,
91128 Palaiseau Cedex, France}
\author{A. Poteryaev}
\affiliation{Centre de Physique Th\'eorique, \'Ecole Polytechnique,
91128 Palaiseau Cedex, France}
\author{S. Biermann}
\affiliation{Centre de Physique Th\'eorique, \'Ecole Polytechnique,
91128 Palaiseau Cedex, France}
\author{M. Posternak}
\affiliation{Institute of Theoretical Physics, Ecole Polytechnique F\'ed\'erale, 
de Lausanne (EPFL), CH-1015 Lausanne, Switzerland}
\author{A. Yamasaki}
\affiliation{Max-Planck-Institut f\"ur Festk\"orperforschung, Heisenbergstrasse 1,
D-70569, Stuttgart, Germany}
\author{O.K. Andersen}
\affiliation{Max-Planck-Institut f\"ur Festk\"orperforschung, Heisenbergstrasse 1,
D-70569, Stuttgart, Germany}
\begin{abstract}
A versatile method for combining density functional theory (DFT) in the local 
density approximation (LDA) with dynamical mean-field theory (DMFT) is presented.
Starting from a general basis-independent formulation, we use Wannier functions 
as an interface between the two theories.
These functions are used for the physical purpose of identifying
the correlated orbitals in a specific material, and also for the more technical
purpose of interfacing DMFT with different kinds of band-structure methods
(with three different techniques being used in the present work). We explore and 
compare two distinct Wannier schemes, namely the 
maximally-localized-Wannier-function (MLWF) and the $N$-th order 
muffin-tin-orbital (NMTO) methods.
Two correlated materials with different degrees of structural and electronic 
complexity, \srvo3 and \bavs3, are investigated as case studies. \srvo3 belongs 
to the canonical class of correlated transition-metal oxides, and is chosen here 
as a test case in view of its simple structure and physical 
properties. In contrast, the sulfide \bavs3 is known for its rich and complex 
physics, associated with strong correlation effects and low-dimensional 
characteristics. New insights into the physics associated with the
metal-insulator transition of this compound are provided, particularly
regarding correlation-induced modifications of its Fermi surface.
Additionally, the necessary formalism for implementing self-consistency
over the electronic charge density in a Wannier basis is discussed.

\end{abstract}

\pacs{71.30.+h, 71.15.Mb, 71.10.Fd, 75.30.Cr}
\maketitle

\section{Introduction}
One of the fundamental points that underlies the rich physics of strongly
correlated electron systems is the competition between the electrons tendency
to localize, and their tendency to delocalize by forming quasiparticle (QP) bands.
Traditional effective single-particle (i.e. band-structure) theories emphasize
the latter aspect, which is appropriate when the kinetic energy dominates. For 
such materials, computational techniques based on electronic density functional 
theory (DFT) (see e.g. Refs.~[\onlinecite{koh99,jon89}] for reviews) have 
nowadays reached a very high degree of accuracy and yield remarkable agreement 
with experiment.

In correlated materials however, the screened Coulomb interaction is
a major aspect of the problem, which cannot be treated perturbatively,
and independent-particle descriptions fail. Albeit the representability of the
electronic charge density by a set of Kohn-Sham~\cite{koh65} (KS) orbitals is 
still guaranteed in most cases, this raises the question of whether such a 
representation is
physically appropriate. Furthermore, the description of excited states of the
many-particle system must be based on other observables than just the charge
density, such as the energy-dependent spectral function. Any appropriate
theoretical framework must then treat band formation (best described in momentum
space) and the tendency to localization (best described in real space) on an
equal footing. For this reason, there is an increasing awareness that
many-body descriptions must also include real-space, orbitally resolved,
descriptions of the solid, close to the quantum chemistry of the
material under consideration~\cite{per01,kri92}. In correlated metals, the
coexistence of coherent QP bands at low energy with high-energy
incoherent Hubbard bands (which originate from atomic-like transitions
persisting in the solid state) is a vivid demonstration that a dual description
(both in momentum space and in real space) is needed. Such a dual description is
at the heart of dynamical mean-field theory (DMFT) (see e.g. 
Refs.~[\onlinecite{geo96,georges_strong,hel02,bie06,kotliar_dmft_physicstoday,kotliar_review}]
for reviews), which in recent years has proven to be a tool of choice for 
treating strong-correlation effects. This theory has been successfully combined 
with electronic-structure methods within the framework of the local density 
approximation~\cite{koh65} (LDA) to 
DFT~\cite{ani97,lichtenstein_lda+dmft_1998} (also labeled as LDA+DMFT), or 
so-called GW formalisms~\cite{bie03,bie03_2,kot01}.

A central physical issue in merging the momentum-space and local descriptions
within those many-body approaches is the identification of a subspace of
orbitals in which correlations are treated using non-perturbative many-body
techniques. Furthermore, an important technical issue is the choice of a
convenient basis set for interfacing the many-body and the band-structure
parts of the calculation. Because the original Wannier construction~\cite{wan37}
is based on a decomposition of the extended Bloch states into a superposition of
rather localized orbitals, it appears that appropriate generalizations of this
construction leading to well-localized basis functions should provide an
appropriate framework for many-body calculations. Exploring this in detail
is the main purpose of this paper. In fact, there has been recently a growing 
activity associated with the Wannier formalism in the context of many-body 
approaches.
The use of Wannier basis sets in the LDA+DMFT context has been lately
pioneered by several groups, using either the $N$-th order muffin-tin orbital 
(NMTO) framework~\cite{pav04,pot04,bie05,pav05} or other types of Wannier 
constructions based on the linear muffin-tin orbital (LMTO) 
framework~\cite{ani05,gavri05,sol06,ani06}. For a detailed presentation of such 
implementations see Ref.~[\onlinecite{ani05}]. Furthermore, the computation of 
many-body interaction parameters has also been discussed~\cite{ku02,schn03}.

In this context, the main motivations of the present article are the following:

\begin{itemize}
\item[i)] We give a presentation of the LDA+DMFT formalism in a way which
should make it easier to interface it with a band-structure method of choice. 
To this aim, we are careful to distinguish between two key concepts:
the orbitals defining the correlated subspace in which a  many-body treatment is
done, and the specific basis set which is used in order to interface the
calculation with a specific band-structure method.
The LDA+DMFT approach is first presented in a manner which makes no reference
to a specific basis set, and then only some technical issues associated with
choosing the basis set for implementation are discussed.

\item[ii)] It is explained how the Wannier-functions formalism provides an elegant
solution to both the physical problem of identifying the correlated orbitals and
to the more technical issue of interfacing DMFT with basically {\sl any kind} of
band-structure method.
So far the LDA+DMFT technique has been implemented with band-structure codes based
on muffin-tin-orbital (MTO)-like representations~\cite{andersen_lmto_1975_prb}.
Although this realization is very successful,
we feel that broadening the range of band-structure methods that can be used
in the LDA+DMFT context may make this method accessible to a larger part
of the band-structure community, hence triggering further progress on a larger 
scale. As an example, one could think of problems involving local structural 
relaxations, which are more difficult to handle within the MTO formalism than
in plane-wave like approaches. 

\item[iii)] In this work, two different Wannier constructions are applied and
the corresponding results are compared in detail. Though there are numerous ways 
of constructing Wannier(-like) functions we have chosen such methods that derive 
such functions in a post-processing step from a DFT calculation.
In this way the method is, at least in principle, independent of the underlying
band-structure code and therefore widely accessible.
First, we used the maximally-localized-Wannier-functions (MLWFs) method 
proposed by Marzari, Vanderbilt and Souza~\cite{mar97,sou01}.
Second, we constructed Wannier functions using the $N$-th order MTO (NMTO) 
framework following Andersen and coworkers~\cite{and00,and00-2,zur05} which has 
first been used in the LDA+DMFT context in Ref.~[\onlinecite{pav04}] and 
actively used since then (e.g. Refs.~[\onlinecite{tan05,pav05,nek06,yam06}]). Note
that the NMTO method also works in principle with any given, not necessarily 
MTO-determined, KS effective potential. However, in practice, this construction 
is presently only available in an MTO environment.

\item[iv)] We also consider the issue of fully self-consistent
calculations in which many-body effects are taken into account in the
computation of the electronic charge density. Appendix~\ref{csc} is
devoted to a technical discussion of implementing charge 
self-consistency, with special attention to the use of Wannier basis sets also
in this context. However, the practical implementation of charge
self-consistency in non-MTO based codes is ongoing work, to be discussed in 
detail in a future publication.

\end{itemize}

Two materials with correlated $3d$ electrons serve as testing grounds for the 
methods developed in this paper, namely the transition-metal oxide \srvo3 and the
sulfide \bavs3. Nominally, both compounds belong to the class of $3d^1$
systems, where due to crystal-field splitting the single $d$ electron is
expected to occupy the $\tg$ states only. The latter form partially filled bands 
in an LDA description. The two compounds have very different physics and exhibit
different degrees of complexity in their electronic structure. The metallic
perovskite \srvo3 has perfect cubic symmetry over the temperature regime of
interest and displays isolated $\tg$-like bands at the Fermi level,
well-separated from bands higher and lower in energy. Its physical properties 
suggest that it is in a regime of intermediate strength of correlations. 
Many experimental results are available for this material (for a detailed list 
see Sec.~\ref{srvo3chap}) and it has also been thoroughly investigated 
theoretically in the LDA+DMFT 
framework~\cite{lie03,sek04,pav04,pav05,nek05,sol06,nek06}.
For all these reasons, \srvo3 is an ideal test case for methodological 
developments.

In contrast, \bavs3 is much more complex in both its electronic structure and
physical properties. The sulfide displays several second-order transitions with 
decreasing temperature, including a metal-insulator transition (MIT) at 
$\sim$70 K. Additionally, the low-energy LDA bands with strong $\tg$ orbital 
character are entangled with other bands, mainly of dominant sulfur 
character, which renders a Wannier construction more challenging. In this 
paper, the Wannier-based formalism is used for \bavs3 to investigate  
correlation-induced changes in orbital populations, and most notably, 
correlation-induced changes in the shape of the different Fermi-surface sheets 
in the metallic regime above the MIT. In the end, these changes are key to a 
satisfactory description of the MIT.

This article is organized as follows. Section~\ref{theory} introduces the
general theoretical formalism. First, the LDA+DMFT approach is briefly
reviewed in a way which does not emphasize a specific basis set. Then, the issue
of choosing a basis set for implementation and interfacing DMFT with a specific
band-structure method is discussed. Finally, the Wannier construction is
shown to provide an elegant solution for both picking the correlated orbitals
and practical implementation. The different Wannier constructions used in 
this paper are briefly described, followed by some remarks on the
calculational schemes employed in this work. In Sect.~\ref{results} the
results for \srvo3 and \bavs3 are presented. To this aim we discuss separately the
LDA band structure, the corresponding Wannier basis sets and the respective
LDA+DMFT results. Appendices are devoted to the basic formalism required to
implement self-consistency over the charge density and total energy
calculations, as well as further technical details on the DFT calculations.

\section{Theoretical framework\label{theoryframe}}
\label{theory}
\subsection{Dynamical mean-field theory and electronic structure}

\subsubsection{Projection onto localized orbitals}

Dynamical mean-field theory provides a general framework for electronic
structure calculations of strongly correlated materials. A main concept in this
approach is a projection onto a set of spatially localized single-particle
orbitals $\{|\chi_{\bR m}\rangle\}$, where the vector $\bR$ labels a site  
in the (generally multi-atom) unit cell and $m$ denotes the orbital degree of 
freedom. These orbitals 
generate a subspace of the total Hilbert space, in which many-body effects will
be treated in a non-perturbative manner. In the following, we shall therefore
refer to this subspace as the ``correlated subspace'' $\mathcal{C}$, and often
make use of the projection operator onto this correlated subspace, defined as:
\begin{equation}
\hP^{(\mathcal{C})}_{\bR} \equiv \sum_{m\in\mathcal{C}}
|\chi_{\bR m}\rangle \langle \chi_{\bR m}|\,.
\label{proop}
\end{equation}
For simplicity, we restrict the present discussion to the basic version of DMFT in
which only a single correlated site is included in this projection. In cluster
generalizations of DMFT, a group of sites is taken into account. Also, it may be
envisioned to generalize the method in such a way that $\bR$ could stand for 
other physically designated real-space entities (e.g. a bond, etc.).

Because the many-body problem is considered in this projected subspace only, and
because it is solved there in an approximate  (though
non-perturbative) manner, different choices of these orbitals will in general
lead to different results. How to properly choose these orbitals is therefore a
key question. Ultimately, one might consider a variational principle which
dictates an optimal choice (cf. Appendix~\ref{appx:energy}). At the present 
stage however, the guiding principles are usually physical intuition based on 
the quantum chemistry of the investigated
material, as well as practical considerations. Many early
implementations of the LDA+DMFT approach have used a linear muffin-tin 
orbital~\cite{andersen_lmto_1975_prb} (LMTO) basis for the correlated orbitals
(e.g. Refs.~[\onlinecite{lichtenstein_lda+dmft_1998,bie04}]). This is very 
natural since
in this framework it is easy to select the correlated subspace $\mathcal{C}$  
regarding the orbital character of the basis functions: e.g., $d$ character 
in a transition-metal oxide, $f$ character in rare-earth materials, etc.. The 
index $m$ then runs over the symmetry-adapted basis functions (or possibly the 
``heads'' of these LMTOs) corresponding to this selected orbital character. 
Exploring
other choices based on different Wannier constructions is the purpose of the
present paper. In this context, the index $m$ should be understood as a mere
label of the orbitals spanning the correlated subset. For simplicity, we shall
assume in the following that the correlated orbitals form an orthonormal set:
$\langle\chi_{\bR m}^{\hfill}|\chi_{\bR' m'}^{\hfill}\rangle$=$\delta_{\bR\bR'}\delta_{mm'}$. This
may not be an optimal choice for the DMFT approximation however, which is
better when interactions are more local. Generalization to non-orthogonal sets
is yet straightforward by introducing an overlap matrix (see e.g.
Ref.~[\onlinecite{kotliar_review}]).

\subsubsection{Local observables}

There are two central observables in the LDA+DMFT approach to electronic 
structure.
The first, as in DFT, is the total electronic charge density $\rho(\br)$. The 
second is the local one-particle Green's function $\bG_{\rm{loc}}(\iomn)$ 
projected onto $\mathcal{C}$, with components $G_{\bR m,\bR m'}(\iomn)$.
Both quantities are related to the full Green's function of the solid 
$G(\br,\br';\iomn)$ by:
\begin{eqnarray}
\label{glocdef}
\rho(\br)\hspace{-0.1cm}&=&\hspace{-0.1cm}
\frac{1}{\beta}\sum_n G(\br,\br;\iomn)\,\mbox{e}^{\iomn 0^+}\\
G_{mm'}^{\rm{loc}}(\iomn)\hspace{-0.1cm}&=&\hspace{-0.1cm}
\iint\hspace{-0.1cm} d\br d\br' \chi_{\bR m}^*(\brR)
\chi_{\bR m'}^{\hfill}(\brrR)\,G(\br,\br';\iomn)\,.\nonumber
\end{eqnarray}
The last expression can be abbreviated as a projection of the full Green's
function operator $\hat{G}$ according to
\begin{equation}
\hat{G}_{\rm{loc}} = \hat{P}_{\bR}^{(\mathcal{C})}\,\hat{G}\,
\hat{P}_{\bR}^{(\mathcal{C})}\,.
\end{equation}
In these expressions, we have used (for convenience) the Matsubara
finite-temperature formalism, with $\omega_n$=$(2n+1)\pi/\beta$ and
$\beta$=$1/k_{\rm{B}}T$. The Matsubara frequencies are related via
Fourier transformation to the imaginary times $\tau$. Note that the factor
e$^{\iomn 0^+}$ in (\ref{glocdef}) ensures the convergence of the Matsubara sum
which otherwise falls of as $1/\omega_n$. We have assumed, for simplicity, that
there is only one inequivalent correlated atom in the unit cell, so that $\Gloc$
does not carry an atom index (generalization is straightforward). In the
following we will drop the index $\bR$ if not explicitly needed.

Taking the KS Green's function $\hat{G}_{\rm{KS}}$ as a
reference, the full Green's function of the solid can be written
in operator form, as:
$\hat{G}^{-1}$=$\hat{G}^{-1}_{\rm{KS}}-\Delta\hat{\Sigma}$, or more 
explicitly (atomic units are used throughout this paper):
\begin{equation}
G(\br,\br';\iomn) = \left\langle\br\left|
\left[\iomn+\mu+\frac{\nabla^2}{2}-\hat{V}_{\rm{KS}}-
\Delta\hat{\Sigma}\right]^{-1}\right|\br'\right\rangle\,.
\label{eq:def_Gsolid}
\end{equation}
Here $\mu$ is the chemical potential and $V_{\rm{KS}}$ the KS effective
potential, which reads:
\begin{eqnarray}
V_{\rm{KS}}(\br)&=&-\sum_{i}\frac{Q_i}{|\br-\mathbf{N}_i|}+
\int\hspace{-0.1cm} d\br' \frac{\rho(\br')}{|\br-\br'|}
+ \frac{\delta E_{\rm{xc}}}{\delta\rho(\br)}\nonumber\\
&=&V_{\rm{ext}}(\br)+V_{\rm{H}}(\br)+V_{\rm{xc}}(\br)\,,
\label{kspot}
\end{eqnarray}
where $Q$, $\mathbf{N}$ are the charges, lattice vectors of the atomic nuclei, 
$i$ runs over the lattice sites, $V_{\rm{ext}}$ is 
the external potential due to the nuclei, $V_{\rm{H}}$ denotes the Hartree 
potential and $V_{\rm{xc}}$ is the exchange-correlation potential, obtained from a
functional derivative of the exchange-correlation energy $E_{\rm{xc}}$.
For the latter, the LDA (or generalized-gradient approximations) may be used. 
Recall that $V_{\rm{KS}}(\br)$ is determined by the true self-consistent
electronic charge density given by Eq.~(\ref{glocdef}) (which is modified by 
correlation effects, and hence differs in general from its LDA value, see below).

The operator$\Delta\hat{\Sigma}$ in Eq. (\ref{eq:def_Gsolid})
describes a many-body (frequency-dependent) self-energy correction.
In the DMFT approach, this self-energy correction is constructed in two steps.
First, $\mathbf{\Delta\Sigma}$ is derived
from an effective local problem~\cite{georges_kotliar_dmft}
(or ``effective quantum impurity model'') within the correlated subspace 
$\mathcal{C}$ via:
\begin{equation}
\Delta\Sigma_{mm'}(\iomn)\equiv\Sigma^{\rm{imp}}_{mm'}(\iomn)-
\Sigma^{\rm{dc}}_{mm'}\,,
\label{deltasigsub}
\end{equation}
whereby $\bsig_{\rm{dc}}$ is a double-counting term that corrects for correlation
effects already included in conventional DFT. The self-energy correction to be
used in (\ref{eq:def_Gsolid}), and subsequently in 
(\ref{impeq},\ref{eq:local_density}), is then obtained by promoting 
(\ref{deltasigsub}) to the lattice, i.e.,
\begin{eqnarray}
&&\hspace{-0.8cm}\Delta\Sigma(\br,\br';\iomn)\equiv\nonumber\\
&&\hspace{-0.7cm}
\sum_{\bT mm'}\chi^*_m(\brRT)\chi_{m'}^{\hfill}(\brrRT)\,
\Delta\Sigma_{mm'}(\iomn)\,,
\label{eq:upfold_Sigma}
\end{eqnarray}
where $\bT$ denotes a direct lattice vector. The key approximation is that the 
self-energy correction is non-zero only inside the (lattice-translated) 
correlated subspace, i.e., 
$\Delta\hat{\Sigma}$=$\Delta\hat{\Sigma}^{(\mathcal{C})}$, hence exhibits only 
on-site components in the chosen orbital set.
\begin{figure*}[t]
\includegraphics*[width=17.75cm]{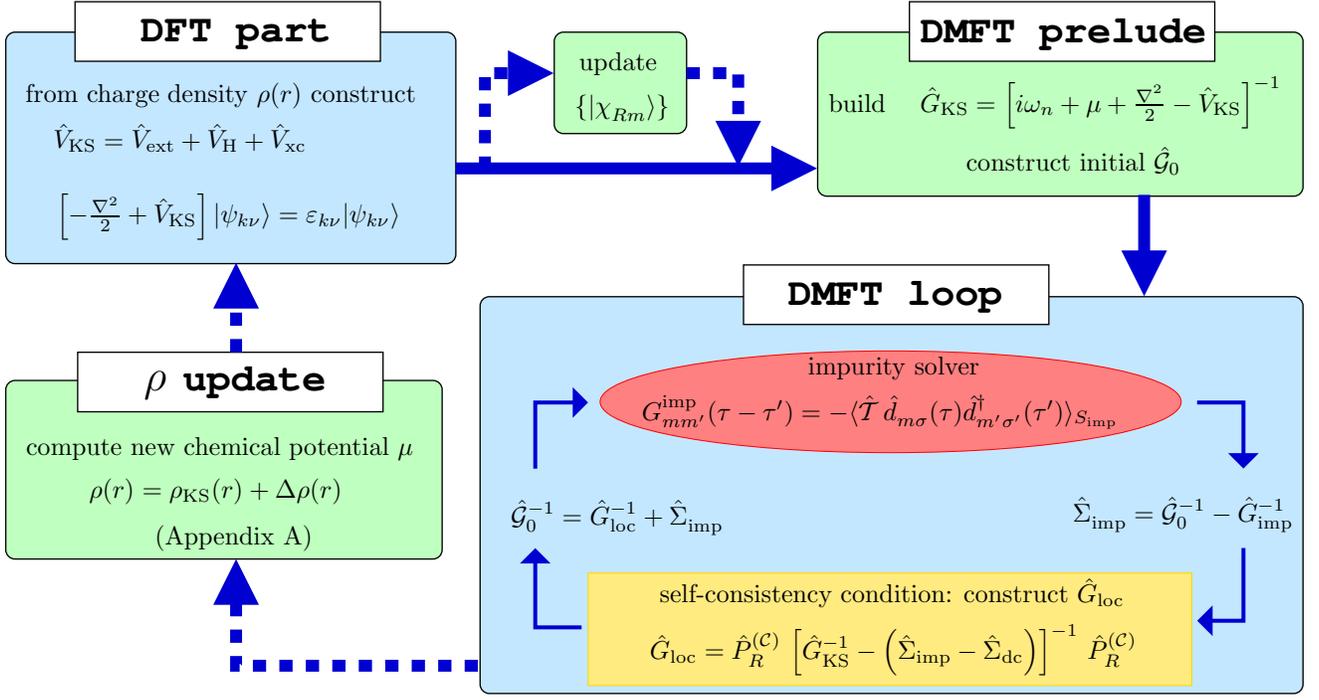}
\caption{Complete self-consistency loop for LDA+DMFT. The charge density $\rho$
determines the KS potential $V_{\rm{KS}}$, from which KS eigenvalues
$\varepsilon_{\bk\nu}$ and eigenfunctions $\psiknu$ follow.
The KS Green's function is then constructed and passed on to the DMFT cycle
(in practice, the KS Hamiltonian $\bH_{\rm{KS}}$ is constructed in the
basis set used to implement the method, and transferred to the DMFT cycle).
The DMFT loop consists in i) solving the effective impurity problem for the
impurity Green's function, hence obtaining an impurity self-energy,
ii) combining the self-energy correction with the KS Green's function in
order to obtain the local Green's function $\bG_{\rm{loc}}$ projected in the 
correlated subset and iii) obtaining an updated Weiss mean-field.
An initial guess for the Weiss dynamical mean-field must be made at the
beginning of the DMFT loop, e.g. by choosing $\hat{{\cal G}}_0^{\rm{init}}=
\hat{P}^{(\mathcal{C})}\hat{G}_{\rm{KS}}\hat{P}^{(\mathcal{C})}$. Is the
DMFT loop converged, the chemical potential is updated in order to ensure
global charge neutrality, and the new charge density (including many-body
effects) is constructed (described in Appendix.~\ref{csc}). This new density 
determines a new KS potential. Note that in addition
one may want to update the set $\{|\chi_m^{\hfill}\rangle\}$ when preparing for
the next DMFT loop (cf. Appendix~\ref{appx:energy}).
The whole process must be iterated until the charge density, the impurity 
self-energy and the chemical potential are converged. In practice, good 
convergence of the DMFT loop is reached before a new $\rho$ is calculated.
Note that in the present paper using a Wannier implementation, the global 
self-consistency on the charge density is not implemented in practice. Thus the
self-consistent LDA Hamiltonian $\bH_{\rm{KS}}$ enters the DMFT loop, which is 
iterated until convergence of the self-energy.
\label{ddloop}}
\end{figure*}
\subsubsection{Effective quantum impurity problem}

The local impurity problem can be viewed as an effective atom
involving the correlated orbitals, coupled to a self-consistent energy-dependent bath.
It can be formulated either in Hamiltonian form, by explicitly introducing the
degrees of freedom of the effective bath, or as an effective action in which the
bath degrees of freedom have been integrated out. In the latter formulation, the
action of the effective impurity model reads:
\begin{eqnarray}
&&\hspace{-0.4cm}S_{\rm{imp}}=-\iint_0^{\beta}\hspace{-0.2cm}d\tau d\tau'
\hspace{-0.15cm}\sum_{mm'\sigma}d^{\dagger}_{m\sigma}(\tau)
[\mathcal{G}_{0}^{-1}]_{mm'}(\tau-\tau')
d^{\hfill}_{m'\sigma}(\tau')\nonumber\\
&&\hspace{1cm}+\int_{0}^{\beta}\hspace{-0.2cm}d\tau\,
H_U\left(\{d^{\dagger}_{m\sigma};d_{m\sigma}\}\right)\quad.
\label{eq:impurity_action}
\end{eqnarray}
In this expression, $d^\dagger_{m\sigma}$, $d^{\hfill}_{m\sigma}$ are the
Grassmann variables corresponding to orbital $\chi^{\hfill}_{ m}$ for spin
$\sigma$, $\hat{H}_U$ is a many-body interaction (to be discussed in section
\ref{ldadmftimp}), and $\hat{\mathcal{G}}_{0}(\iomn)$ is the dynamical
mean-field, determined self-consistently (see below), which encodes the
coupling of the embedded atom to the effective bath. This quantity is the
natural generalization to quantum many-body problems of the Weiss mean-field of
classical statistical mechanics. Its frequency dependence is the essential 
feature which renders DMFT distinct from static approaches such as e.g. the
LDA+U method~\cite{anisimov_lda+u_1991_prb}. The frequency dependence
allows for the inlcusion of all (local) quantum fluctuations. Thereby the
relevant (possibly multiple) energy scales are properly taken into account, 
as well as the description of the transfers of spectral weight.
One should note that the dynamical mean-field $\hat{\mathcal{G}}_{0}(\iomn)$
formally appears as the bare propagator in the definition of the effective
action for the impurity (\ref{eq:impurity_action}). However, its actual value is 
only known after iteration of a self-consistency cycle (detailed below) and 
hence depends on many-body effects for the material under consideration.

The self-energy correction is obtained from the impurity model as:
\begin{equation}
\bsig_{\rm{imp}}(\iomn)\equiv 
\mbox{\boldmath$\mathcal{G}$\unboldmath}_{0}^{-1}(\iomn)-
\bG_{\rm{imp}}^{-1}(\iomn)\quad.
\label{eq:def_Sigma_imp}
\end{equation}
in which $\bG_{\rm{imp}}$ is the impurity model Green's function, associated with
the effective action (\ref{eq:impurity_action}) and defined as:
\begin{equation}
G^{\rm{imp}}_{mm'}(\tau-\tau') \equiv
- \langle\hat{\mathcal{T}}\,\hat{d}^{\hfill}_{m\sigma}(\tau)
\hat{d}^\dagger_{m'\sigma'}(\tau')\rangle_{S_{\rm{imp}}}\quad,
\label{eq:def_Gimp}
\end{equation}
where $\mathcal{T}$ stands for time-ordering.
Note that computing this Green's function, given a specific Weiss dynamical 
mean-field $\hat{\mathcal{G}}_{0}(\iomn)$ is in fact the most demanding step in 
the solution of the DMFT equations.

\subsubsection{Self-consistency conditions\label{selfcons}}

In order to have a full set of self-consistent equations, one still needs to 
relate the effective impurity problem  to the whole solid. Obviously, the 
dynamical mean-field
\boldmath$\mathcal{G}$\unboldmath$_0(\iomn)$ is the relevant link, but we have not
yet specified how to determine it. The central point in DMFT is to evaluate
\boldmath$\mathcal{G}$\unboldmath$_0(\iomn)$ in a self-consistent manner by
requesting that the impurity Green's function coincides with the components of the
lattice Green's function projected onto the correlated subspace $\mathcal{C}$, 
namely that:
\begin{equation}
\bG_{\rm{imp}}=\bG_{\rm{loc}}\quad,
\label{eq:Gloc_scc}
\end{equation}
or, in explicit form, using (\ref{glocdef},\ref{eq:def_Gsolid},\ref{deltasigsub})
and (\ref{eq:upfold_Sigma}):
\begin{eqnarray}
&&\hspace{-0.3cm}G^{\rm{imp}}_{mm'}(\iomn)=\iint d\br d\br'
\chi_m^*(\brR)\chi_{m'}^{\hfill}(\brrR)\times\label{impeq}\\
&&\hspace{-0.3cm}\times\left\langle\br\left|\left[
\iomn+\mu+\frac{\nabla^2}{2}-\hat{V}_{\rm{KS}}-
\left(\hat{\Sigma}_{\rm{imp}}-\hat{\Sigma}_{\rm{dc}}\right)\right]^{-1}
\right|\br'\right\rangle\,.
\nonumber
\end{eqnarray}
In this representation, it is clear that the self-consistency condition 
involves only impurity quantities, and therefore yields a relation between the 
dynamical mean-field $\hat{{\cal G}}_0$ and $\Gimp$ which, together with the 
solution of the impurity problem 
Eqs. (\ref{eq:impurity_action}-\ref{eq:def_Gimp}) fully determines both 
quantities in a self-consistent way.

The effective impurity problem (\ref{eq:impurity_action}) can in fact be thought
of as a {\it reference system} allowing one to represent the local Green's
function. This notion of a reference system is analogous to the KS
construction, in which the charge density is represented as the solution of a
single-electron problem in an effective potential (with the difference that here,
the reference system is an interacting one).

Finally, by combining (\ref{glocdef}) and (\ref{eq:def_Gsolid}) the electronic
charge density is related to the KS potential by:
\begin{equation}
\rho(\br) = \frac{1}{\beta}\sum_n
\left\langle\br\left|\left[
\iomn+\mu+\frac{\nabla^2}{2}-\hat{V}_{\rm{KS}}-\Delta\hat{\Sigma}
\right]^{-1}\right|\br\right\rangle\,\mbox{e}^{\iomn 0^+}\,,
\label{eq:local_density}
\end{equation}

Expression (\ref{eq:local_density}) calls for two remarks. Firstly, many-body
effects in $\mathcal{C}$ affect via $\mathbf{\Delta\Sigma}$ the
determination of the charge density, which will thus differ at self-consistency 
from its LDA value. Secondly, the familiar KS representation of $\rho(\br)$ in
terms of virtually independent electrons in an effective static potential is 
modified in LDA+DMFT in favor of a non-local and energy-dependent (retarded) 
potential given by
$V_{\rm{KS}}(\br)\delta(\br-~\br')\delta(\tau-\tau')+\Delta\Sigma(\br,\br';\tau-\tau')$.
In Appendix~\ref{csc}, we give a more detailed discussion of the technical
aspects involved in calculating the charge density from expression 
(\ref{eq:local_density}). However, we have not yet implemented this calculation 
in practice in our Wannier-based code: the computations presented in this paper 
are performed for the converged $\rho(\br)$ obtained at the LDA level.
Finally, let us mention that the LDA+DMFT formalism and equations presented
above can be derived from a (free-energy) functional~\cite{sav04} of both the 
charge density and the projected local Green's function, 
$\Gamma[\rho,\bG_{\rm{loc}}]$. This is reviewed in Appendix~\ref{appx:energy}, 
where the corresponding formula for the total energy is also discussed.

Fig.~\ref{ddloop} gives a synthetic overview of the key steps involved in
performing a fully self-consistent LDA+DMFT calculation, irrespective of the 
specific basis set and band-structure code chosen to implement the method.

\subsubsection{Double-counting correction}

We briefly want to comment on the double-counting (DC) correction term.
Since electronic correlations are already partially taken into account within 
the DFT approach through the LDA/GGA exchange-correlation potential,
the double-counting correction $\bsig_{\rm{dc}}$ has to correct for this
in LDA+DMFT. The problem of precisely defining DC is hard to 
solve in the framework of conventional DFT~\cite{czy94,pet03}. Indeed,
DFT is not an orbitally-resolved theory and furthermore the LDA/GGA 
does not have a diagrammatic interpretation (like simple Hartree-Fock)
which would allow to subtract the corresponding terms from the DMFT
many-body correction.
Simply substracting the matrix elements of $V_{\rm{H}}$ and $V_{\rm{xc}}$ 
in the correlated orbital subset $\mathcal{C}$ from the KS Green's function to 
which the many-body self-energy is applied to is not a physically reasonable
strategy. Indeed, the DMFT approach (with a static, frequency-independent
Hubbard interaction) is meant to treat the low-energy, screened interaction,
so that the Hartree approximation is not an appropriate starting point.
Instead, one wants to benefit from the spatially-resolved screening effects 
which are already partially captured at the LDA level. In practice, the 
DC terms introduced for LDA+U, i.e., ``fully-localized limit''~\cite{ani93} and
``around mean field''~\cite{anisimov_lda+u_1991_prb,czy94}, appear to
be reasonable also in the LDA+DMFT framework. It was recently 
shown~\cite{ani06}, that the fully-localized-limit form can be derived from 
the demand for discontinuity of the DFT exchange-correlation potential at 
integer filling.

The DC issue in fact has a better chance to be resolved in a satisfactory manner,
from both the physical and formal points of view, when the concept of
local interaction parameters is extended to frequency-dependent
quantities (e.g. a frequency-dependent Hubbard interaction $U(\omega)$),
varying from the bare unscreened value at high frequency to a
screened value at low energy, and determined from first principles.
The GW+DMFT construction, and the extended DMFT framework,
in which this quantity plays a central role and is determined self-consistently on
the same footing as the one-particle dynamical mean field, may prove
to be a fruitful approach in this respect.

\subsubsection{Implementation: choice of basis sets and Hilbert spaces}

In the previous sections, care has been taken to write the basic equations
of the LDA+DMFT formalism in a {\it basis-independent} manner.
In this section, we express these equations in a general basis set, which is 
essential for practical implementations, and
discuss advantages and drawbacks of different choices for the basis set.
At this point, a word of caution is in order: it is important to clearly 
distinguish between the set of {\it local orbitals} $\{|\chi_{ m}\rangle\}$ 
which specifies the correlated subspace, and the {\it basis functions} which one 
will have to use in order to implement the method in practice within a 
electronic-structure code.
Different choices for $\{|\chi_{ m}\rangle\}$ will lead to different results, 
since DMFT involves a local approximation which has a different degree of 
accuracy in diverse orbital sets. In contrast, once the correlated orbital set 
is fixed, any choice of basis set can be used in order to implement the method, 
with in principle identical results.

Let us denote a general basis set by $\{|\Bka\rangle\}$, in which ${\bf k}$ runs
over the Brillouin zone (BZ) and $\alpha$ is a label for the basis functions.
For example if the KS (Bloch) wave functions are used as a basis set,
$\alpha$=$\nu$ is a band index and the basis functions are
$\psi_{\bk\nu}(\br)$=e$^{i{\bf k}.\br}u_{\bk\nu}(\br)$. In the case of a pure
plane-wave basis set, $\alpha$=$\mathbf{G}$ runs over reciprocal lattice vectors.
For a Bloch sum of LMTOs $\alpha$=$(\bR lm)$ runs over sites in the primitive 
cell and orbital (angular momentum) quantum number. For hybrid basis sets one may 
have $\alpha$=$(\bR lm,\mathbf{G})$. As an example for the latter serves the
linear augmented-plane-wave~\cite{andersen_lmto_1975_prb} (LAPW) basis set, 
though here in the end it is summed over the orbital indices and the
basis is finally labelled by $\mathbf{G}$ only.

Consider now the DMFT self-consistency condition (\ref{eq:Gloc_scc}).
In the (yet arbitrary) basis set $\{|\Bka\rangle\}$, its explicit expression in
reciprocal space reads (correct normalization of the $k$-sum is understood):
\begin{eqnarray}\label{eq:scc_basis}
G^{\rm{imp}}_{mm'}(\iomn) &=&
\sum_{\bk}\sum_{\alpha\alpha'}
\langle\chi_{m}^{\bk}|\Bka\rangle\langle\Bkap|
\chi_{ m'}^{\bk}\rangle 
\times \nonumber \\
&&\hspace{-1.5cm}\times\left\{\left[\iomn+\mu-\bH_{\rm{KS}}(\bk)-
\mathbf{\Delta\Sigma}(\bk,\iomn)\right]^{-1}\right\}_{\alpha\alpha'}\,,
\end{eqnarray}
where $|\chi_{m}^{\bk}\rangle$=
$\sum_{\bT}\mbox{e}^{i\bk\cdot(\bT+\bR)}|\chi_m^{\hfill}\rangle$ denotes the Bloch 
transform of the local orbitals. In this expression, $\hks(\bk)$ is the KS 
Hamiltonian at a given $k$-point, expressed in the basis set of interest:
\begin{equation}
\HKS(\bk)=\sum_{\alpha\alpha'}|\Bka\rangle\langle\Bkap|
\left[\sum_\nu\varepsilon_{\bk\nu}\langle\Bka|\psi_{\bk\nu}
\rangle\langle\psi_{\bk\nu}|\Bkap\rangle\right]\,,
\end{equation}
with $\{\varepsilon_{\bk\nu},\psi_{\bk\nu}\}$ the set of KS eigenvalues and wave
functions:
\begin{equation}
\left[-\frac{\nabla^2}{2}+\hat{V}_{\rm{KS}}\right]|\psi_{\bk\nu}\rangle
=\varepsilon_{\bk\nu}|\psi_{\bk\nu}\rangle
\label{eq:KS_schro}
\end{equation}
The self-energy correction, in the chosen basis set, reads:
\begin{eqnarray}
\Delta\Sigma_{\alpha\alpha'}(\bk,\iomn)&=&
\sum_{mm'} \langle\Bka|\chi_{m}^{\bk}\rangle\langle\chi_{m'}^{\bk}|
\Bkap\rangle\times\nonumber\\
&&\hspace{-1cm}\times\left[\Sigma_{mm'}^{\rm{imp}}(\iomn)-
\Sigma_{mm'}^{\rm{dc}}\right]
\end{eqnarray}
and it should be noted that, although purely local when expressed in the set of
correlated orbitals, it acquires in general momentum-dependence when expressed
in an arbitrary basis set.

The self-consistency condition (\ref{eq:scc_basis}) is a central step in
interfacing DMFT with a chosen band-structure method. Given a charge
density $\rho(\br)$, the effective potential $V_{\rm{KS}}(\br)$ is constructed,
and the corresponding KS equations (\ref{eq:KS_schro}) are solved
(Fig.~\ref{ddloop}), in a
manner which depends on the band-structure method. Each specific technique makes
use of a specific basis set $\{|\Bka\rangle\}$. The KS Hamiltonian serves as an
input to the DMFT calculation for $\mathcal{C}$, which is used
in (\ref{eq:scc_basis}) to recalculate a new local Green's function from
the impurity self-energy, and hence a new dynamical mean-field from
\boldmath$\mathcal{G}$\unboldmath$_0^{-1}$=$\bG_{\rm{loc}}^{-1}+\bsig_{\rm{imp}}$.

A remark which is important for practical implementation must now be made.
Although $\bG_{\rm{loc}}(\iomn)$, i.e., the right-hand side
of (\ref{eq:scc_basis}), can be
evaluated in principle within any basis set $\{|\Bka\rangle\}$,
the computational effort may vary dramatically
depending on the number $N_B$ of basis functions in the set.
According to (\ref{eq:scc_basis}), this computation involves an inversion of a
$N_B$$\times$$N_B$ matrix {\it at each $k$-point and at each
frequency} $\iomn$, followed by a summation over $k$-points for each frequency.
Since the number of discrete frequencies is usually of the
order of a few thousands, this procedure is surely feasible within a minimal
basis set such as, e.g., LMTOs. In the latter case, the correlated orbitals may
furthermore be chosen as a specific subset of the basis functions (e.g. with $d$
character in a transition-metal oxide) -or possibly as the normalized ``heads''
corresponding to this subset-, making such basis sets quite naturally
tailored to the problem.
In contrast, computational efficiency is harder to reach for
plane-wave like basis sets in the LDA+DMFT context. For such large basis sets,
the frequency dependence substantially increases the already large numerical 
effort involved in static schemes such as LDA or LDA+U.
Furthermore, another more physical issue in using plane-wave based
codes in the DMFT context is how to choose the local orbitals
$\{|\chi_m\rangle\}$ which define the correlated subset. Because the
free-electron like basis functions usually do not have a direct physical
connection to the quantum chemistry of the material problem at hand, these
orbitals must be chosen quite independently from the basis set itself.

To summarize, when implementing LDA+DMFT in practice, a decision must be made on 
the following two issues:

-i) The first issue is a physical one, namely how to choose the local orbitals
$\chi_{ m}$ spanning the correlated subspace $\mathcal{C}$. The quality of 
the DMFT approximation will in general depend on the choice of
$\mathcal{C}$, and different choices may lead to different results. Obviously, one
would like to pick $\mathcal{C}$ in such a way that the DMFT approximation is
better justified, which is intuitively associated with well-localized orbitals.

-ii) The second point is a technical, albeit important, one. It is the choice of
basis functions $\{|\Bka\rangle\}$ used for implementing the self-consistency
condition (\ref{eq:scc_basis}). As discussed above and as clear from
(\ref{eq:scc_basis}), computational efficiency requires that as many matrix 
elements $\langle\Bka|\chi_{ m}^{\bk}\rangle$ as possible are zero 
(or very small),  i.e., such that $\chi_{ m}^{\hfill}$ has overlap with only 
few basis functions.

As discussed above, both issues demand particular attention when using 
band-structure methods based on plane-wave techniques, because those
methods do not come with an obvious choice for the orbitals 
$\chi_{ m}^{\hfill}$ and because the demand for well-localized 
$\chi_{ m}^{\hfill}$ implies that they will 
overlap with a very large number of plane waves.

In this paper, we explore the use of WFs as an elegant way of 
addressing both issues i) and ii), leading to a convenient and efficient 
interfacing of DMFT with any kind of band-structure method.

\subsection{Wannier functions and DMFT}

\subsubsection{General framework and Wannier basics}

Let us outline the general strategy that may be used for implementing
LDA+DMFT using Wannier functions (WFs), postponing technical details to
later in this section. First, it is important to realize that a Wannier 
construction needs not be applied to all Bloch bands spanning the full Hilbert 
space, but only to a smaller set $\mathcal{W}$ corresponding to a certain 
energy range, defining a subset of valence bands relevant to the material 
under consideration.
To be concrete, in a transition-metal oxide for example, it may be advisable to
keep bands with oxygen $2p$ and transition-metal $3d$ character in the valence 
set $\mathcal{W}$. WFs spanning the set $\mathcal{W}$ may be 
obtained by performing a ($\bk$-dependent) unitary transformation on the selected 
set of Bloch functions. This unitary transformation should ensure a 
strongly localized character of the emerging WFs.
Among the localized WFs spanning $\mathcal{W}$, a subset is 
selected which defines the correlated subspace 
$\mathcal{C}$$\subseteq$$\mathcal{W}$. For transition-metal oxides, 
$\mathcal{C}$ will in general correspond to the WFs with $d$ character.

The correlated orbitals $\chi_{ m}^{\hfill}$ are thus identified with a certain
set of WFs generating $\mathcal{C}$. It is then recommendable (albeit not 
compulsory) to choose the (in general larger) set of WFs generating the valence 
set $\mathcal{W}$, as basis functions in which
to implement the self-consistency condition (\ref{eq:scc_basis}).
Indeed, the KS Hamiltonian can then be written as a matrix with
diagonal entries corresponding to Bloch bands outside $\mathcal{W}$, and
only one non-diagonal block corresponding to $\mathcal{W}$. It follows that
the self-consistency condition (\ref{eq:scc_basis}) may be expressed in
a form which involves only the knowledge of the KS Hamiltonian
within $\mathcal{W}$ and requires only a matrix inversion
within this subspace, as detailed below.
Hence, using WFs is an elegant answer to both points i) and ii) above: 
it allows to build correlated orbitals defining the set $\mathcal{C}$ with 
tailored localization properties, and by construction only the matrix elements 
$\langle\chi_{ m}^{\hfill}|w_{\alpha}\rangle$ with
$w_{\alpha}$ a WF in the set $\mathcal{W}$ are non-zero.

We now describe in more details how WFs are constructed.
Within the Born-von K\'arm\'an periodic boundary conditions, the effective
single-particle description of the electronic structure is usually based on
extended Bloch functions $\psiknu$, which are classified with two quantum
numbers, the  band index $\nu$ and the crystal momentum $\bk$. An alternative
description can be derived in terms of localized WFs~\cite{wan37},
which are defined in real space via an unitary transformation performed on the
Bloch functions. They are also labeled with two quantum numbers: the 
index $\alpha$ which describes orbital character and position, as well as the 
direct lattice vector $\bT$, indicating the unit cell they belong to. The relation
between WFs and Bloch functions can be considered as the generalization to solids 
of the relation between ``Boys orbitals''\cite{boy60} and localized molecular
orbitals for finite systems. It is crucial to realize, that the unitary 
transformation is not unique.
In the case of an isolated band in one dimension, this was emphasized long ago
by W.~Kohn~\cite{koh59}. He stated that infinitely many WFs can be constructed by
introducing a $\bk$-dependent phase $\varphi(\bk)$,
yet there is only {\it one} real high-symmetry WF that falls off exponentially. 
Hence in general $\varphi(\bk)$ may be optimized in order to improve the spatial 
localization of the WF in realistic cases. This observation was generalized and 
put in practice for a group of several bands in Ref.~[\onlinecite{mar97}]. 

Let us consider the previously defined group ${\cal W}$ of bands of interest.
A general set of WFs corresponding to this group can be constructed
as~\cite{mar97}
\begin{equation}
w_\alpha(\brT)=\frac{V}{(2\pi)^3}\int_{\rm{BZ}}\hspace{-0.2cm}d\bk\,
\mbox{e}^{-i\bk\cdot\bT}\sum_{\nu\in\,{\cal W}} U_{\alpha\nu}^{(\bk)}\,
\psiknu(\br)\,,\label{wann}
\end{equation}
$V$ denoting the volume of the primitive cell. The WF 
$\langle\br|w_{\bT\alpha}\rangle$ only depends on $\br$$-$$\bT$, since
$\psiknu(\br)$=e$^{i\bk\cdot\br}u_{\bk\nu}(\br)$, with $u_{\bk\nu}$ a periodic
function on the lattice. The unitary matrix $U_{\alpha\nu}^{(\bk)}$ reflects the
fact that, in addition to the gauge freedom with respect to a $\bk$-dependent
phase, there is the possibility of unitary mixing of several crystal wave
functions in the determination of a desired WF. Optimization of
these degrees of freedom allows one to enforce certain properties on the WFs,
including the demand for maximal localization (see next paragraph). Of
course, the extent of the WF still depends on the specific material problem.
Due to the orthonormality of the Bloch functions, the WFs also form an 
orthonormal basis:
$\langle w_{\bT\alpha}|w_{\bT'\alpha'}\rangle$=
$\delta_{\bT\bT'}\delta_{\alpha\alpha'}$. More on the general properties and
specific details of these functions may be found in the original
literature~\cite{wan37,koh59,blo62,clo63}, or 
Refs.~[\onlinecite{mar97,sou01,ani05}] and references therein.

Here, LDA+DMFT will be implemented by selecting a certain subset
$\{w_{m}, m\in {\cal C}\}$ of the WFs
$\{w_{\alpha}, \alpha\in {\cal W}\}$ as generating the correlated subset. 
Thus we directly identify $\{|\chi_{ m}^{\hfill}\rangle\}$ with a specific 
set of WFs. Note again that this is a certain {\sl choice}, and that other 
choices are possible (such as identifying $\{|\chi_{ m}^{\hfill}\rangle\}$ 
from only parts of the full WFs through a projection). With our choice, 
the functions
\begin{equation}
|w_{\bk\alpha}\rangle\equiv\sum_{{\mathbf T}} \mbox{e}^{i\bk\cdot\bT}
|w_{\bT\alpha}\rangle = \sum_{\nu\in\,{\cal W}} U_{\alpha\nu}^{(\bk)}
|\psiknu\rangle\,,
\label{wanneq}
\end{equation}
will be used in order to express the KS Hamiltonian and to implement
the self-consistency condition (\ref{eq:scc_basis}).
Because the unitary transformation acts only inside $\mathcal{W}$,
only the block of the KS Hamiltonian corresponding to this subspace needs
to be considered when implementing the self-consistency condition, hence
leading to a quite economical and well-defined implementation. The KS 
Hamiltonian in the space $\mathcal{W}$ reads:
\begin{eqnarray}
&&\hat{H}_{\rm{KS}}^{({\cal W})}(\bk) =
\sum_{\alpha\alpha'\in\,{\cal W}} H_{\alpha\alpha'}(\bk)|w_{\bk\alpha}\rangle
\langle w_{\bk\alpha'}|\,,\nonumber \\
&&H_{\alpha\alpha'}(\bk)=\sum_{\nu\in\,{\cal W}}\varepsilon_{\bk\nu}
U_{\alpha\nu}^{(\bk)*}U_{\alpha'\nu}^{(\bk)}
\end{eqnarray}
while the self-energy correction reads:
\begin{equation}
\DSig^{({\cal C})} =
\sum_{mm'\in\,{\cal C}}[\Sigma^{\rm{imp}}(\iomn)-\Sigma^{\rm{dc}}]_{mm'}
\sum_\bk |w_{\bk m}\rangle\langle w_{\bk m'}|\,.
\end{equation}
Accordingly, the DMFT self-consistency condition takes the form:
\begin{eqnarray}\label{gimpwann}
&&\hspace{-0.4cm}G^{\rm{imp}}_{mm'}(\iomn)=\\
&&\sum_{\bk}\left\{\left[(i\omega_n+\mu)\mathbf{\openone}
-\hks^{({\cal W})}(\bk)-\mathbf{\Delta\Sigma}^{({\cal C})}(\iomn)\right]^{-1}
\right\}_{mm'}\nonumber
\end{eqnarray}
In this expression, the matrix inversion has to be done for the full
$\mathcal{W}$-space matrix, while only the block corresponding to
$\mathcal{C}$ has to be summed over $\bk$ in order to produce the
local Green's function $\mathbf{G}_{\rm{loc}}$ in the correlated subspace,
i.e. the r.h.s of (\ref{gimpwann}).
In practice, the latter is inverted and added to $\mathbf{\Sigma}_{\rm{imp}}$
in order to produce an updated dynamical mean-field according to:
\boldmath$\mathcal{G}$\unboldmath$_0^{-1}$=$\bG_{\rm{loc}}^{-1}+\bsig_{\rm{imp}}$.
This new dynamical mean-field is injected into the impurity solver, and the 
iteration of this DMFT loop leads to a converged solution of (\ref{gimpwann})
(cf.~Fig.~\ref{ddloop}).

In all the above, we have been careful to distinguish the (larger) space 
$\mathcal{W}$ in which the Wannier construction is performed, and the (smaller) 
subset $\mathcal{C}$ generated by the Wannier functions associated with 
correlated states. In some cases however, it may be possible to work within an 
energy window encompassing only the ``correlated'' bands (e.g when they are well 
separated from all other bands), and choose $\mathcal{W}$=$\mathcal{C}$.
This of course leads to more extended Wannier functions than when the Wannier 
construction is made in a larger energy window. For the two materials considered 
in this paper, we shall nonetheless adopt this ``massive downfolding'' route, and
work with $\mathcal{W}$=$\mathcal{C}$.
For the correlated perovskite \srvo3, the bands originating from the ligand 
orbitals are well separated from the transition-metal ones. In other words, the
size of the many-body interaction, say the Hubbard $U$, is expected to be 
significantly smaller than the former level separation. In that case the
minimal choice of a subset ${\cal W}$=${\cal C}$ involving only the $d$-like WFs 
of the $t_{2g}$ panel is quite natural (see below). The situation is more 
involved for the \bavs3 compound, since the S$(3p)$ bands are strongly entangled 
with the $t_{2g}$ bands. Despite this stronger hybridization, it is not expected
that the S($3p$)-V($3d$) level separation is the relevant energy scale, but 
still $U$. Hence we continue to concentrate on a disentangled 
$t_{2g}$-like panel, thereby integrating out explicit sulfur degrees of freedom. 
The resulting minimal basis is only ``Wannier-like'', but nonetheless should 
provide a meaningful description of the low-energy sector of this material.
It should be kept in mind however that the minimal choice 
$\mathcal{W}$=$\mathcal{C}$ may become a critical approximation at some point.
For late transition-metal oxides in particular, the fact that $p$- and 
$d$-like bands are rather close in energy almost certainly implies that 
$\mathcal{W}$ must retain O($2p$) states, as well as transition-metal $3d$ states 
(while $\mathcal{C}$ will involve the $3d$ states only)~\cite{zsa_1985_prl}.

\subsubsection{Maximally-localized Wannier functions}

The maximally-localized Wannier functions~\cite{mar97,sou01}
(MLWFs) are directly based on Eq. (\ref{wann}). In order to ensure a
maximally-localized Wannier(-like) basis, the unitary matrix
$U^{(\bk)}_{\alpha\nu}$
is obtained from a minimization of the sum of the quadratic spreads of the
Wannier probability distributions, defined as
\begin{equation}
\Omega\equiv\sum_{\alpha}\left(\langle r^2\rangle_{\alpha}-
\langle\br\rangle^2_{\alpha}\right)\,,
\,\,\langle O\rangle_{\alpha}=
\int\hspace{-0.1cm}d\br\hspace{0.1cm}O|w_{{\bf 0}\alpha}(\br)|^2\,.
\end{equation}
Thus the quantity $\Omega$ may be understood as a functional of the Wannier basis
set, i.e., $\Omega$=$\Omega[\{w_{\alpha}\}]$. Starting from some inital guess for
the Wannier basis, the formalism uses steepest-decent or conjugate-gradient
methods to optimize $U^{(\bk)}_{\alpha\nu}$. Thereby, the gradient of $\Omega$
is expressed in reciprocal space with the help of the overlap matrix
\begin{equation}
M_{\nu\nu'}^{(\bk,\bq)}=\langle
u_{\bk\nu}|u_{\bk+\bq\,\nu'}\rangle\quad,
\label{mmat}
\end{equation}
where $\bq$ is connecting $\bk$ vectors on a chosen mesh in reciprocal space.
Hence this scheme needs as an input the KS Bloch eigenfunctions $\psiknu$, or
rather their periodical part $u_{\bk\nu}$.
In the formalism, all relevant observables may be written in terms of 
$M_{\nu\nu'}^{(\bk,\bq)}$.
The resulting MLWFs turn out to be real functions, although there is
no available general proof for this property.

In the following, two cases of interest shall be separately discussed:

\paragraph{Bands of interest form a group of isolated bands.}
This is the case e.g for \srvo3 discussed in this paper.
The matrix $\bM^{(\bk,\bq)(0)}$ has to be initially calculated from the KS
Bloch eigenvectors $\psiknu$, where $\nu$ runs over the bands defining the 
isolated group. Starting from $U^{(\bk)}_{\alpha\nu}$ according to the initial 
Wannier guess, the unitary transformation matrix will be updated
iteratively~\cite{mar97}. Correspondingly, the $\bM$ matrices evolve as
\begin{equation}
\bM^{(\bk,\bq)}=\bU^{(\bk)\dagger}\bM^{(\bk,\bq)(0)}\bU^{(\bk+\bq)}\quad.
\end{equation}
The minimization procedure not only determines the individual spreads of the
WFs, but also their respective centers. Thus generally the centers do not
have to coincide with the lattice sites as in most tight-binding representations.
For instance, performing this Wannier construction for the four valence bands of 
silicon leads to WFs which are exactly centered in between the atoms along the 
bonding axes~\cite{mar97}.

\paragraph{Bands of interest are entangled with other bands.}
The handling of \bavs3 discussed in this paper falls into this category.
This case is not so straightforward, since before evaluating the MLWFs one has
to decide on the specific bands subject to the Wannier construction. Lets assume
there are $N_b$ target bands, e.g. a \t2g-like manifold, strongly hybridized with
$N_b'$ other bands of mainly different character, e.g. $s$- or $p$-like bands. 
Then first the matrix (\ref{mmat}) has to be calculated initially for the enlarged
set of $N_b$+$N_b'$ bands. Within the latter set, the orbital character
corresponding to the aimed at WFs may jump significantly. Thus
new {\sl effective} bands, associated with eigenvectors $\psiknut$, have to be
constructed in the energy window of interest according to a physically meaningful
description.

To this aim, the functional $\Omega[\{w_{\alpha}\}]$ was decomposed in
Ref.~[\onlinecite{sou01}] into two non-negative contributions, i.e.,
$\Omega$=$\Omega_I+\tilde{\Omega}$. Here $\Omega_I$ describes the
spillage~\cite{san95} of the WFs between different regions in reciprocal
space. The second part $\tilde{\Omega}$ measures to what extent the MLWFs fail
to be eigenfunctions of the band-projected position operators. In the case of
an isolated set of bands $\Omega_I$ is gauge invariant. However it plays a major
role in the case of entangled bands~\cite{sou01}, since here it may define a
guiding quantity for ``downfolding'' the maximally ($N_b$+$N_b'$)-dimensional 
Hilbert space at each $k$-point to a corresponding Hilbert space with maximal
dimension $N_b$. The reason for this is that an initial minimization of 
$\Omega_I$ provides effective target bands with the property of ``global 
smoothness of connection''~\cite{sou01}. Since $\Omega_I$ 
measures the spillage, minimizing it corresponds to choosing paths in reciprocal
space with minimal mismatch within the reduced set of $N_b$. In a second step
$\tilde{\Omega}$ is minimized for these effective bands, corresponding to the
``traditional'' procedure outlined for the isolated-bands case. Hence
$U^{(\bk)}_{\alpha\nu}$ is now applied to the $\psiknut$. Note however that no 
true WFs in the sense of (\ref{wann}) result from this procedure due
to the intermediate creation of effective bands. Yet the obtained Wannier-like 
functions are still orthonormal and stem from Bloch-like functions.

\subsubsection{$N$-th order muffin-tin-orbital Wannier functions}

In this paper, we also consider another established route for the construction
of localized Wannier(-like) functions, namely the $N$-th order 
muffin-tin-orbital (NMTO) method~\cite{and00,and00-2,zur05}. This method is
the latest development of the linear muffin-tin-orbital (LMTO)
method~\cite{andersen_lmto_1975_prb,and84}. It uses multiple-scattering theory
for an overlapping muffin-tin potential to construct a local-orbital minimal
basis set, chosen to be exact at some mesh of $N$+1 energies,
$\epsilon_0,\ldots,\epsilon_N$. This NMTO set is therefore a polynomial
approximation (PA) in the energy variable to the Hilbert space formed by all
solutions of Schr\"odinger's equation for an effective single-particle potential.
In the present case this potential is given by the overlapping muffin-tin
approximation to the KS potential (Eq. (\ref{kspot})). Hence in
contrast to the maximally-localized procedure, the NMTO-WFs for correlated
bands may be generated without explicitly calculating the corresponding
Bloch functions.

Apart from its energy mesh, an NMTO set is specified by its members $\{\bR lm\}$,
where $lm$ denotes an angular-momentum character around site $\bR$, in any
primitive cell. The
$\bR lm$-NMTO is thus centered mainly at $\bR$ and has mainly $lm$ character.
Moreover, for the NMTO set to be complete for the energies on the mesh, each
NMTO must be constructed in such a way that its projections onto the 
$\bR lm$-channels
{\sl not} belonging to the $\{\bR lm\}$-set are regular solutions of
Schr\"odinger's equation~\footnote{This holds for each so-called
{\sl kinked partial wave}, which is a solution of Schr\"odinger's equation for an
energy on the mesh, but only approximately for the superposition of
kinked partial waves forming the NMTO; see Appendix A of
Ref.~[\onlinecite{pav05}].}. Finally, in order to {\sl confine} the $\bR lm$-NMTO,
it is constructed in such a way that its projections onto all {\sl other}
channels belonging to the $\{\bR lm\}$-set {\sl vanish}.

For example~\cite{pav05}, the three isolated \t2g bands of cubic \srvo3
are spanned quite accurately by the quadratic ($N$=2) muffin-tin orbital set
which consists of the three (congruent) $d_{xy}$, $d_{yz}$ and $d_{xz}$
NMTOs placed on each V site in the crystal. Locally, the
$d_{xy}$ orbital has $xy$ character as well as minute other characters
compatible with the local symmetry, but {\sl no} $yz$ or $xz$ characters.
On the O sites, the V $d_{xy}$ orbital has antibonding O($p$) and
other characters compatible with the energy and the symmetry, in particular
$p_{x}$ character on O along the $y$-axis and $p_y$ character on the O along
the $x$ axis. On the Sr sites, there are small contributions
which bond to O($p$). Finally, on the {\sl other} V sites, there can be no
\t2g character, but minute other characters are allowed by the local symmetry.
Note that when the symmetry is lowered, as is the case for the distorted
perovskites CaVO$_3$, LaTiO$_3$, and YTiO$_3$, there are less symmetry
restrictions on the downfolded channels and the cation character of the V or
Ti \t2g NMTOs will increase~\cite{pav04,pav05}. This describes a measurable effect
of cation covalency, and is not an artefact of the NMTO construction.

The main steps in the NMTO construction are thus: (a) numerical solution of the
radial Schr\"odinger (or Dirac) equation for each energy on the mesh and for each
$l$ channel with a non-zero phase shift; (b) screening (or downfolding)
transformation of the Korringa-Kohn-Rostocker (KKR) matrix for each
energy on the mesh; and (c) formation of divided differences on the mesh of the
inverse screened matrix to form the Lagrange matrix of the PA, as well as the
Hamiltonian and overlap matrices in the NMTO representation. It should be noted
that this procedure of downfolding plus PA differs from standard
L\"owdin downfolding~\cite{low51} and is more accurate when $N$$>$1.

For an isolated set of bands and with an energy mesh spanning these bands, the
NMTO set converges fast with $N$. The converged set spans the same Hilbert space
as any Wannier set, and may even be more localized because the NMTO set is not
forced to be orthonormal. Symmetrical orthonormalization of the converged NMTO
set yields a set of WFs $w_{\bR lm}$, which are atom-centered and
localized. However this does not imply that the centre of gravity is the
centre of the atom (see e.g. Fig. 5 and 6 of Ref.~[\onlinecite{pav05}]). Note 
that NMTO-WFs have {\sl not} been chosen to minimize the spread
$\langle w_{\bR lm}|r^2|w_{\bR lm}\rangle$, but to satisfy 
the above-mentioned criterion of confinement. Using localized NMTOs, it does 
not require a major computational effort to form linear combinations 
which maximize any other suitable measure of localization.

\subsection{Calculational scheme}

\subsubsection{Band-structure calculations and Wannier construction}

In the following we briefly name the different first-principles techniques that 
were used in the DFT part of the work. More technical details on the specific 
setups may be found in Appendix~\ref{bsappendix}.

The MLWF scheme was interfaced in this work with a mixed-basis~\cite{lou79} 
pseudopotential~\cite{har60,hei70} (MBPP) code~\cite{mbpp_code}. 
This band-structure program utilizes scalar-relativistic norm-conserving 
pseudopotentials~\cite{van85} and a basis of plane waves 
supplemented by non-overlapping localized functions centered at appropriate 
atomic sites. The localized functions, usually atomic functions for a 
given reference configuration, are necessary to allow for a reasonable 
plane-wave cutoff when treating electronic states with substantial local 
character. No shape approximations to the potential or the charge density are
introduced and no MT spheres are utilized in this formalism. 

In addition, we also interfaced an already existing MLWF scheme~\cite{pos02} with
the all-electron, full-potential-linearized-augmented-plane-wave (FLAPW) 
method~\cite{andersen_lmto_1975_prb,jan84,mas93}. This technique is fully 
self-consistent, i.e., all electrons are treated within the self-consistency 
procedure, and no shape approximations are made for the charge 
density and the potential. The core electrons are treated fully relativistically 
and the valence electrons scalar-relativistically. The LAPW basis consists of
atomic-like functions within MT spheres at the atomic sites and plane waves in 
the interstitial region. The conventional basis set is 
furthermore expanded with local orbitals~\cite{sin91} where appropriate. 
Inclusion of local orbitals in addition to the normal FLAPW basis enforces mutual
state orthogonality and increases variational freedom. 

The explicit MLWF construction was performed with the corresponding 
publicly available code~\cite{mlwf-code}. Several minor additions to the
exisiting code were performed in this work in order to account for the specifc
interfacing requirements within LDA+DMFT.

The NMTO construction was performed on the basis of scalar-relativistic 
LMTO\cite{and84} calculations in the atomic-sphere approximation (ASA) with 
combined corrections. Also LMTO is an all-electron method, i.e., it is 
fully self-consistent for core and valence electrons. We utilized the Stuttgart 
TB-LMTO-ASA code~\cite{lmto}.

\subsubsection{Impurity-model solver\label{ldadmftimp}}

The crucial part of the DMFT framework is the solution of the
effective quantum impurity problem. Depending on the symmetries of
the specific case at hand, and the
demands for accuracy, several different techniques are available to solve this
problem in practice (for reviews see 
Ref.~[\onlinecite{geo96,kotliar_review}]). First
the on-site interaction vertex has to be defined. In both cases, i.e., \srvo3 and
\bavs3, we are facing a realistic three-band problem. We keep only
density-density interactions in $\hat{H}_U$, thus no spin-flip or pair-hopping
terms are included. When neglecting explicit orbital dependence of the
interaction integrals, $\hat{H}_U$ reads then as
\begin{eqnarray}
\label{hubmod}
\hat{H}_U=&&U\sum_m\hat{n}_{m\uparrow}
\hat{n}_{m\downarrow}+\frac{U'}{2}
\mathop{\sum_{mm'\sigma}}_{m\ne m'}
\hat{n}_{m\sigma}\hat{n}_{m'\bar{\sigma}}\nonumber\\
&&+\frac{U''}{2}\mathop{\sum_{mm'\sigma}}_{m\ne m'}
\hat{n}_{m\sigma}\hat{n}_{m'\sigma}\quad.
\end{eqnarray}
Here $\hat{n}_{m\sigma}$=$\hat{d}_{m\sigma}^{\dagger}
\hat{d}_{m\sigma}^{\hfill}$, where $m$,$\sigma$ denote orbital and spin index.
The following parametrization of $U'$ and $U''$ has been proven to be
reliable~\cite{cas78,fre97} in the case of \t2g-based systems:
$U'$=$U$$-$$2J$ and $U''$=$U$$-$$3J$. No explicit double-counting term
$\bsig_{\rm{dc}}$ was introduced in our specific calculations. This is due to the
fact that we used ${\cal C}$=${\cal W}$, i.e., our correlation subspace was chosen
to be identical with the set of Wannier bands. In that case the double counting
may be absorbed in the overall chemical potential.

The solution of the quantum impurity problem corresponds to the evaluation of the
impurity Green's function $\bG_{\rm{imp}}$ for a given input of the dynamical 
mean-field (Eq.~(\ref{eq:def_Gimp})), which may be expressed within the
path-integral formalism via
\begin{eqnarray}
\label{pathi}
&&G^{\rm{imp}}_{mm'}(\tau-\tau')\equiv-\langle\hat{\mathcal{T}}
\hat{d}^{\hfill}_{m}(\tau)\hat{d}_{m}^{\dagger}(\tau')
\rangle_{S_{\rm{imp}}}\\
&&=-\frac{1}{\mathcal{Z}_{\rm{imp}}}\int\prod_{\sigma}\mathcal{D}\{d^{\dagger};d\}
\hspace{0.1cm}d^{\hfill}_{m\sigma}(\tau)d_{m'\sigma}^{\dagger}(\tau')
\hspace{0.1cm}\mbox{e}^{-S_{\rm{imp}}}\quad,\nonumber\\
&&\mbox{with}\quad\mathcal{Z}_{\rm{imp}}=
\int\prod_{\sigma}\mathcal{D}\{d^{\dagger};d\}
\hspace{0.1cm}\hspace{0.1cm}\mbox{e}^{-S_{\rm{imp}}}\quad,\nonumber
\end{eqnarray}
where $S_{\rm{imp}}$ is the effective action defined in 
Eq.~(\ref{eq:impurity_action}).
We utilize the auxiliary-field Quantum Monte Carlo (QMC) method following
Hirsch-Fye~\cite{hirsch_fye} to compute (\ref{pathi}).
In this method the path integral is evaluated by a stochastic integration.
Therefore $S_{\rm{imp}}$ is represented on $L$ discretized imaginary time slices 
of size $\Delta\tau$=$\beta/L$. Since the vertex $H_U$ is quartic
in the fermionic degrees of freedom, a decoupling using an exact discrete
Hubbard-Stratonovich transformation is needed. For $M$ orbitals involved,
a number of $M(2M-1)$ so-called ``Ising fields'' emerge from this decoupling for
each time slice. In the end, the number of time slices $L$ and the number of
Monte-Carlo sweeps $N_{\rm{MC}}$ are the sole convergence parameters of the
problem. The QMC technique has no formal approximations, however the numerical
effort scales badly with $M$ and $\beta$.

Note again that although we so far outlined LDA+DMFT as a fully self-consistent
scheme, i.e., including charge-density updates, the results in the following
sections were obtained from a simpler post-processing approach. Thereby the
self-consistent LDA Wannier Hamiltonian was used in (\ref{gimpwann}) and no
charge-density updates were performed.

\section{Results\label{results}}

\subsection{\srvo3}

\subsubsection{Characterization and band-structure calculations\label{srvo3chap}}

A quadratic temperature behavior of the resistivity up to
room temperature\cite{ono91}, albeit with a large prefactor,
qualifies the electronic structure of the $3d(t_{2g})^1$ compound \srvo3 as a
Fermi liquid with intermediate strength of the electron-electron interactions.
Still, a direct comparison of the photoemission spectral function with the 
one-particle density of states (DOS), calculated e.g. within DFT-LDA, yields poor
agreement, indicating a strong need for an explicit many-body treatment of 
correlations effects. DFT-LDA also yields a specific-heat coefficient 
(slope of $C/T$ at low-$T$) which is too small by approximately a factor of two, 
i.e., the electronic effective mass is enhanced due to correlation effects.

A perfectly cubic perovskite structure and the absence of magnetic ordering down
to low temperatures makes \srvo3 an ideal test material for first-principles 
many-body techniques. 
It has thus been the subject of many experimental and theoretical investigations
(using LDA+DMFT)~\cite{imada_mit_review,fujimori_pes_oxides,maiti_2001,maiti_phd,
inoue_casrvo3_1995_prl,sek04,lie03,nek05,pav04,yos05,wad06,sol06,nek06,egu06}.
In the present work, we use this material as a benchmark for our Wannier 
implementation of LDA+DMFT, with results very similar to previous theoretical
studies.

We start the investigation of \srvo3 with a brief DFT-LDA study. The crystal
structure of the transition-metal oxide \srvo3 is rather simple, exhibiting full
cubic symmetry (space group $Pm\bar{3}m$) with a measured~\cite{cha71} lattice
constant of 7.2605 a.u.. The V ion is placed in the center of an ideal
octahedron formed by the surrounding O ions see Fig.~\ref{srvo3t2gwbands}. The
O ions are at the face centers of a cube having V at its center and Sr at its
corners.

\begin{figure}[t]
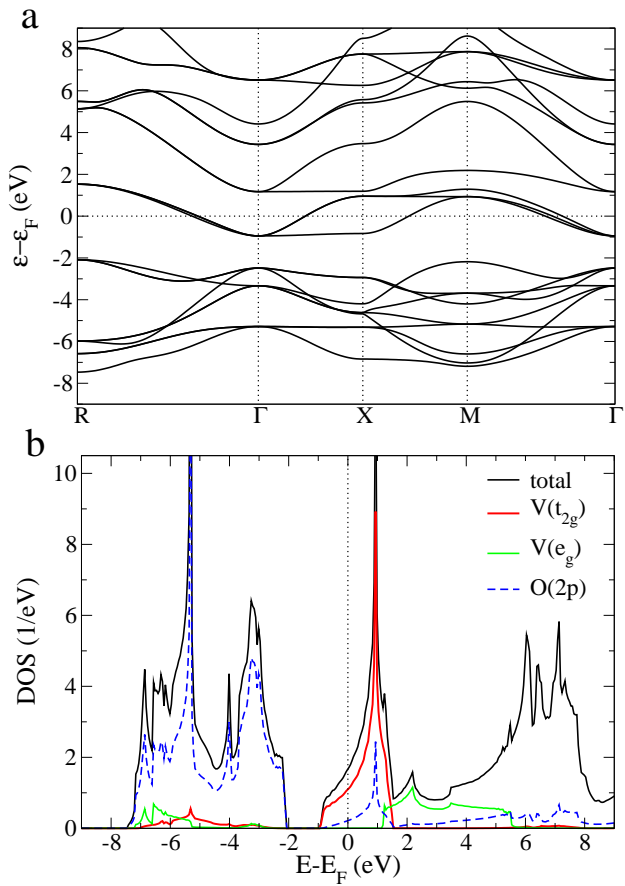

\includegraphics*[width=8.2cm]{srvo3-ldabands.eps}\\
\includegraphics*[width=8.0cm]{srvo3-ldados.eps}
\caption{(Color online) (a) LDA data for \srvo3 calculated with the MBPP code.
(a) Band structure. (b) DOS. For the local V$(3d)$/O$(2p)$-DOS the cutoff radius
was half the nearest-neighbor distance, respectively.\label{srvo3lda}}
\end{figure}

Figure~\ref{srvo3lda} shows the band structure and the DOS within LDA. The data 
reveals that there is an isolated group of partially occupied bands at the Fermi 
level, with a total bandwidth of 2.5 eV. For an ion at site $\bR$ the local 
orbital density matrix
$n^{(l)}_{mm'}$$\sim$$\sum_{\bk\nu}f_{\bk\nu}\langle\psi_{\bk\nu}|
K^{(l)}_m\rangle\langle K^{(l)}_{m'}|\psi_{\bk\nu}\rangle$ is
a measure for the occupation probabilities within the set of, say cubic,
harmonics $\{K^{(l)}_m\}$. In the case of \srvo3 this matrix is diagonal, and
it is seen in Fig.~\ref{srvo3lda}b that from such a projection the bands at
\fermi may be described as stemming dominantly from V$(\tg)$ orbitals. Since the
three \t2g orbitals, i.e, $\{d_{xy},d_{xz},d_{yz}\}$ are degenerate, they have
equal contribution to the bands. Due to the full cubic symmetry the distinct
\t2g orbitals are nearly exclusively restricted to perpendicular planes
which explains the prominent 2D-like logarithmic-peak shape of the DOS. The
V$(e_g)$ states have major weight above the Fermi level, whereas the O$(2p)$
states dominantly form a block of bands below \fermi. The energy gap
between the O$(2p)$ and $\tg$ block amounts to 1.1 eV. In spite of the
``block'' characterization, there is still significant hybridization between the
most relevant orbitals, i.e., V$(3d)$ and O$(2p)$, over a broad energy range.

\subsubsection{Wannier functions}

The low-energy physics of \srvo3 is mainly determined by the isolated set
of three \t2g-like bands around the Fermi level. This suggests the
construction of an effective three-band Wannier Hamiltonian as the relevant
minimal low-energy model. In the following, we construct Wannier functions
associated with this group of bands, and also pick these three Wannier functions
as generating the correlated subset $\mathcal{C}$, so that
$\mathcal{W}$=$\mathcal{C}$ in the notations of the previous section. This choice 
of course implies that the resulting Wannier functions, though centered on a 
vanadium site, have also significant weight on neighbouring oxygen sites. More 
localized functions can indeed be obtained by keeping more bands in the Wannier 
constructions (i.e. by enlarging the energy window) and thus keeping 
$\mathcal{W}$ larger than $\mathcal{C}$, as described at the end of this 
subsection.
However, we choose here to explore this minimal construction as a basis for a
DMFT treatment and show that it actually gives a reasonable description of this 
material.
\begin{figure}[t]
\includegraphics*[width=8cm]{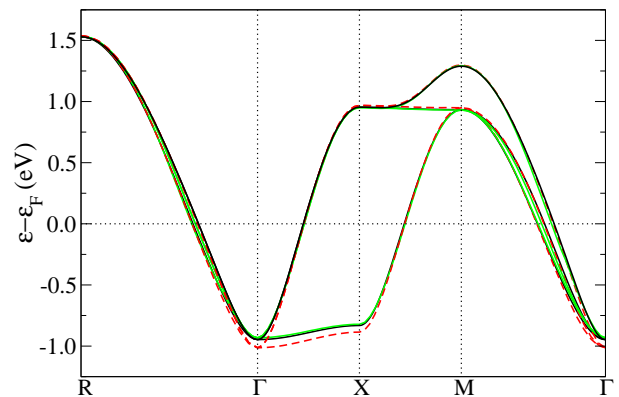}
\caption{(Color online) \t2g bands for \srvo3 using different schemes to compute 
the \t2g Wannier functions (and the underlying LDA band structure or potential).
dark: MLWF(MBPP), dashed-red (dashed-gray): MLWF(FLAPW) and green (light gray): 
NMTO(LMTO-ASA). The \t2g bandwidth is marginally larger in FLAPW, leading to 
small differences. 
\label{srvo3wbands}}
\end{figure}

Figure~\ref{srvo3wbands} exhibits the Wannier bands obtained within our
three utilized schemes: maximally-localized WFs from the MBPP and FLAPW
codes (abbreviated in the following respectively by MLWF(MBPP) and
MLWF(FLAPW)) and the NMTO scheme used as a postprocessing tool on top
of the LMTO-ASA code (denoted as NMTO(LMTO-ASA)).
For the MLWF construction a starting guess for the WFs was provided by utilizing
atomic-like functions with \t2g symmetry centered on the V site. Some details
on the construction of the NMTO-WFs are provided in Appendix~\ref{bsappendix}.
Both MLWF and NMTO schemes yield bands identical to the LDA bands. The small 
discrepancies seen in Fig.~\ref{srvo3wbands} are due to differences in the 
self-consistent LDA potentials. This overall agreement between the different 
methods reflects the coherent LDA description for this material.

\begin{figure}[t]
\includegraphics*[width=8cm]{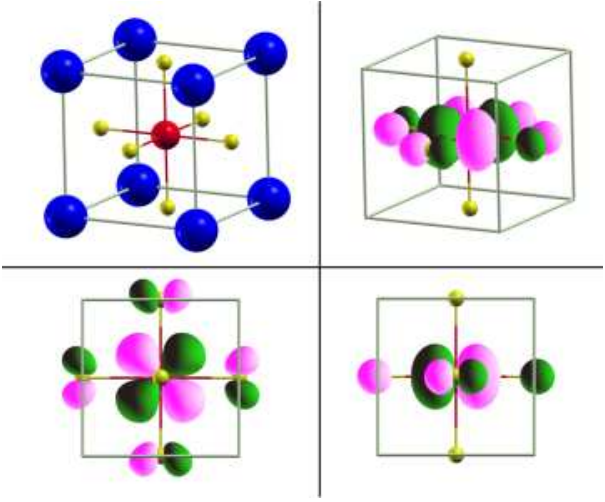}
\caption{(Color online) \t2g-like MLWF $w_{xy}$ for \srvo3 derived from the MBPP 
code. First row: \srvo3 structure with Sr (large blue/dark), V (red/gray) and 
O (small yellow/light gray) and perspective view on $w_{xy}$. 
Second row: $w_{xy}$ viewed along the $c$ axis and along $a$ axis. The contour 
value $w_{xy}^{(0)}$ was chosen as 0.05 (a.u.)$^{-3/2}$.\label{srvo3t2gwbands}}
\end{figure}
\begin{figure}[b]
\includegraphics*[width=8cm]{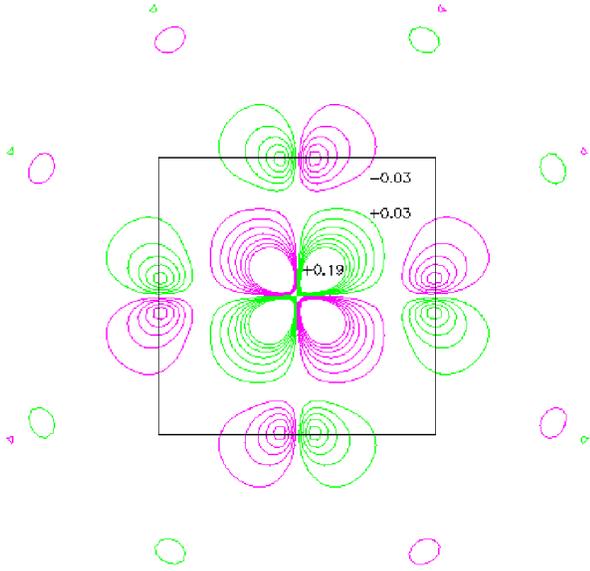}
\caption{(Color online) Contour-lines plot of the \t2g-like MLWF $w_{xy}$ for 
\srvo3. Distinct contour values (in (a.u.)$^{-3/2}$) are given in the 
plot.\label{srvo3t2gcontour}}
\end{figure}
\begin{figure}[t]
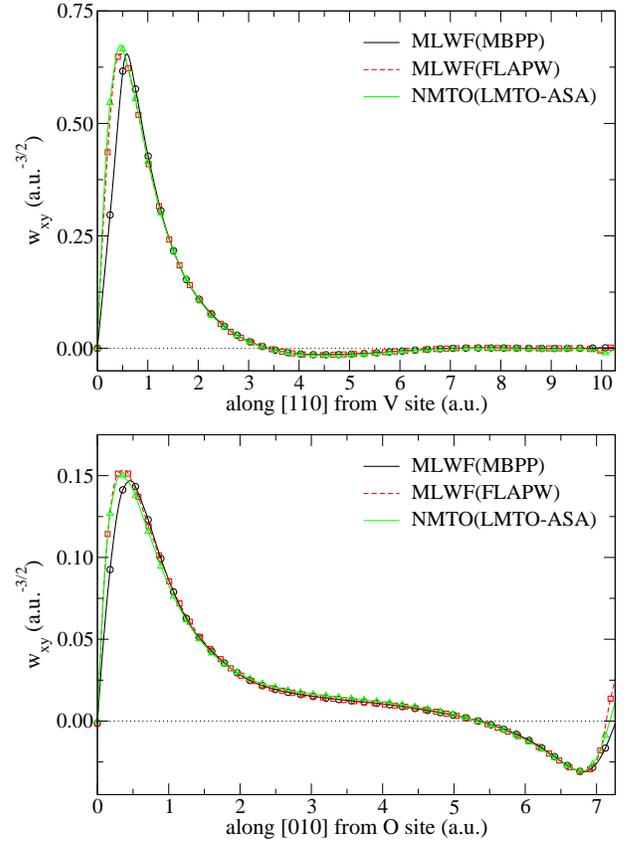

\includegraphics*[width=8cm]{psi.srvo3.v.eps}\\[0.2cm]
\includegraphics*[width=8cm]{psi.srvo3.o.eps}
\caption{(Color online) (a) \t2g-like $w_{xy}$ WF. (a) From V to V along [110],
and (b) from O to O along [010].\label{srvo3t2cuto}}
\end{figure}

Although all three sets of WFs span the same Hilbert space, and the bands are
therefore the same, the MLWFs and the WFs obtained by symmetrically
orthonormalizing the NMTO set are not necessarily identical. In order to compare
the Wannier orbitals, we generated the set $\{w_m(\br)\}$ within a
(3$\times$3$\times$3) supercell on a (120$\times$120$\times$120) real-space mesh.
As an example, Fig.~\ref{srvo3t2gwbands} shows the \t2g-like $w_{xy}(\br)$
Wannier orbital for a chosen constant value $w^{(0)}_{xy}$ as obtained from the
MLWF(MBPP) construction.
By symmetry, all three Wannier orbitals come out to be centered on the V site.
A general contour plot for $w_{xy}(\br)$ is given in Fig.~\ref{srvo3t2gcontour}.
The Wannier orbitals show clear \t2g symmetry, but in
addition have substantial oxygen character, $\pi$-O($2p$) in particular.
The important hybridization between the V(\t2g) and O($2p$) atomic-like orbitals
seen in Fig.~\ref{srvo3lda} is explicitly transfered in the Wannier orbital.
By comparing the three different sets of Wannier orbitals we find remarkably
close agreement. Thus the MLWF and NMTO constructions provide nearly identical
vanadium \t2g Wannier orbitals in the case of cubic \srvo3. A detailed
comparison is shown in Fig.~\ref{srvo3t2cuto} where the WFs are plotted along
specific directions.
\begingroup
\begin{table}[t]
\caption{Spread $\langle r^2\rangle$ of the V-centered \t2g WFs
for \srvo3. We employed a (8$\times$8$\times$8) $k$-point mesh for the
construction of the MLWFs. The spreads were calculated in two ways. First from
the $k$-space integration performed in
in the MLWF code, and second via an $r$-space integration within a
(3$\times$3$\times$3) supercell. Additionally shown is the respective
normalization of the WFs within this supercell.
Finally, for completeness the $\Omega$ value minimized within the MLWF
construction is also given, even if due to the complete degeneracy it
reduces to three times the spread $\langle r^2\rangle$.
\label{table-sr1}}
\begin{ruledtabular}
\begin{tabular}{l|c|c|c}
 scheme & \hspace{0.5cm}$\langle r^2\rangle$ (a.u.$^2$) & norm & $\Omega$ (a.u.$^2$)\\ \cline{2-4}
  & $\bk$ space\hspace{0.5cm}$\br$ space & $\br$ space & $\bk$ space \\ \hline
MLWF(MBPP)              & 6.86\hspace{1cm}6.64 & 0.998 & 20.57 \\
MLWF(FLAPW)             & 6.96\hspace{1cm}6.75 & 0.997 & 20.93 \\
NMTO(LMTO-ASA)          &   - \hspace{1.37cm}6.82 & 0.995 &  -   \\ \hline
Ref.~[\onlinecite{sol06}] & \hspace{0.2cm}8.46 &   -   &  -   \\
\end{tabular}
\end{ruledtabular}
\end{table}
\endgroup
From these graphs it may be seen that the MLWF(MBPP) slightly disagrees with
the WFs from the two other schemes close to the nuclei. This discrepancy
is due to the pseudization of the crystal wave functions close to the
nucleus. Although the $3d$ wave function is nodeless, the pseudo wave function
is modified in order to provide an optimized normconserving
pseudopotential. However, this difference in the WFs has no observable effect on
the description of the bonding properties as outlined in general pseudopotential 
theory~\cite{har60,hei70} (see also Tab.~\ref{table-sr2}). Only marginal
differences between the different WFs can be observed away from the nuclei.
Generally, the fast decay of the WFs is documented in
Fig.~\ref{srvo3t2cuto}. In this respect, Table~\ref{table-sr1} exhibits the
values for the spread $\langle r^2\rangle$ of the WFs from the different
schemes. The MBPP and FLAPW implementations of the MLWFs have spreads which differ
by 2$\%$. Since for the MLWFs the spread has been minimized, that of the 
NMTO-WFs should be larger, and it indeed is, but merely by a few per cent.
So in this case the
NMTO-WFs may be seen as maximally localized, also in the sense of
Ref.~[\onlinecite{mar97}]. A substantially larger value for the spread is
however obtained from the orthonormal LMTOs, as seen from
Ref.~[\onlinecite{sol06}].

To finally conclude this part of the comparison, we deduced the relevant
near-neighbor hopping integrals from the real-space Hamiltonian in the respective
Wannier basis, given in Table.~\ref{table-sr2}. The dominance of the
nearest-neighbor hopping in connection with the fast decay of the remaining
hoppings clearly demonstrates the strong short-range bonding in \srvo3. The
close agreement of the hoppings between the three different Wannier schemes
again underlines their coherent description of this material.
It can be concluded that although conceptually rather different, MLWF and NMTO
provide a nearly identical minimal Wannier description for \srvo3. The small
numerical differences seem to stem mainly from the differences in the
electronic-structure description within the distinct band-structure methods.
\begin{table}[t]
\caption{Symmetry-inequivalent intersite hopping integrals $H_{yz,yz}$ for
\srvo3. Energies are in meV.\label{table-sr2}}
\begin{ruledtabular}
\begin{tabular}{l|c c c c c c c}
$xyz$          &   001  & 100   & 011   & 101 &  111 & 002 & 200 \\ \hline
MLWF(MBPP)     & -260.5 & -28.2 & -83.1 & 6.5 & -6.0 & 8.4 & 0.1 \\
MLWF(FLAPW)    & -266.8 & -29.2 & -87.6 & 6.4 & -6.1 & 8.3 & 0.1 \\
NMTO(LMTO-ASA) & -264.6 & -27.2 & -84.4 & 7.3 & -7.6 & 12.9 & 3.5 \\
\end{tabular}
\end{ruledtabular}
\end{table}
\begin{figure}[t]
\includegraphics*[width=7cm]{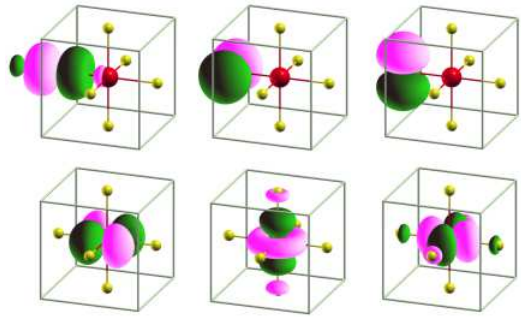}
\caption{(Color online) Distinct WFs for \srvo3 obtained from the MLWF
construction using the MBPP code. First row: O($p_x$), O($p_y$) and O($p_z$) for
a chosen oxygen site.  Second row: V($t_{2g},xy$) as well as
V($e_g$,$3z^2$-$r^2$) and V($e_g$,$x^2$-$y^2$). The contour value for each of the
MLWFs was chosen as 0.05 (a.u.)$^{-3/2}$.\label{srvo3fullwbands}}
\end{figure}

At the end of this subsection we want to draw attention to the fact that the
performed minimal Wannier construction {\sl solely} for the \t2g bands is
of course not the only one possible, as already mentioned above.
Depicted in Fig.~\ref{srvo3fullwbands} are
the WFs obtained by downfolding the LDA electronic structure of \srvo3 to V($3d$)
{\it and} O($2p$) states.
Hence this corresponds to describing \srvo3 via an 14-band model, i.e., three $p$
orbitals for three O ions and five $d$ orbitals for the single V ion in the unit
cell. Due to minor degeneracies with higher lying bands (see Fig.~\ref{srvo3lda})
the disentangling procedure for the MLWF construction has to be used, but no
relevant impact is detected in this case. Now there are distinct WFs
for O($2p$) and V($3d$) with significantly smaller spreads.
Individually the latter are in a.u.$^2$: 2.61 for V(\t2g) and 2.32 for V(\eg),
and 2.68 for $\sigma$-O($2p$) and 3.39 for $\pi$-O($2p$), resulting in a total
spread of $\Omega$=40.75 a.u.$^2$.

\subsubsection{LDA+DMFT calculations}

Thanks to the simplicity of the perfectly cubic perovskite structure
and the resulting degeneracy of the three \t2g orbitals, \srvo3 is a simple
testing ground for first-principles dynamical mean-field
techniques.
In fact \srvo3 is quite a unique case in which
the calculation of the local Green's function
(\ref{gimpwann}), which usually involves a
$\bk$ summation, can be reduced to the simpler calculation of
a Hilbert transform of the LDA DOS.
Indeed, because of the perfect cubic symmetry, all local quantities in
the $t_{2g}$ subspace are proportional to the unit matrix:
$G_{mm'}^{\rm{loc}}(\iomn)=G_{\rm{loc}}(\iomn)\delta_{mm'}$,
$\Delta\Sigma_{mm'}(\iomn)=\Delta\Sigma(\iomn)\delta_{mm'}$ (as well as
the LDA DOS
$D_{mm'}(\varepsilon)=D(\varepsilon)\delta_{mm'}$ projected onto the
orbitals $\chi_m$), so that
(\ref{gimpwann}) reduces to:
$G_{\rm{loc}}(\iomn)$=$\int\frac{d\varepsilon D(\varepsilon)}
{\iomn+\mu-\varepsilon-\Delta\Sigma}$. Note however that this {\it does not}
hold in general for other materials, as soon as the local quantities are
no longer proportional to the unit matrix. Although many actual LDA+DMFT 
calculations in literature use this representation as an approximation to the 
correct form given by Eq.~(\ref{gimpwann}). In the calculations documented in this
work we always used the more generic Hamiltonian representation and
$\bk$ summations.

Taking into account the strong correlations within the \t2g manifold
results in substantial changes of the local spectral function compared
to the LDA DOS, namely a narrowing of the QP bands close
to the Fermi level while the remaining spectral weight is shifted to
Hubbard bands at higher energies.
This general physical picture of the correlated metal can be
understood already in the framework of the multi-orbital Hubbard
model as the coexistence of QP bands with atomic-like
excitations at higher energy.
It directly carries through to the realistic case of \srvo3 as
studied in several previous works~\cite{lie03,sek04,pav04,nek05}.

Moreover, an important feature of LDA+DMFT that emerges in the present
case of a completely orbitally degenerate self-energy has been put
to test against experiments. Indeed, in this special case Fermi-liquid
behavior in conjunction with a $\bk$-independent self-energy leads
to the value of the local spectral function $\rho(\varepsilon)$ at the Fermi
level being equal to its non-interacting counterpart just in the same way as in
the one-band Hubbard model~\cite{mue89}.

\begin{figure}[t]
\includegraphics*[width=8cm]{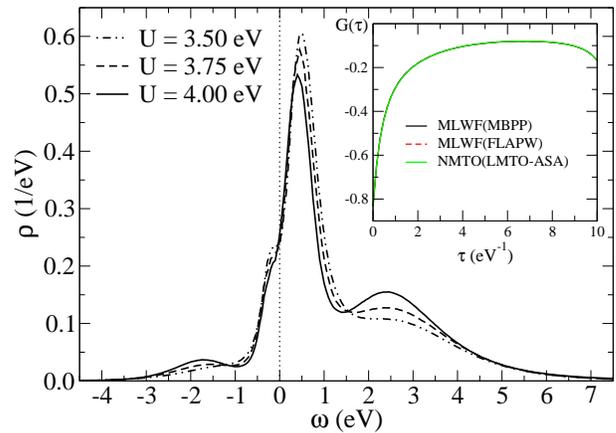}
\caption{(Color online) Spectral function for SrVO3 for different values of $U$ 
resulting from LDA+DMFT in the MBPP implementation. Inset: comparison of
Green's functions from MBPP, FLAPW and NMTO implementations of LDA+DMFT for
$U$=4 eV. In all calculations $J$=0.65 eV was used.
\label{dosG.ps}}
\end{figure}

In this work we performed LDA+DMFT calculations for \srvo3 by using the
self-consistent LDA Wannier Hamiltonian $\bH_{\rm{KS}}$ derived from the
different
band-structure codes, i.e., MBPP, LMTO-ASA and FLAPW, described above.
As expected from the good agreement of the band structure and the Hamiltonians
the resulting Green's functions are identical within the statistical errors bars
(see the inset of Fig.~\ref{dosG.ps}). Fig.~\ref{dosG.ps} also displays the
local spectral functions based on the MLWF(MBPP) scheme and calculated for
 different values of $U$. The ``pinning'' of $\rho(0)$, independently of
the value of the interactions is clearly visible, despite the finite
temperature of the calculations. This indicates that the
calculations have indeed been performed at a temperature smaller than the
QP (Fermi-liquid) coherence scale of this material.

\begin{figure}[b]
\includegraphics*[width=8cm]{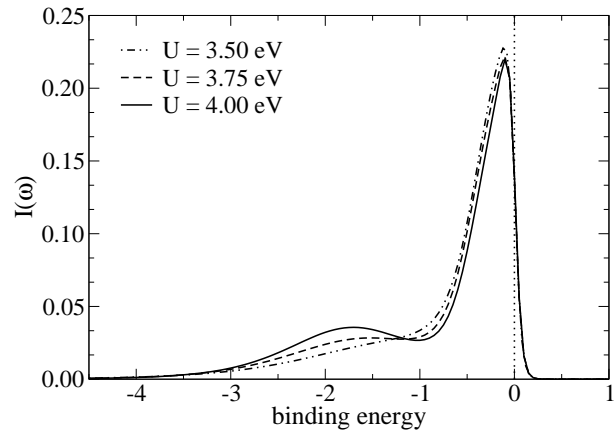}
\caption{\srvo3 spectral function convoluted with a Gaussian experimental
resolution (assumed to be 0.15 eV) and with the Fermi function
at $\beta$=30 eV$^{-1}$.
\label{theoPES_SVO}}
\end{figure}

Figure~\ref{theoPES_SVO} displays the local spectral function convoluted
with an assumed experimental resolution of 0.15 eV and multiplied
by the Fermi function. This quantity represents thus a direct
comparison to angle-integrated photoemission spectra (albeit
neglecting matrix elements, which can in certain circumstances
appreciably depend e.g. on the polarization of the photons,
see Ref.~[\onlinecite{mai05}]).

The general agreement with recent experimental data
\cite{sek04,yos05,mai05,wad06,egu06} is reasonable. Photoemission experiments 
locate the lower and upper Hubbard bands at energies about -2 eV to -1.5 eV
\cite{sek04,mai05} and 2.5 eV \cite{morikawa95} respectively.
In our calculations the lower Hubbard band extends between
-2 eV to -1.5 eV, while the maximum of the upper Hubbard band
is located at about 2.5 eV, for values of the Coulomb interaction
$U$ of about 4 eV.
However, we also confirm the findings of Ref.~[\onlinecite{mai05}]
who point out that LDA+DMFT calculations generally locate the lower
Hubbard band at slightly higher (in absolute value) binding
energies than -1.5 eV, the energy where their data exhibits
its maximum.

Concerning the choice of the Coulomb interaction $U$ different
points of view can be adopted. First, one can of course choose
to try to calculate $U$ itself from first principles by e.g.
constrained LDA~\cite{ded84,mcm88,gun89,ani91,coc05} or
RPA-based techniques~\cite{ary04,sol05}.
Another option is to use it as an adjustable parameter and to determine it
thus indirectly from experiments.
While in the present case the order of magnitude of the interaction
($U$$\sim$3.5-5.5 eV) \cite{sek04,sol06,ary06} is indeed known from
first-principles approaches, the exact values determined from different methods
still present a too large spread to be satisfactory for precise quantitative
predictions.
We therefore adopt the second point of view here, noting that
$U$ values of around 4 eV reproduce well the experimentally
observed~\cite{yos05,wad06,inoue98} mass enhancement of $\sim$1.8 to 2.
The agreement concerning the position of the Hubbard bands
seems to be fair, given the theoretical uncertainty linked to
the analytical continuation procedure by maximum-entropy techniques
and the spread in available experimental data.
Still, it is conceivable that the determination of
the precise position of the Hubbard bands could require more
sophisticated methods than LDA+DMFT done with a static $U$ parameter,
and that in fact we are facing the consequences of subtle screening
effects which, within an effective three-band model, could only
be described by a frequency-dependent interaction\cite{ary04}.

\subsection{$Cmc2_1$-B\lowercase{a}VS$_3$}
\label{sec:bavs3}

\subsubsection{Structure and physical properties}
\begin{figure}[b]
\includegraphics*[width=6cm]{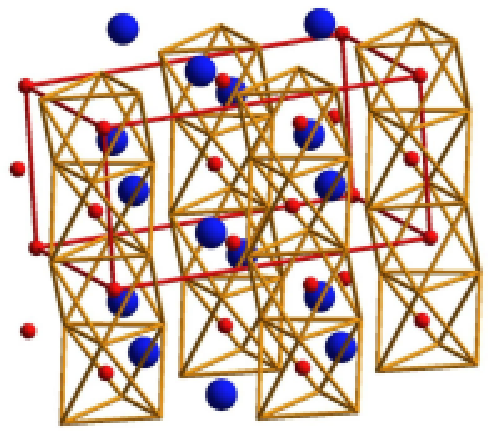}\\[0.2cm]
\includegraphics*[width=6cm]{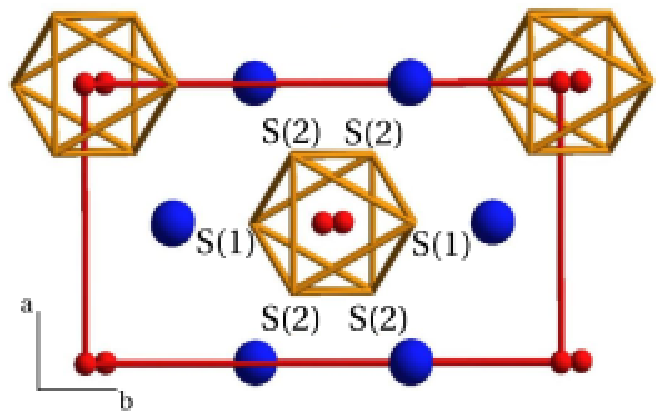}
\caption{(Color online) BaVS$_3$ in the $Cmc2_1$ structure. The V ions are shown 
as smaller (red/gray) spheres, the Ba ions as larger (blue/dark) 
spheres.\label{bavs3crystal}}
\end{figure}
The transition-metal sulfide \bavs3 is also a $3d(t_{2g})^1$ system, but its 
physical properties are far more 
complex~\cite{boo99,whangbo_bavs3_puzzling_jsschem_2003} than
those for the cubic perovskite \srvo3 considered above.
In a recent work~\cite{lec05,lecproc}, three of us  have suggested that a 
correlation-induced redistribution of orbital populations is the key mechanism 
making the transition into a charge-density wave (CDW) insulating phase possible.
Here, we use our Wannier formalism to make a much more refined study of this
phenomenon and to calculate how correlations modify the Fermi-surface sheets of 
the metal. Also, this material is a challenging testing ground for the Wannier 
construction because of the strong hybridization between the transition-metal and 
ligand bands.

We first give a very brief summary of some of the physical properties of
\bavs3 of relevance to the present paper. At room temperature \bavs3 exists in a
hexagonal crystal structure (space group $P6_3/mmc$), with two formula units
of \bavs3 in the primitive cell. There are straight chains of face sharing VS$_6$
octahedra along the $c$ axis, and Ba ions in between. A continuous structural
phase transition at $T_{\rm{S}}$$\sim$240 K reduces the crystal symmetry to
orthorhombic, thereby stabilizing the $Cmc2_1$ structure, again with two formula
units in the primitive cell. Now the VS$_3$ chains are zigzag-distorted in the
$bc$ plane.
In this phase, \bavs3 is a quite bad metal, with unusual properties such as
a Curie-Weiss susceptibility from which the presence of local moments
can be inferred. At $\sim$70 K a second continuous phase transition takes
place~\cite{graf_bavs3_pressure_prb_1995,mih00}, namely
a metal-insulator transition (MIT) below which \bavs3 becomes a
paramagnetic insulator. A doubling of the primitive unit
cell~\cite{inami_bavs3_prb_2002,fagot_bavs3_prl_2003,fag05} is accompanying
the MIT. Together with large one-dimensional structural fluctuations along the
chains~\cite{fagot_bavs3_prl_2003} and additional precursive behavior for the
Hall constant~\cite{boo99} just above $T_{\rm{MIT}}$, the transition
scenario is reminescent of a Peierls transition into a charge-density wave (CDW)
state. Finally a third second order transition appears to occur at $\sim$30 K.
This so-called ``X'' transition is of magnetic kind and shall announce the onset
of incommensurate antiferromagnetic order~\cite{Nak00} in the insulator.

Here we want to focus on the orthorhombic ($Cmc2_1$) structure
(see Fig.~\ref{bavs3crystal}) at $T$=100 K, i.e., just above the MIT. Ten ions
are incorporated in the primitive cell. Whereas the two Ba and two V ions occupy
(4a) sites, there are two types of sulfur ions. Two S(1) ions are positioned at
(4a) apical sites on the $b$ axis, while four S(2) ions occupy (8b) sites. The
lattice parameters are:~\cite{ghedira_bavs3_neutrons_jpc_1986} $a$=12.7693 a.u.,
$b$=21.7065 a.u. and $c$=10.5813 a.u..

\subsubsection{Band structure}
\begin{figure}[t]
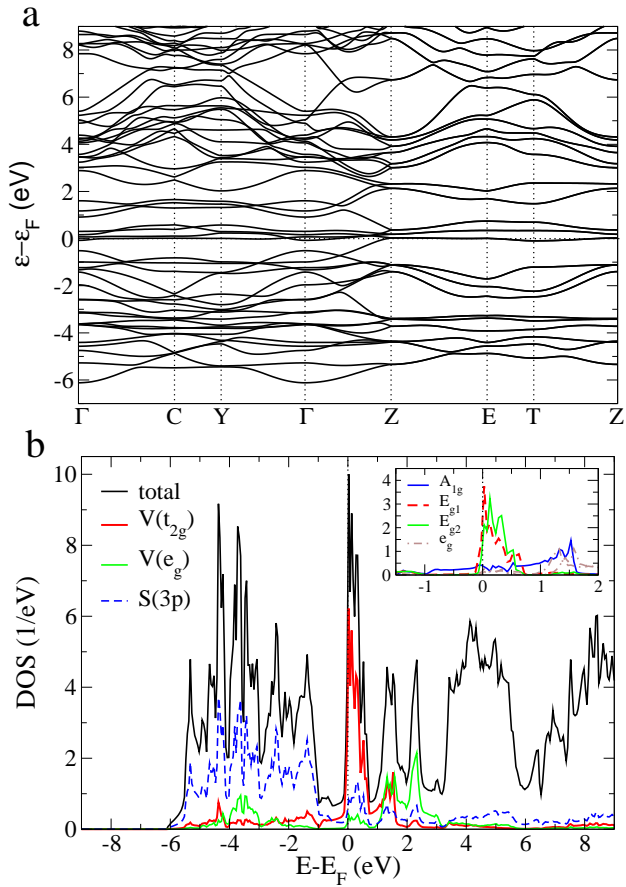

\includegraphics*[width=8.2cm]{bavs3-ldabands.eps}\\
\includegraphics*[width=8cm]{bavs3-ldados.eps}
\caption{(Color online) LDA data for \bavs3 after the MBPP method. (a) band 
structure and(b) DOS. For the local V$(3d)$/S$(3p)$-DOS the cutoff radius was 
half the minimum nearest-neighbor distance, respectively. The inset in (b) shows 
the symmetry-adapted local V$(3d)$-DOS close to the Fermi level.\label{bavs3lda}}
\end{figure}
Figure~\ref{bavs3lda} depicts the LDA band structure and DOS of
$Cmc2_1$-BaVS$_3$. To allow for orbital resolution, the local DOS was again
projected onto symmetry-adapted cubic harmonics by diagonalizing $n^{(d)}_{mm'}$.
It is seen that the bands at \fermi have dominant \t2g character, however they
still carry sizeable S($3p$) weight. Furthermore, the \t2g-like bands are now
not isolated but {\it strongly entangled} with S($3p$)-like bands. Due to the 
reduction
of symmetry from hexagonal to orthorhombic, the \t2g manifold splits into
\A1g and \eeg. The two distinct \eeg states will be denoted in the following
$E_{g1}$ and $E_{g2}$. Being directed along the $c$ axis, the \A1g orbital points
towards neigboring V ions within a chain and the corresponding band (see
Fig.~\ref{bavs3wbands}a) shows a folded structure because of the existence of
two symmetry-equivalent V ions in the unit cell. The folded \A1g band has a
bandwith of 2.7 eV, while \eeg states form very narrow (0.66 eV) bands right at 
the Fermi level.

From the LDA DOS it seems that a projection onto \t2g-like orbitals close to the
Fermi level by diagonalizing $n^{(d)}_{mm'}$ is, at least to a first
approximation, meaningful\cite{lec05}. However there is a substantial S($3p$)
contribution close to \fermi and generally large charge contributions in the
interstitial. Hence establishing a very accurate correspondance between relevant
bands and orbitals is not possible in such a way. In contrast, the Wannier schemes
discussed above are quite suitable for dealing with this situation. To be 
specific, we applied the MLWF scheme in the new MBPP implementation to this 
problem.

Besides providing a test for the MLWF scheme, the present study will allow us to
make considerably more precise the findings of Ref.~[\onlinecite{lec05}] 
regarding the crucial role of correlation-induced changes in the orbital 
populations, and most notably to clarify how these changes can modify the Fermi 
surface (FS) of this material in such a way that favorable conditions for a 
CDW transition indeed hold.
Key to the physics of \bavs3 is the simultaneous presence of two quite distinct
low-energy states, the rather delocalized \A1g and quite localized \eeg,
among which the electronic density with one electron per vanadium has to
divide itself. Depending on temperature, the associated orbital populations 
correspond to the best compromise between gain of kinetic energy and cost of 
potential energy. As it appears, this compromise seems to be realized by a CDW 
state below the MIT. However, as revealed in several electronic structure
studies~\cite{mattheiss_bavs3_1995,whangbo_bavs3_puzzling_jsschem_2003,lec05},
a DFT-LDA description of \bavs3 does not explain the occurence of a CDW
instability. Though the mainly \A1g-like band appears to be a promising
candidate, a nesting scenario in agreement with the critical wave vector
${\bq}_c$=$0.5{\bf c}^*$ from experiment~\cite{fagot_bavs3_prl_2003} is not
realizable. In Fig.~\ref{bavs3wbands}a we elaborated a so-called
``fatband''~\cite{jep95} resolution of the LDA band structure close to the Fermi
level, which is helpful
to reveal the respective band character to a good approximation. Thereby the
Bloch function associated with a given $k$-point and eigenvalue is projected
onto orthonormal symmetry-adapted local orbitals (determined as usual by
diagonalizing the local orbital density matrix $n^{(l)}_{mm'}$). The resulting
magnitude of the overlap is depicted as a broadening of the corresponding band. 
Here it is seen that \A1g-like band cuts the Fermi level close to the boundary 
of the BZ along $\Gamma$-Z, i.e., along the $c^*$ axis in reciprocal
space. Since the Z point is located at $k_z$=$0.5c^*$, in numbers this amounts 
to 2$k_F$=0.94 for the \A1g-like band within LDA, nearly twice the experimental 
value determined for the nesting vector. Furthermore, also other parts of the 
LDA FS are out of reach for ${\bq}_c$, as the \A1g sheet is too extended and
additionally strongly warped (see also Fig.~\ref{bavs3-fs}b). In other words,
LDA apparently overestimates the population of the more itinerant \A1g state.

Moreover the role of the electrons with strong \eeg character at the MIT is not
obvious. When approaching $T_{\rm{MIT}}$ these nearly localized electrons should
surely contribute to the Curie-Weiss form of the magnetic
susceptibility~\cite{mih00,lec05}. In fact the ``bad-metal''
regime~\cite{graf_bavs3_pressure_prb_1995} above the MIT, including significant
changes in the Hall coefficient~\cite{boo99}, might largely originate from
scattering processes involving the \eeg electrons. But even if the \A1g bands
become gapped at the MIT, from an effective single-particle LDA viewpoint the
remaining \eeg bands may still ensure the metallicity of the system.
We therefore believe for several reasons that correlation effects beyond LDA are
important~\cite{lec05} for an understanding of the physics of \bavs3. We will
further outline relevant mechanisms, now based on a more elaborate Wannier
scheme, in section~\ref{secdmftbavs3}.

\subsubsection{Wannier functions\label{bavs3wannsec}}
\begin{figure}[t]
\includegraphics*[width=8cm]{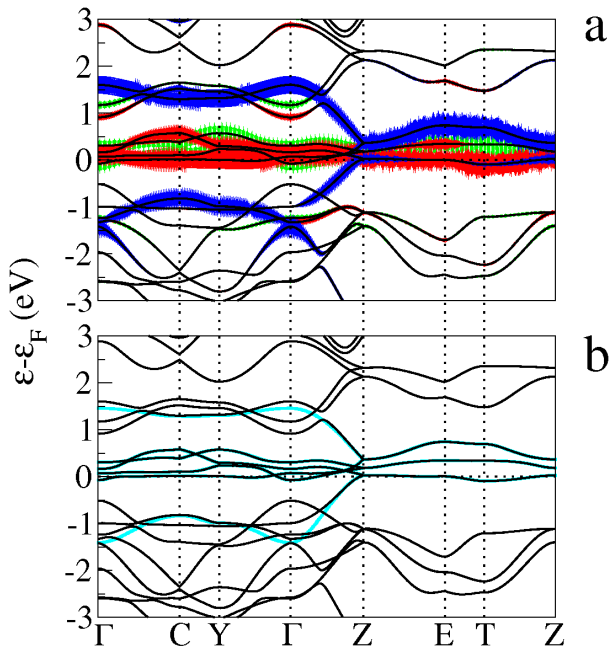}
\caption{(Color online) (a) \t2g fatband resolved band structure of \bavs3. The 
color code is as follows: \A1g (blue/dark), $E_{g1}$ (red/gray) and $E_{g2}$ 
(green/light gray). (b) downfolded \t2g Wannier bands (light blue/light gray) for 
\bavs3 obtained from the MLWF construction using the MBPP code.\label{bavs3wbands}}
\end{figure}
The central difficulty in constructing \t2g-like WFs for \bavs3 is the strong
hybridization between V($3d$) and S($3p$), leading to a substantial entanglement
between the two band manifolds. In detail, whereas the two \eeg states form 
four very narrow bands, mainly confined to the Fermi level, the folded \A1g band 
extends into the dominantly S($3p$)/V($e_g$) band manifolds lower/higher in 
energy. This entanglement is documented in Fig.~\ref{bavs3wbands}a by 
significant ``jumps'' of the corresponding \A1g fatband between different bands.
One may of course downfold the \bavs3 band structure including not only V(\t2g)
but also S($3p$) and V(\eg) orbitals. However, in this work we wanted to
investigate the properties and reliability of the minimal, i.e., \t2g-only,
model. In the following we discuss the results obtained via the MLWF
construction. Corresponding studies were also performed with an NMTO basis set
leading to the same physical picture. But a detailed comparison would at this
point shift the attention from the investigated physical mechanisms.

\begingroup
\begin{table}[b]
\caption{Wannier centers $\bR_w$ and spread $\langle r^2\rangle$ of \t2g-like
MLWFs for \bavs3 constructed from a (6$\times$6$\times$6) $k$-point mesh. The
positions of the symmetrically equivalent V sites in cartesian coordinates read
$\bR_{\rm{V(1)}}$=(0.00,0.46,-0.01) a.u. and
$\bR_{\rm{V(2)}}$=(0.00,-0.46,5.28) a.u.. The V(2) site is symmetry-related
to the V(1) site by the symmetry operation $C_2^{(z)}\bR_{\rm{V(1)}}$+0.5.
\label{table-ba1a}}
\begin{ruledtabular}
\begin{tabular}{c|c|c|c}
WF & $\bR_w$ (a.u.) & $\bR_w-\bR_{\rm{V}}$ (a.u.) & $\langle r^2\rangle$
(a.u.$^2$) \\ \hline
\A1g, V(1) & 0.00,  0.75, -0.20  & 0.00,  0.30, -0.19 & 16.57    \\
$E_{g1}$, V(1) & 0.00,  0.64,\hspace{0.1cm}  0.38  & 0.00,  0.18,\hspace{0.1cm}  0.39 & 17.55  \\
$E_{g2}$, V(1) & 0.00,  1.02, -0.32  & 0.00,  0.56, -0.31 & 17.53
\end{tabular}
\end{ruledtabular}
\caption{Wannier centers $\bR_w$ and spread $\langle r^2\rangle$ as in
Tab.~\ref{table-ba1a}, now in the crystal-field basis.\label{table-ba1b}}
\begin{ruledtabular}
\begin{tabular}{c|c|c|c}
WF & $\bR_w$ (a.u.) & $\bR_w-\bR_{\rm{V}}$ (a.u.) & $\langle r^2\rangle$
(a.u.$^2$) \\ \hline
\A1g, V(1) & 0.00,  0.75, -0.17  & 0.00,  0.30, -0.16 & 16.60    \\
$E_{g1}$, V(1) & 0.00,  0.65,\hspace{0.1cm}  0.34  & 0.00,  0.19,\hspace{0.1cm}  0.35 & 17.55  \\
$E_{g2}$, V(1) & 0.00,  1.02, -0.32  & 0.00,  0.56, -0.31 & 17.53
\end{tabular}
\end{ruledtabular}
\end{table}
\endgroup

In order to downfold onto $\{A_{1g},E_{g1},E_{g2}\}$ we employed the
disentangling procedure~\cite{sou01} of the MLWF construction.
The WFs were initialized via cubic harmonics adapted to an ideal local hexagonal
symmetry. To the aim of correct disentangling of the six Wannier target bands we
provided twenty bands in an outer energy window around the Fermi level for the
construction of $M_{\nu\nu'}^{(\bk,\bq)}$. In order to reproduce the LDA
FS and the band dispersions close to the Fermi level correctly, we additionally
forced the Wannier bands in an inner energy window near \fermi to coincide with
the true LDA bands~\cite{sou01}. The initial WFs correspond to an optimized
$\Omega_I$=101.58 a.u.$^2$ and a starting value $\tilde{\Omega}$=3.62 a.u.$^2$,
hence a total $\Omega$ of 105.19 a.u.$^2$ for the chosen energy windows. After
$\sim$50000 iteration steps $\Omega$ finally converged to 103.30 a.u.$^2$.
During the minimization process, adaptation of the WFs to the true orthorhombic
symmetry was clearly observed by the occurence of distinct steps in $\Omega$. The
resulting \t2g Wannier bands are shown in Fig.~\ref{bavs3wbands}b in comparison
with the original LDA band structure. It is seen that the Wannier bands at the
Fermi level are truly pinned to the original LDA bands. Furthermore the
interpolated lowest/highest Wannier band follows nicely the former \A1g fatbands.
The same \t2g dispersion is also obtained within an NMTO contruction.

\begin{figure}[t]
\includegraphics*[width=8cm]{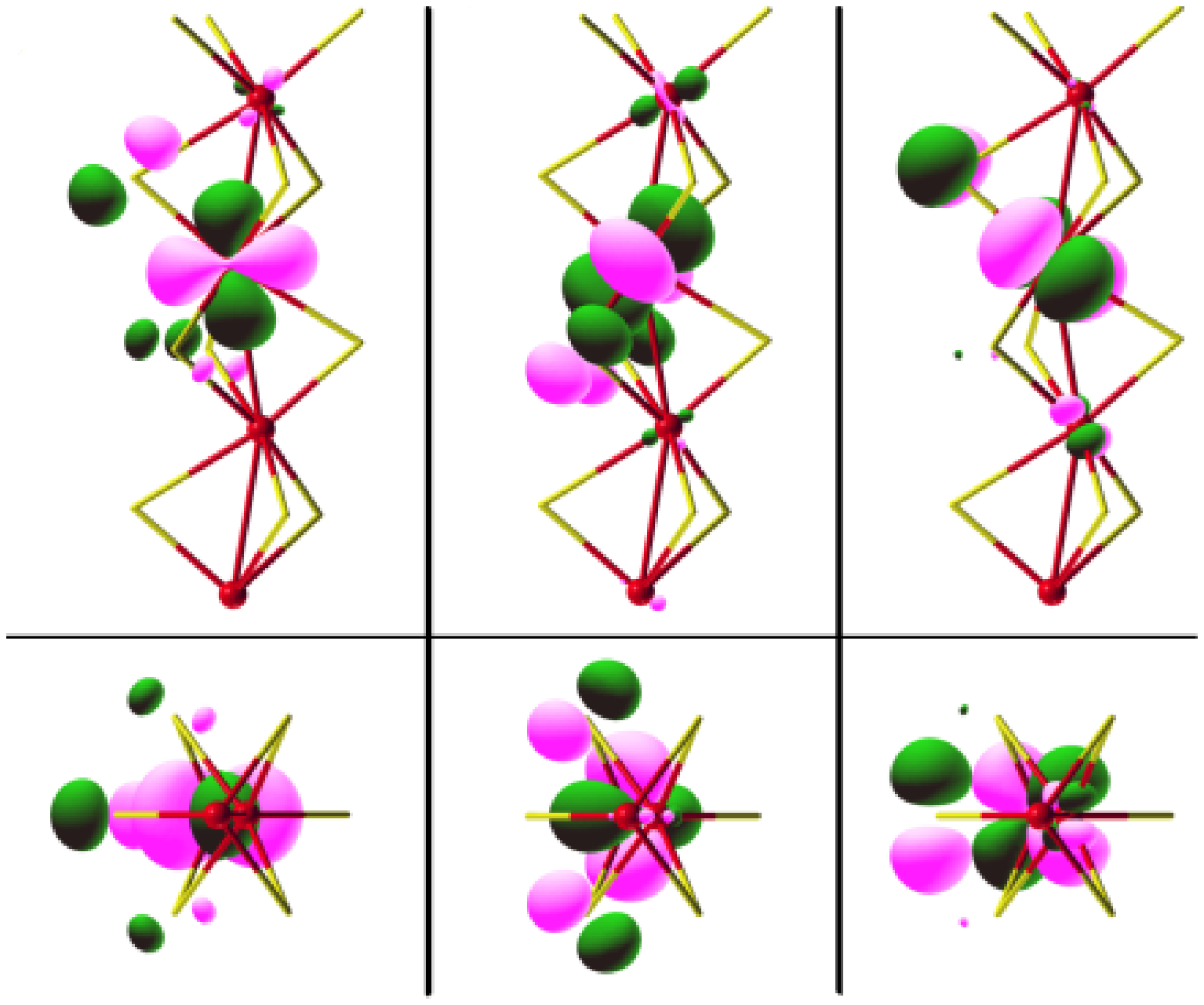}
\caption{(Color online) \t2g Wannier orbitals for BaVS$_3$ forming the 
maximally-localized basis.
The columns from left to right show the \A1g, $E_{g1}$ and $E_{g2}$ orbitals,
while the second row displays the orbitals viewed along the $c$ axis. There the
$a$ axis is vertically oriented, while the $b$ axis horizontally.
The contour value for each of the orbitals was chosen as
0.045 (a.u.)$^{-3/2}$.\label{bavs3worbnorm}}
\includegraphics*[width=8cm]{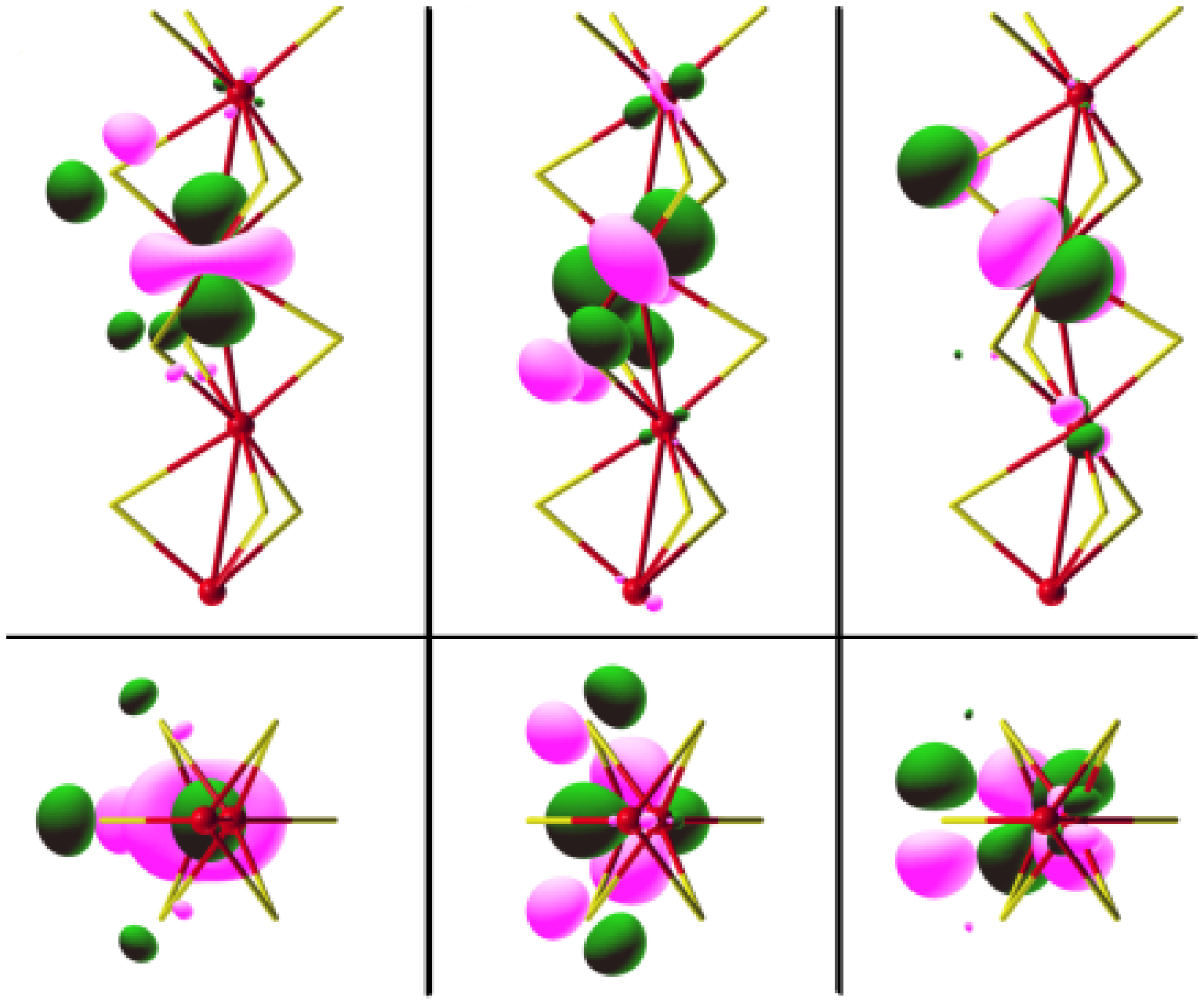}
\caption{(Color online) \t2g Wannier orbitals for BaVS$_3$ forming the 
crystal-field basis. The
order of the columns and row same as in Fig.\ref{bavs3worbnorm}. Note that the
$E_{g2}$ orbital remains invariant under this transformation. The contour value
for each of the orbitals was equally chosen as 0.045 (a.u.)$^{-3/2}$.
\label{bavs3worbdiag}}
\end{figure}
\begingroup
\begin{table}[b]
\caption{Hopping integrals between the \t2g Wannier orbitals of \bavs3 in the
maximally-localized basis. The term '00$\frac{1}{2}$' shall
denote the hopping to the nearest-neighbor V site within the unit cell. One of
the nearest-neighbor V ions in the $ab$ plane is located at '100', while '110'
and '$\bar{1}$10' are closest V ions along $a$ and $b$, respectively.
Energies in meV.
\label{table-bahop1}}
\begin{ruledtabular}
\begin{tabular}{c|r|r|r|r|r|r}
           & \A1g-\A1g  & $E_{g1}$-$E_{g1}$ & $E_{g2}$-$E_{g2}$ & \A1g-$E_{g1}$ & \A1g-$E_{g2}$ & $E_{g1}$-$E_{g2}$ \\ \hline
000            &  414.4   &  218.0 &  235.6 &   40.8 &  0.0 &   0.0  \\
00$\frac{1}{2}$& -441.5   &  -24.7 &  -12.4 & -242.6 &  0.0 &   0.0  \\
001            &  -66.0   &   -5.4 &    2.7 &   -8.9 &  0.0 &   0.0  \\
100            &  -30.4   &    8.5 &  -26.1 &  -15.8 & 16.4 & -10.9  \\
110            &  -17.4   &  -84.4 &   29.2 &  -16.5 & -0.7 &  11.6  \\
$\bar{1}$10    &    1.6   &    1.7 &   -5.5 &   -1.8 &  0.0 &   0.0  \\
\end{tabular}
\end{ruledtabular}
\caption{Hopping integrals as in Tab.~\ref{table-bahop1} but now in the
crystal-field basis.\label{table-bahop2}}
\begin{ruledtabular}
\begin{tabular}{c|r|r|r|r|r|r}
           & \A1g-\A1g  & $E_{g1}$-$E_{g1}$ & $E_{g2}$-$E_{g2}$ & \A1g-$E_{g1}$ & \A1g-$E_{g2}$ & $E_{g1}$-$E_{g2}$ \\ \hline
000            &  422.6  & 209.8 & 235.6 &    0.0 &   0.0 &   0.0 \\
00$\frac{1}{2}$& -510.5  &  44.3 & -12.4 & -145.7 &   0.0 &   0.0 \\
001            &  -85.5  &  14.2 &   2.7 &    7.1 &   0.0 &   0.0 \\
100            &  -35.4  &  13.6 & -26.1 &   -7.0 & -14.0 &  13.9 \\
110            &  -26.3  & -75.5 &  29.2 &  -28.1 &  -1.6 & -11.5 \\
$\bar{1}$10    &    1.2  &   2.1 &  -5.5 &   -1.7 &   0.0 &   0.0 \\
\end{tabular}
\end{ruledtabular}
\end{table}
\endgroup
Table~\ref{table-ba1a} reveals that the Wannier spreads are
significantly larger than for the \t2g WFs in \srvo3. The spread for the \A1g WF
is slightly smaller than those for the \eeg orbitals. Moreover, it is seen from
Tab.~\ref{table-ba1a} that the Wannier centers are shifted from the V sites.
This may be explained by the low symmetry of the $Cmc2_1$-\bavs3 structure, since
already from the missing inversion symmetry in the site symmetry for the V ion
there is no need for the WFs to be centered on the V sites~\cite{mar97}.
Additionally, the explicit inclusion of the sulfur contribution to the \t2g
WFs leads to directed Wannier orbitals. The latter may be inspected in
Fig.~\ref{bavs3worbnorm}, where we plotted the three Wannier orbitals
associated with the first V ion in the unit cell (the other three are directly
related by symmetry due to the equivalence of the V ions). Indeed the
$A_{1g}$-like orbital is directed along the $c$ axis, whereby the zigzag
distortion of the chains causes some tilting. Note that the orbitals
have some weight on neighboring V sites. As especially observable for the
$E_{g1}$-like orbital, this weight has local $e_g$ symmetry. Thus the MLWF
construction reproduces here the intuitively formulated symmetry constraints
imposed in the NMTO construction~\cite{and00,and00-2,zur05}. This symmetry
relation between WFs on neigboring sites has also been noted for the V$_2$O$_3$
compound~\cite{tanusri-tbp}, half-metallic ferromagnets~\cite{yam06_2} and, 
most dramatically, for Na$_x$CoO$_2$ compounds~\cite{and-coo2-tbp}. 
The $E_{g1}$ orbital
is mainly oriented in the plane defined by the S(2) ions, while $E_{g2}$ remains
in the corresponding perpendicular plane. Accordingly, $E_{g1}$ has stronger
weight on the S(2) ions, whereas $E_{g2}$ connects to the apical S(1). For
\A1g the sulfur distribution is in-between, yet favoring the nearest-neighbor
S(1).
\begin{figure}[t]
\includegraphics*[width=8cm]{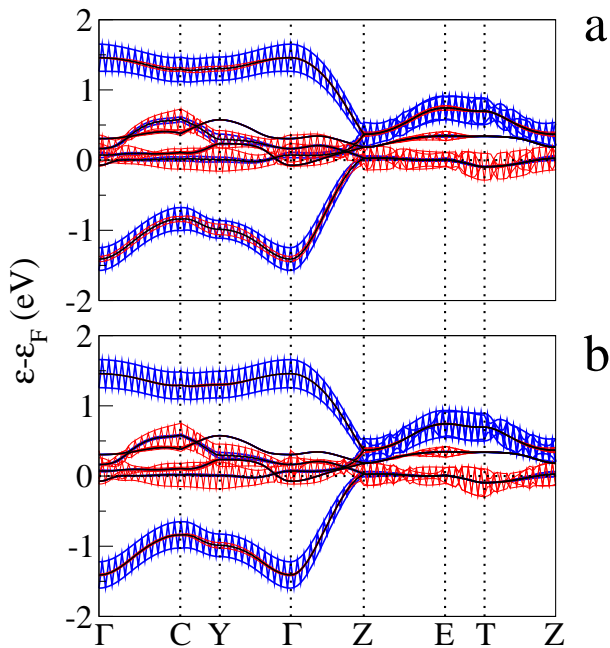}
\caption{(Color online) \t2g Wannier fatbands for \bavs3, (a) with respect to the
maximally-localized basis, and (b) with respect to the crystal-field basis
(see text). Color coding: \A1g-like WF (blue/dark), $E_{g1}$-like WF (red/gray).
\label{bavs3wbfatands}}
\end{figure}

In order to investigate the properties of these WFs in more detail,
Tab.~\ref{table-bahop1} displays the hopping integrals for relevant paths on the
lattice. One immediately realizes that while the $E_{g2}$ orbital remains
isolated, there is a sizable hybridization between the $A_{1g}$ and
$E_{g1}$ Wannier orbitals. The latter hybridization clearly couples $A_{1g}$ and
$E_{g1}$ in view of the zigzag distortion along the $b$ axis
(see Fig.~\ref{bavs3worbnorm}) This coupling should play an important role for
the understanding of the whole \bavs3 phase diagram, also crucial for the
hexagonal-to-orthorhombic structural transition~\cite{lecto}. Note also the
dominant $E_{g1}$-$E_{g1}$ hopping along the $a$ axis, i.e. [110], which is in
accordance with orbital extension of the corresponding WFs
(see Fig.~\ref{bavs3worbnorm}).

In Figure~\ref{bavs3wbfatands}a we plot the disentangled (effective) \t2g bands
with a fatband resolution to reveal the respective contribution of the
obtained MLWFs (for simplicity we do not plot the $E_{g2}$ fatband).
One can see that the \A1g band carries substantial $E_{g1}$
weight (and vice versa) due to their sizeable hybridization in the
maximally-localized basis. This is is somehow counter-intuitive to the original
low-energy picture of a broader \A1g band and narrower $E_{g1}$ band. Though some
minor \A1g-$E_{g1}$ hybridization in the $\Gamma$-C-Y plane is in line with the
original LDA fatbands, it seems that this hybridization is slightly
over-represented in the maximally-localized basis. Surely, the latter
basis is not {\sl a priori} physically designated, and it might be that the
the straightforward MLWF construction does not provide the most suitable
physical Wannier basis for the description of \bavs3.

With the aim of reducing this strong \A1g-$E_{g1}$ hybridization in the
maximally-localized basis, we diagonalized the on-site Wannier Hamiltonian
$\bH_{\rm{KS}}^{(\bT=\mathbf{0})}$=$\sum_{\bk}\bH_{\rm{KS}}(\bk)$ and
transformed $\bH_{\rm{KS}}(\bk)$ into the so-called {\sl crystal-field basis}
(e.g. also utilized in Ref.~[\onlinecite{pav05}]). The (again unitary)
transformation is explicitly written as:
\[
\begin{pmatrix}
w^{\rm{(cf)}}_{A_{1g}} \\ w^{\rm{(cf)}}_{E_{g1}} \\ w^{\rm{(cf)}}_{E_{g2}} 
\end{pmatrix} =
\begin{pmatrix} 0.981 & 0.196 &  0.000 \\
               -0.196 & 0.981 &  0.000 \\
                0.000 & 0.000 &  1.000 \\
\end{pmatrix}
\begin{pmatrix}
w^{\rm{(ml)}}_{A_{1g}} \\ w^{\rm{(ml)}}_{E_{g1}} \\ w^{\rm{(ml)}}_{E_{g2}} 
\end{pmatrix}\,,
\]
where the superscript 'cf' marks the crystal-field basis and 'ml' the
maximally-localized basis. This procedure obviously decouples \A1g and $E_{g1}$
on average and provides a true adaption to the local symmetry at the V site.
Within this new basis the \A1g hopping along the $c$ axis is strenghtend at the
cost of a reduced \A1g-$E_{g1}$ hybridization (see Tab.~\ref{table-bahop2}).
In addition, the sign of the near-neighbor $E_{g1}$-$E_{g1}$ hopping is changed 
from negative to positive. The Wannier fatbands promote now more the elucidated
picture of the $E_{g1}$ bands being confined to \fermi (see
Fig.~\ref{bavs3wbfatands}b). Only minor changes may be observed however in the
same contour plot for the transformed Wannier orbitals, as seen in
Fig.~\ref{bavs3worbdiag}. The \A1g Wannier orbital is now more reminescent
of a $d_{3z^2-r^2}$ orbital perpendicular to $c$ and $E_{g1}$ is slightly tilted
out of the plane defined by the S(2) ions. Correspondingly, the numbers for the
Wannier centers and spreads have changed only marginally, as seen
in Tab.~\ref{table-ba1b}.

By summing the spreads one easily checks that $\Omega$ is now marginally larger
than in the original maximally-localized basis, which is of course consistent
with the fact that the original set was constructed by minimizing the spread.

\begin{figure}[t]
\includegraphics*[width=8cm]{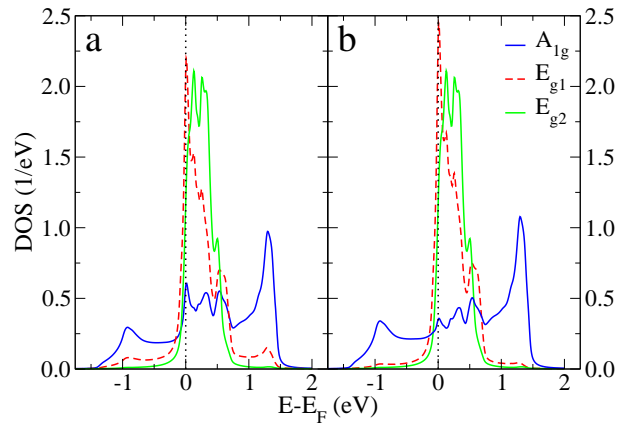}
\caption{(Color online) \t2g Wannier DOS for BaVS$_3$. (a) In the 
maximally-localized basis, and (b) in the crystal-field basis.\label{bavs3wdos}}
\end{figure}

Although the real-space quantities truly do not differ much, the electronic
structure representation is very sensitive to rather minor changes in the
basis. This is not only seen for the hoppings (cf.
Tabs~\ref{table-bahop1},\ref{table-bahop2}) but may also be observed when
comparing the different orbitally-resolved DOS originating from the two Wannier
basis sets (see Fig.~\ref{bavs3wdos}). There, in general the overall
lowdimensional character of the \A1g band is emphasized within the minimal
Wannier set. Note the reduced DOS magnitude close to the Fermi energy for the
latter band in the crystal-field basis. On the other hand the \A1g DOS is
reinforced below the Fermi energy at the cost of a reduced $E_{g1}$ DOS. This
effect precisely reflects the low-energy confinement of $E_{g1}$ in the
crystal-field basis.

\subsubsection{LDA+DMFT calculations\label{secdmftbavs3}}
\begin{figure}[b]
\includegraphics*[width=8cm]{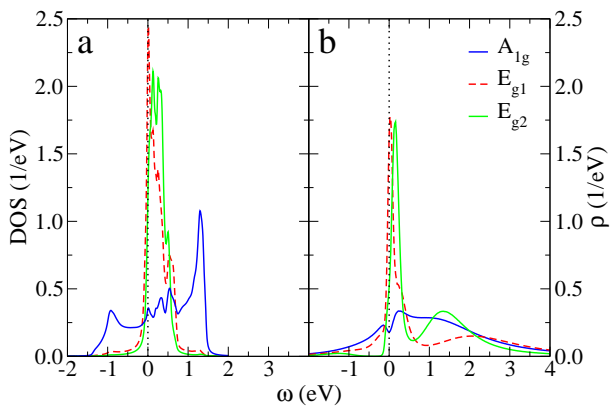}
\caption{(Color online) (a) Wannier DOS for \bavs3 in the crystal-field basis. (b)
corresponding $\bk$-integrated spectral function for \bavs3 within LDA+DMFT
for $\beta$=30 eV$^{-1}$.
\label{bavs3spec}}
\end{figure}
It was pointed out in Ref.~[\onlinecite{lec05}] that correlation effects are
important for an understanding of the MIT in \bavs3. From LDA+DMFT calculations
using as an input the symmetry-adapted local DOS for
$\{A_{1g},E_{g1},E_{g2}\}$, a substantial charge transfer from \A1g to $E_g$ was
revealed. The latter was associated with a reduced Fermi wave vector $k_F$ for
the \A1g band, allowing for the possibility of the observed CDW instability.
In this section we will check, and also considerably refine these earlier results.
Indeed, because the Wannier construction allows us to study this intricate
material within a Hamiltonian formalism which allows us to address $\bk$-resolved
issues, we are now in a position to study in detail the correlation-induced 
changes of the Fermi surface of \bavs3. Since we believe that the crystal-field 
basis constructed in the last section is more closely adapted to the physics of 
\bavs3 than the direct MLWF basis, we will use this basis in the following.
We made cross checks using the MLWF basis, and will comment on relevant 
differences.

For the new LDA+DMFT calculations we chose $U$=3.5 eV, $J$=0.7 eV. Although no
concrete knowledge about the value of $U$ for \bavs3 is known, we believe that
the latter values are in agreement with realistic values for this
compound~\cite{lec05}. The single-site impurity problem was solved for
$\beta$=30 eV$^{-1}$ ($\sim$390 K). A number of 128 time slices and up to
10$^6$ Monte-Carlo sweeps were used in the actual calculations.

In Fig.~\ref{bavs3spec} the orbital-resolved $\bk$-integrated spectral function
is shown in comparision to the corresponding LDA DOS in the Wannier basis. The
band-narrowing effect is clearly visible. Additionally, the tendency towards the
opening of a pseudo-like gap in the \A1g band may be oberved. This latter feature
is missing when using the maximally-localized Wannier basis.

\begin{table}[b]
\caption{Band fillings for \bavs3 from LDA+DMFT within the crystal-field
Wannier basis.\label{table-ba2}}
\begin{ruledtabular}
\begin{tabular}{l|l|ccc}
$\beta$ (eV$^{-1}$) & $U$,$J$ (eV) & $A_{1g}$ & $E_{g1}$ & $E_{g2}$ \\ \hline
 $\rightarrow\infty$ & 0.0, 0.0 & 0.59 & 0.31 & 0.10 \\
 30                  & 0.0, 0.0 & 0.58 & 0.30 & 0.12 \\
 30                  & 3.5, 0.7 & 0.41 & 0.45 & 0.14 \\
\end{tabular}
\end{ruledtabular}
\end{table}
Table~\ref{table-ba2} shows final occupations of the Wannier orbitals varying
with temperature and interaction strength. The values for
($\beta\rightarrow\infty$,$U$=$J$=0) correspond to the LDA limit. Comparing
with LDA values obtained in Ref.~[\onlinecite{lec05}], the current LDA
orbital polarization is not so severe in the Wannier description. This is due
to the fact that in Ref.[\onlinecite{lec05}] we used an empirical downfolding
for the local DOS, merging most sulfur character with the \A1g band. Since in the
Wannier description the downfolding of sulfur now also puts some weight on
the $E_g$ states, their filling is somehow increased. Nonetheless, turning on
$U$ does transfer sizable charge between the orbitals,
hence the correlation effects envisioned in Ref.~[\onlinecite{lec05}] are indeed
confirmed in the more elaborate Hamiltonian
framework with WFs. However, due to the now resolved
\A1g-$E_{g1}$ hybridization, the charge transfer dominantly takes place between
these two orbitals, leaving $E_{g2}$ as a mere ``spectator''.
These interorbital charge transfers suggest that the FS of this 
material might actually be quite different than predicted by LDA, namely that the
relative size of the various FS sheets may be significantly changed.
A word of caution is in order however: the Luttinger theorem~\cite{lut60} only 
constrains the
{\it total} $k$-space volume encompassed by all sheets of the FS, and stipulates
that it should correspond to one electron per vanadium for the present material,
independently of interactions, in the metallic phase.
There is no a priori theoretical relation between the volume of each individual
sheet and the orbital populations as calculated above. Nonetheless, the reduction
of the $A_{1g}$ population in favor of $E_{1g}$ may provide a hint that there
is a corresponding shrinking of the $A_{1g}$ FS sheet. We now address in details
whether this shrinking does occur and along which directions in reciprocal space.

In order to investigate $\bk$-resolved effects of correlations, one must
in principle determine the real-frequency self-energy. Since we are however mainly
interested in the Fermi surface of the interacting system, we can extract the
low-energy expansion of $\Sigma(\omega)$ from our QMC calculation in the form:
\begin{equation}
\Re\Sigma_{mm'}(\omega+i0^+)\simeq\Re\Sigma_{mm'}(0)+
\left(1-[\mathbf{Z}^{-1}]_{mm'}\right)\omega+\cdots,
\end{equation}
where $\mathbf{Z}$ describes the matrix of QP weights. The QP dispersion relation
is then obtained from the poles of the Green's function:
\begin{equation}
\mbox{det}[\omega_{\bk}\mathbf{\openone}-\mathbf{Z}\left(\bH_{\rm{KS}}(\bk)+
\Re\mathbf{\Sigma}(0)-\mu\mathbf{\openone}\right)]=0\quad.\label{poleq}
\end{equation}
Correspondingly, the FS in the interacting system is defined by
\begin{equation}
\mbox{det}[\mu\mathbf{\openone}-\bH_{\rm{KS}}(\bk)-\Re\Sigma(0)]=0\quad.
\label{intferm}
\end{equation}
From Eqs. (\ref{poleq}-\ref{intferm}) one understands that
$\Re\mathbf{\Sigma}(0)$ provides an energy shift to the LDA bands. The direction
and magnitude of this shift depends at each $k$-point on the amount of the
contributing orbital character, since $\Re\mathbf{\Sigma}(0)$ is explicitly
orbital-dependent. Hence although our self-energy is explicitly
$\bk$ independent within single-site DMFT, one may still evaluate some
$\bk$-dependent effects due to the explicit $\bH_{\rm{KS}}(\bk)$ inclusion.

From the converged self-energy matrix $\mathbf{\Sigma}(i\omega_n)$ we derived
$\Re\mathbf{\Sigma}(0)$ and $\mathbf{Z}$ via Pade approximation.
Table~\ref{table-ba3} displays the two matrices.
\begin{table}[t]
\caption{$\Re\mathbf{\Sigma}(0)$ (in eV) and $\mathbf{Z}$ for \bavs3 in the
Wannier crystal-field basis from LDA+DMFT.\label{table-ba3}}
\begin{ruledtabular}
\begin{tabular}{c|rrr|rrr}
     &    & $\Re\mathbf{\Sigma}(0)$ & & & $\mathbf{Z}$ & \\
     & $A_{1g}$ & $E_{g1}$ & $E_{g2}$ & $A_{1g}$ & $E_{g1}$ & $E_{g2}$\\ \hline
$A_{1g}$ &  0.35 & -0.01 & 0.00 &  0.57 & -0.02 & 0.00 \\
$E_{g1}$ & -0.01 & -0.03 & 0.00 & -0.02 &  0.52 & 0.00 \\
$E_{g2}$ &  0.00 &  0.00 & 0.03 &  0.00 &  0.00 & 0.58 \\
\end{tabular}
\end{ruledtabular}
\end{table}
First, the symmetry of these matrices follows the earlier observations, hence
there is some \A1g-$E_{g1}$ coupling also reflected in the self-energy. The band
renormalization cast into $\mathbf{Z}$ are roughly of the same order, with a 
slight
maximum band narrowing close to a factor of two for the $E_{g1}$ band. Since
we want to elucidate FS deformations due to correlations,
looking for an explanation for the CDW instability, the $\Re\mathbf{\Sigma}(0)$
matrix is of high relevance. From that, states with strong \A1g character should
considerably shifted upwards in energy, while for dominantly $E_g$ states it
depends on their symmetry. Bands with $E_{g1}$ character should be shifted down,
whereas those with $E_{g2}$ character should be shifted up. The latter
discrepancy may be relevant to understand the opening of the gap at the MIT
within the $E_g$ states. Since the dispersion of the \A1g-like band is highly
anisotropic, i.e. 1D-like, the rather strong shifting should result in a major
FS deformation, invoking the possibility for an arising CDW instability.
\begin{figure}[b]
\includegraphics*[width=8cm]{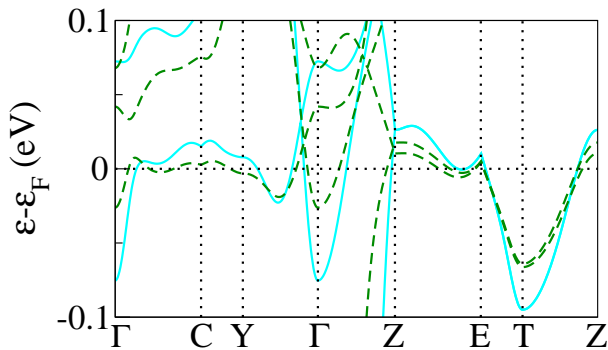}
\caption{(Color online) $\bk$-resolved correlation effect for \bavs3. QP band
structure from LDA+DMFT (darkgreen dashed/dark dashed) in comparision with the 
Wannier band structure based on LDA (cyan/light gray) close to the Fermi level.
\label{bavs3-qp}}
\end{figure}

Fig.~\ref{bavs3-qp} displays the QP band structure according to Eq.
(\ref{poleq}) close to the Fermi level. Using the linearized self-energy should
be valid within the small energy window around \fermi. Note first the overall
narrowing of the bands since $Z$$<$1. The former statements
concerning the respective shifts of the bands are accordingly reproduced. The
electron pocket at the $\Gamma$ point was identified as $E_{g2}$ like
(compare Fig.~\ref{bavs3wbands}a) and hence is now shifted upwards. 
Roughly speaking, the system is getting rid of the $E_{g2}$ states when turning
on correlations. Also the $E_{g1}$-like bands are considerably shifted downwards.
In fact it is revealed in studies of the monoclinic phase below $T_{\rm{MIT}}$, 
that the internal $E_g$ splitting seems indeed realized in the way outlined 
above~\cite{lecto}.
Moreover, the off-diagonal \A1g-$E_{g1}$ self-energy terms within LDA+DMFT lift 
the band degeneracy in the E-T-Z plane of the BZ. Originally it was 
argued~\cite{mattheiss_bavs3_1995} within the LDA picture that the presence of 
the doubly-degenerated bands at this zone boundary ensures the metallic 
properties of $Cmc2_1$-\bavs3. Concerning the \A1g-like
band along $\Gamma$-Z, there is a small shift to a lower $k_F$, however not as
strong as expected from Tab.~\ref{table-ba3}. This may be explained by the fact
that the character of this band gains substantial $E_{g1}$ weight when
approaching the Fermi level (to be observed in Fig.~\ref{bavs3wbfatands}). Thus
the states away from \fermi are strongly shifted upwards but this shift weakens
for the very low-energy regime. Recall that also in Ref.~[\onlinecite{lec05}]
and recent angle-resolved photoemission (ARPES) measurements~\cite{mit05,mo05} the
shift to a lower $k_F$(\A1g) along $\Gamma$-Z was also not too strong.

\begin{figure}[t]
\hspace*{-2.75cm}{\large a}\\
\includegraphics*[width=3.6cm]{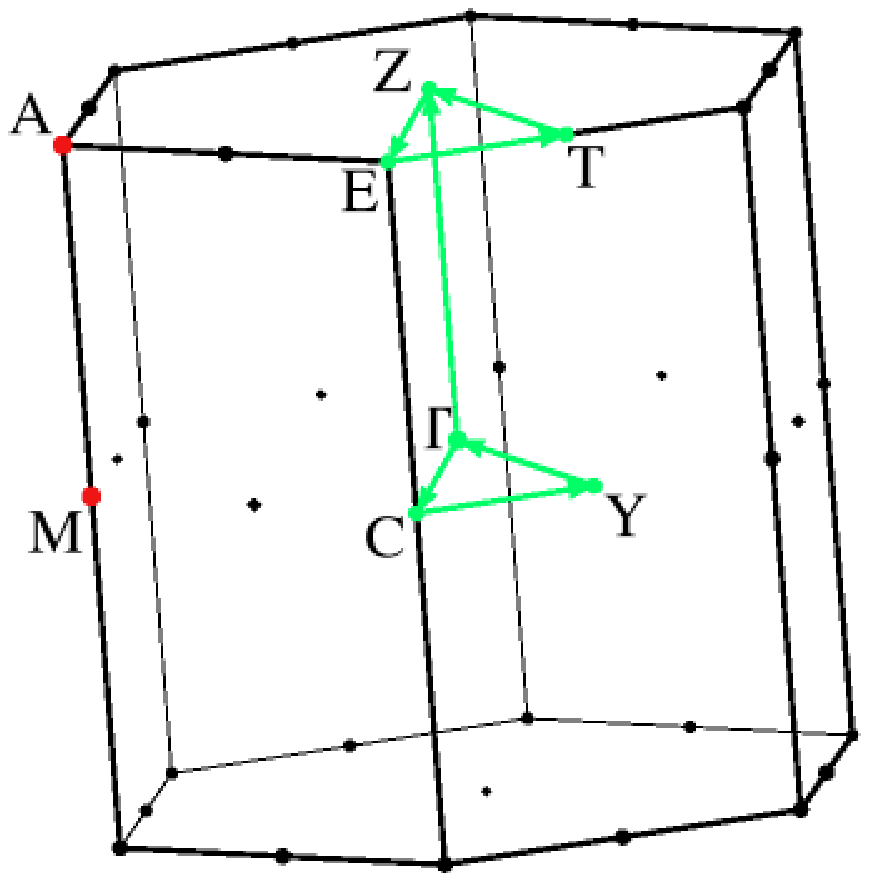}\\[0.2cm]
\hspace*{-0.3cm}{\large b.$\quad$LDA}
\hspace{2.4cm}{\large c.$\quad$LDA+DMFT}\\[0.2cm]
\includegraphics*[width=8cm]{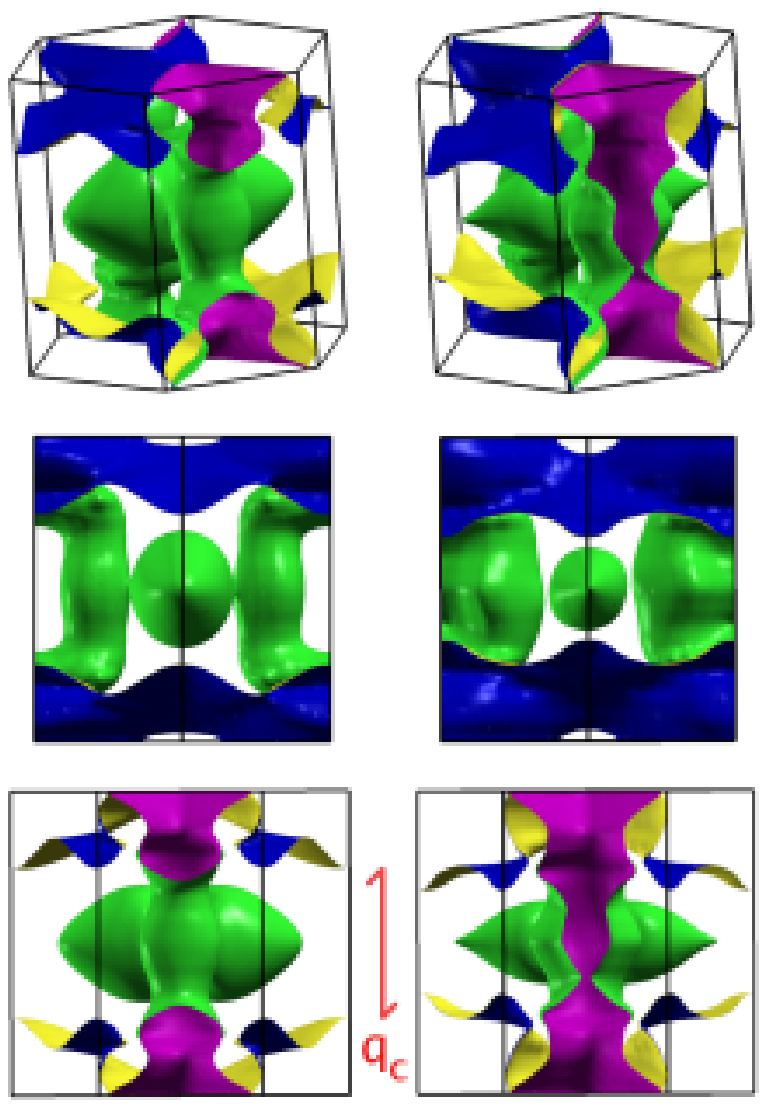}
\caption{(Color online) (a) 1. BZ of the $Cmc2_1$ structure, including the
high-symmetry lines used to plot the band structures. The $a^*$ axis runs
towards M, the $b^*$ axis towards $Y$. (b column) LDA Fermi
surface, and (c column) QP Fermi surface from LDA+DMFT. The
second row of the columns show the BZ along $a^*$, the third row along
$b^*$. In the latter column the experimental CDW nesting vector $\bq_c$ is
displayed.\label{bavs3-fs}}
\end{figure}
\begin{figure}[t]
\includegraphics*[width=8cm]{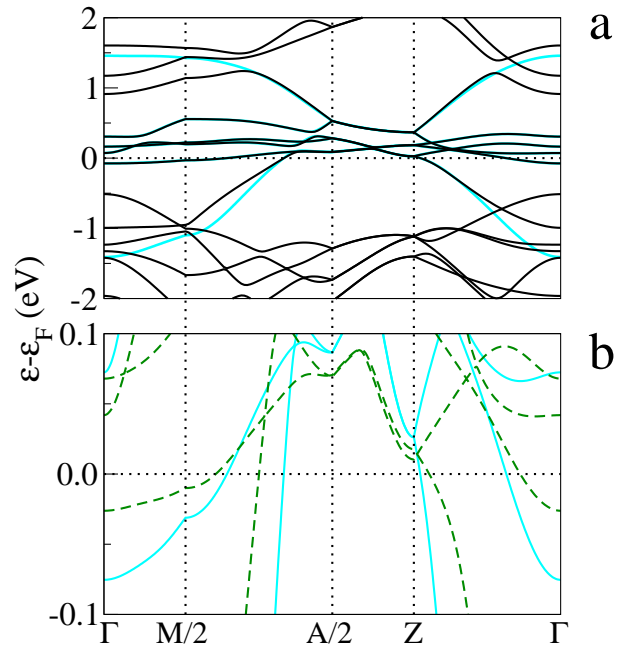}
\caption{(Color online) (a) LDA band structure (dark) and according Wannier bands 
(cyan/light gray) along a closed  path in the 1. BZ including the line 
connecting ``M/2''and ``A/2'' (see text). (b) pure (cyan/light gray) and 
renormalized (darkgreen dashed/dark dashed) Wannier bands.
\label{bavs3-newbands}}
\end{figure}

To proceed it is important to realize that studying only the single
high-symmetry line along $\Gamma$-Z in order to reveal a possible CDW instability
may not be sufficient. In fact FS nesting can only be thoroughly
investigated by taking into account the complete FS in the full BZ.
With this aim we have plotted the LDA FS and the deformed
LDA+DMFT FS in Fig.~\ref{bavs3-fs}. In LDA the FS consists of two
sheets which for the main part can truly be associated to \A1g and $E_g$. The
\A1g-like one shows the expected strong 1D behavior, i.e., corresponds to two
main surfaces extending dominantly along ($k_x$,$k_y$), however with no real
flattening. As already stated, the distance between these two surface parts is
too large to account for the experimentally observed nesting. The second sheet
incorporates the $E_{g2}$ electron pocket at $\Gamma$ as well as two
$E_{g1}$ ``pillars'' on the $b^*$ axis (where also $\Gamma$-Y runs), extending
along $c^*$ and opening towards the zone boundaries. Within LDA+DMFT the
$E_{g2}$ pocket shrinks and the $E_{g1}$ pillars thicken. Interestingly, the
first sheet is indeed only little modified in the immediate neighborhood of the
$c^*$ axis, however the further parts along $a^*$ tend to flap down towards lower
absolute values. Finally these latter parts are now not only within the distance
of the experimental nesting vector ${\bq}_c$=$0.5{\bf c}^*$, but are furthermore
somehow more flattened. That this shift indeed brings these outer
parts very close to half filling may be seen in Fig.~\ref{bavs3-newbands}. There
the Wannier band-structure plot includes a line with non-high symmetry endpoints
named ``M/2'' and ``A/2''. These latter points are halfway from $\Gamma$ to M and
halfway from Z to A (cf. Fig.~\ref{bavs3-fs}).
A true quantitative determination of the influence of the FS shape and of
nesting properties on the CDW instability will require a full calculation of 
the Lindhard function using the presently calculated self-energy. Although it is 
conceivable that a truly accurate account of the CDW transition for this material
would require going beyond a $\bk$-independent self-energy, we feel that the
interorbital charge transfer and corresponding FS deformation found in
the present work qualitatively points to the correct mechanism.

Let us finally comment on the role of the $E_g$ electrons in the MIT.
It appears from our findings on the FS that the CDW in this material is in fact
associated with the $A_{1g}$ electrons.
The question then arises of the fate of the remaining $E_g$ electrons through
the MIT. Of course, from a many-body viewpoint,
and in view of our present and earlier results~\cite{lec05}, one has to question
the band-like character of these states from early on.
The very narrow $E_g$ bandwidth causes a very low coherence
scale for the corresponding QPs. Thus above $T_{\rm{MIT}}$
the $E_g$ QPs have not yet reached their coherence scale~\cite{lec05},
which leaves them essentially localized already in the ``metal''. In a sense, 
this is a realisation of an ``orbital selective'' phase~\cite{ani02} at 
intermediate temperatures, i.e., a coexistence region of correlation-induced 
localized states and still itinerant states for $T$$>$$T_{\mbox{\tiny{MIT}}}$.
In such a regime, the $A_{1g}$ electrons acquire a large scattering rate due to 
the presence of almost localized $E_g$ degrees of freedom, as studied in
Refs.~[\onlinecite{lie05,bie05}]. This is a likely explanation of
the ``bad-metal'' behavior observed above the 
MIT\cite{lie05,bie05}\footnote{One should realize however that due to the quarter 
filling of the \eeg orbitals such an orbital-selective scenario is necessarily more
complex than so far studied in simple models.}, and is also
in good agreement with recent optical studies~\cite{kez06}. We also performed
supporting two-site cluster-DMFT calculations, in order to allow for explicit 
intersite V-V self-energy terms, revealing no essential differences to the 
outlined picture.
A closer investigation of the CDW state, i.e., the electronic structure of the 
system in the paramagnetic insulating regime below $T_{\rm{MIT}}$ is however
necessary~\cite{fag05,fag06,lecto}, including their spin degree of freedom, to
reveal more details about the role of the \eeg elecrons.

\section{Conclusion and prospects}

In this paper, we have explored in detail the use of WFs as a
flexible technique to perform electronic-structure calculations within the
LDA+DMFT framework. WFs are useful for two different purposes.
First, as a physically meaningful way of defining the correlated orbitals
to which a many-body treatment will be applied.
Second, as a convenient choice of basis functions
for interfacing the many-body (DMFT) part of the calculation with virtually any
kind of band-structure method. In this paper, three different methods have
been used, namely a pseudopotential-based method (MBPP), an FLAPW method and
an LMTO-based method. We have applied this approach to two transition-metal
compounds. The first one, \srvo3, was chosen in view of its simplicity as
a testing ground for performing a detailed comparison between two different
Wannier constructions: the maximally-localized-Wannier-function construction
(MLWF) and the N-th order muffin-tin-orbital (NMTO) method. For this simple
material, very close agreement between the two methods was found.
The second material, \bavs3, was chosen in view of its physical interest and
of some open key questions. We have been able, in particular,
to make new precise statements about the correlation-induced changes of the Fermi
surface for this material, using our Wannier based method.

There are several directions in which the present work can be extended, and 
several
open issues which need further attention. Here, we outline just a few of them.
First, we emphasized above that the localisation properties of the Wannier
functions depend of course on the energy range (or alternatively the set of
bands) defining the subspace $\mathcal{W}$ in which the Wannier construction is
performed. The correlated orbitals are then picked as a subset
$\mathcal{C}$$\subseteq$$\mathcal{W}$, in general smaller than $\mathcal{W}$ 
itself. In the actual calculations performed in this paper, the minimal choice
$\mathcal{C}$=$\mathcal{W}$ was made, associated for the two materials that were 
considered to exhibit low-energy bands with dominant $\tg$ character. This of 
course, means that the correlated orbitals defining $\mathcal{C}$ had sizeable 
weights on ligand atoms. In contrast, one may want
to enlarge $\mathcal{W}$ (including in particular ligand bands) and define the
correlated subset from Wannier orbitals which would be more localized on the
transition-metal site. Exploring these various choices, and comparing them to
other choices in which the correlated orbitals are not constructed from WFs
(e.g are taken to be heads of LMTOs, or truncated WFs, or even atomic 
wavefunctions) is certainly worth further
investigations, particularly in the context of late transition-metal oxides.
An important related issue is the appropriate way of calculating the
local-interaction (Hubbard) parameters from first principles, for each of the
possible choices of correlated orbitals. Working with WFs will
make it easier to address this issue in a manner which is independent of the
underlying band-structure method and basis set used for performing the 
calculations in practice.
Finally, one should emphasize that the accuracy of the DMFT approximation depends 
on the choice made for the local orbitals defining $\mathcal{C}$. In fact, it has 
been recently suggested~\cite{pau06} that the local orbitals could be chosen
in such a way as to make the DMFT approximation optimal (according to some 
criterion on the magnitude of the local interactions). 

Another line of development that we are currently pursuing, is the practical
implementation of the self-consistency over the charge density (and of total 
energy), along the lines of Appendix~\ref{csc}, within non-LMTO based electronic 
structure methods.
A more remote perspective for such developments would be the possibility of 
allowing for local structural relaxation for correlated materials within LDA+DMFT.
Additionally, the Wannier-based formalism is also well suited for the 
calculation of response functions (e.g. optics). We hope to be able to address 
these issues in future work.

\appendix
\section{Self-consistency over the charge density in LDA+DMFT\label{csc}}

In this appendix, we briefly discuss the implementation of self-consistency over 
the charge density in the LDA+DMFT framework. Until now, this has been 
implemented in practice only in the 
MTO~\cite{savrasov_kotliar_pu_nature_2001,chi03,sav04,pouto} or 
KKR~\cite{min05} context. Here, we discuss the use of other basis sets, with 
particularly the Wannier framework in mind.

As outlined in section~(\ref{selfcons}), charge self-consistency means that the 
KS cycle and the DMFT loop are explicitly coupled, i.e., the charge density is 
calculated at the end of a DMFT cycle (including self-energy effects). From the 
new charge density, a new KS potential is obtained, and so on.

The charge density is calculated from the full Green's function of the solid. 
Without yet introducing a specific basis set, it is given by
 Eq.~(\ref{eq:local_density}), namely:
\begin{equation}
\rho(\br)=\frac{1}{\beta}\sum_n \left\langle\br\left|\left[
\iomn+\mu+\frac{\nabla^2}{2}-\hat{V}_{\rm{KS}}-\Delta\hat{\Sigma}
\right]^{-1}\right|\br\right\rangle\mbox{e}^{\iomn 0^+}\,. \label{locdens}
\end{equation}
We shall find it convenient to split $\rho(\br)$ into
\begin{eqnarray}
\rho(\br)&=&\rho_{\rm{KS}}(\br)+
\left[\rho(\br)-\rho_{\rm{KS}}(\br)\right]\nonumber\\
&\equiv&\rho_{\rm{KS}}(\br)+\Delta\rho(\br)\quad,\\
&&\hspace{-2.5cm}\mbox{with}\quad\rho_{\rm{KS}}(\br)=
\left\langle\br\left|\hat{G}_{\rm{KS}}\right|\br\right\rangle\quad.
\end{eqnarray}
It is important to realize here that the demand for charge neutrality is 
{\it not imposed} on $\rho_{\rm{KS}}(\br)$ but rather on $\rho(\br)$. The 
chemical potential $\mu$ must be therefore explicitly determined (at the end of 
a DMFT loop) in such a way that the total number of electrons is the correct one 
(i.e. charge neutrality is preserved): $N_e$=$\tr\hat{G}$=$\int d\br\rho(\br)$. 
This value of the chemical potential will in general {\it not} be such that 
$\tr\hat{G}_{\rm{KS}}$=$\int d\br\rho_{\rm{KS}}(\br)$ equals the total number of 
electrons $N_e$. This is quite natural, since the KS representation of the 
charge density by independent KS wavefunctions no longer holds in the LDA+DMFT 
formalism.

Formally, the charge-density correction $\Delta\rho$ may be expressed as
\begin{eqnarray}
\Delta\rho(\br)&=&\left\langle\br\left|\hat{G}-\hat{G}_{\rm{KS}}
\right|\br\right\rangle\nonumber\\
&=&\left\langle\br\left|\hat{G}_{\rm{KS}} 
\left(\hat{G}_{\rm{KS}}^{-1}-\hat{G}^{-1}\right)\hat{G}
\right|\br\right\rangle\nonumber\\
&=&\left\langle\br\left|\hat{G}_{\rm{KS}}\,\hat{\Delta\Sigma}\,\hat{G} 
\right|\br\right\rangle\quad.
\label{deltarho}
\end{eqnarray}

In a concrete implementation, this equation must be written in the specific 
basis set of interest. We shall do this with the Wannier formalism in mind, 
i.e. use the basis functions (\ref{wanneq}) corresponding to the set
$\mathcal{W}$, which are unitarily related to the Bloch functions by
\begin{equation}
|w_{\bk\alpha}\rangle\equiv\sum_{{\mathbf T}} 
\mbox{e}^{i\bk\cdot\bT} |w_{\bT\alpha}\rangle =
\sum_{\nu\in\,{\cal W}} U_{\alpha\nu}^{(\bk)} |\psiknu\rangle\,. \label{wanneq2}
\end{equation}
Since in the Wannier basis set $\bG_{\rm{KS}}$ and $\bG$ are block diagonal, 
they have common non-zero matrix elements only in the chosen subspace 
$\mathcal{W}$. Let us define the operator $\Delta\hat{N}$ in $\mathcal{W}$
by its matrix elements:
\begin{equation}
\Delta N_{\alpha\alpha'}^{(\bk)}\equiv\frac{1}{\beta}\sum_{n\,mm'}
G_{\alpha m}^{\rm{KS}}(\bk,\iomn)\,\Delta\Sigma_{mm'}(\iomn)\, 
G_{m'\alpha'}(\bk,\iomn)\,, \label{deltanop}
\end{equation}
Note that the indices $\alpha\alpha'$ run over $\mathcal{W}$, while the 
summation over $mm'$ runs only within the correlated subset $\mathcal{C}$. Note 
also that the frequncy summation in this expression converges rather quickly, 
since the terms in the sum decay as $1/\omega_n^2$ at large frequencies (hence 
no convergence factor e$^{\iomn0^+}$ is needed in the sum). The desired 
charge-density correction therefore reads:
\begin{equation}
\Delta\rho(\br)=\sum_{\bk\alpha\alpha'} \langle\br|w_{\bk\alpha}\rangle\, 
\Delta N_{\alpha\alpha'}^{(\bk)} \langle w_{\bk\alpha'}|\br\rangle\,
\label{deltarho2}
\end{equation}
Alternatively, (\ref{deltarho2}) can also be written in the Bloch basis as:
\begin{eqnarray}
\label{deltan} &&\Delta\rho(\br)= \sum_{\bk}
\sum_{\nu\nu'\in\mathcal{W}}\psi^{\hfill}_{\bk\nu}(\br)
\psi_{\bk\nu'}^*(\br)\,\Delta N_{\nu\nu'}^{(\bk)}\,,\\
&&\hspace{-1cm}\mbox{with}\quad\Delta N_{\nu\nu'}^{(\bk)}= 
\sum_{\alpha\alpha'}U^{(\bk)}_{\nu\alpha}\,\Delta
N_{\alpha\alpha'}^{(\bk)}\, U^{(\bk)*}_{\alpha'\nu'}
\end{eqnarray}
The KS part of the charge density is easily calculated within the given 
band-structure code:
\begin{equation}
\rho_{\rm{KS}}(\br)=\mathop{\sum_{\bk\nu}}_{\varepsilon_{\bk\nu}\le\mu}|
\psiknu(\br)|^2\quad. \label{rhoks}
\end{equation}
Note the difference between (\ref{deltan}) and (\ref{rhoks}), as in 
(\ref{deltan}) additional terms, off-diagonal in the band indices, contribute to 
$\Delta\rho(\br)$. In this respect it proves to be convenient to introduce the 
density matrix of KS orbitals:
\begin{equation}
D_{\nu'\nu}^{(\bk)}(\br)=\psi^{\hfill}_{\bk\nu}(\br)\psi^{*}_{\bk\nu'}(\br)\,,
\end{equation}
The two contributions to the charge-density may now be compactly written as:
\begin{eqnarray}
\rho_{\rm{KS}}(\br)&=&\mathop{\sum_{\bk\nu}}_{\varepsilon_{\bk\nu}\le\mu}
D^{(\bk)}_{\nu\nu}(\br)\quad,\\
\Delta\rho(\br)&=&\sum_{\bk}\tr_{\nu}\left[ \bD^{(\bk)}
\cdot{\bf \Delta N}^{(\bk)}\right]\quad.
\end{eqnarray}
Hence finally, we can write the full charge density, i.e., eq.~(\ref{locdens}), 
in the following explicit form:
\begin{equation}
\rho(\br)=\sum_{\bk\nu\nu'} D_{\nu'\nu}^{(\bk)}(\br) \Delta N^{(\bk)}_{\nu\nu'}+
\sum_\nu\Theta(\mu-\varepsilon_{\bk\nu}) D_{\nu\nu}^{(\bk)}(\br)\,. \label{exprho}
\end{equation}
Generalization to finite-$T$ DFT is straightforward.
Realize that in the Wannier implementation, the first (double) sum in
(\ref{exprho}) would only run 
over Bloch bands in the $\mathcal{W}$ set (since only there $\Delta N$ is 
non-zero), while the second sum runs over all filled Bloch states. Again, it is 
important that the chemical potential has been correctly updated at the end of 
the DMFT loop, so that the full charge density (\ref{exprho}) sums up to the 
correct total number of electrons, hence insuring charge neutrality.

Expression (\ref{exprho}) has in fact a general degree of validity, not limited 
to the Wannier implementation, provided the matrix $\Delta N$, originally 
calculated in the basis set in which the DMFT calculation has been performed, is 
correctly transformed to the Bloch basis set.

\section{Free-energy functional and total energy\label{appx:energy}}

For the sake of completeness, we briefly summarize in this appendix how the 
LDA+DMFT formalism for electronic structure, described in 
Sec.~\ref{theoryframe} can be derived from a free-energy functional. As pointed 
out by Savrasov and Kotliar~\cite{sav04}, a (``spectral-density-'')
functional of both the electron charge density $\rho(\br)$ and the on-site 
Green's function in the correlated subset $G^{\rm{loc}}_{mm'}$, can be 
constructed for this purpose. Let us emphasize that these are
independent quantities: because $\mathbf{G}_{\rm{loc}}$ is restricted to on-site 
components and to the correlated subset $\mathcal{C}$, the charge density 
$\rho(\br)$ cannot be reconstructed from it. A compact formula for the total 
energy can also be obtained from this functional 
formulation~\cite{ama06,georges_strong,sav04}, given at the end of this appendix.

The functional is constructed by introducing source terms 
$V_{\rm{KS}}(\br)$$-$$V_{\rm{ext}}(\br)$ and $\Delta\Sigma_{mm'}(\iomn)$, 
coupling to the operators $\hat{\psi}^\dagger(\br)\hat{\psi}(\br)$ and 
$\sum_{\bT}\chi^*_m(\brRT)\hat{\psi}(\br,\tau)\hat{\psi}^\dagger(\br',\tau')\chi_{m'}(\brrRT)$, respectively. It reads:
\begin{eqnarray}
&&\Omega[\rho,G^{loc}_{mm'};V_{\rm{KS}},\Delta\Sigma_{mm'}]=\\ \nonumber
&&-\frac{1}{\beta}\rm{tr}\ln\left[\iomn+\mu+\frac{\nabla^2}{2}-
\hat{V}_{\rm{KS}}-\Delta\hat{\Sigma}\right]\nonumber\\
&&-\int d\br\,\left(V_{\rm{KS}}(\br)-V_{\rm{ext}}(\br)\right)\rho(\br)
-\rm{tr}[\hat{G}_{loc}\Delta\hat{\Sigma}]\nonumber\\
&&+\frac{1}{2}\int d\br d\br'\frac{\rho(\br)\rho(\br')}{|\br-\br'|}
+ E_{\rm{xc}}[\rho]\nonumber\\
&&+\sum_\bT\left(\Phi_{\rm{imp}}[G^{loc}_{mm'}]-
\Phi_{\rm{DC}}[G^{loc}_{mm'}]\right)\label{bigom}
\end{eqnarray}
In this expression, $\Phi_{\rm{imp}}$ is the Luttinger-Ward functional of the 
quantum impurity model, and $\Phi_{\rm{DC}}$ a corresponding functional 
generating the double-counting correction. Note that, as usual, it is understood
that $\Delta\hat{\Sigma}$ in the logarithm term of (\ref{bigom}) represents the
self-energy correction on the lattice.

Variations of this functional with respect to the sources 
$\delta\Omega/\delta\,V_{\rm{KS}}$=0 and 
$\delta\Omega/\delta\Sigma_{mm'}$=0 yield the standard expression of the charge
density and local Green's function in terms of the full Green's function in the 
solid, Eq.~(\ref{glocdef}). The Legendre multiplier functions $V_{\rm{KS}}$ and 
$\Delta\Sigma$ can be eliminated in terms of $\rho$ and $G_{\rm{loc}}$, so that 
a functional only of the local observables is obtained:
\begin{equation}
\Gamma[\rho,G_{\rm{loc}}]=\Omega\left[\rho,G^{loc}_{mm'};
V_{\rm{KS}}[\rho,G_{\rm{loc}}],\Delta\Sigma[\rho,G_{\rm{loc}}]\right]\,.
\end{equation} 
Extremalization of this functional with respect to $\rho$ 
($\delta\Gamma/\delta\rho$=0) and 
$G_{\rm{loc}}$ ($\delta\Gamma/\delta\,G_{\rm{loc}}$=0) yields the expression of 
the KS potential and self-energy correction at self-consistency, i.e., the
self-consistency conditions (over both the local projected Green's function and 
the charge density) of the LDA+DMFT formalism.

Using the above expressions, the free-energy can be written as:
\begin{eqnarray}\nonumber
\Omega=&&\hspace{-0.2cm}\Omega_{\rm{DFT}}+\tr\ln\,\hat{G}_{\rm{KS}}^{-1}
-\tr\ln\,\hat{G}^{-1}-\tr[\hat{G}_{\rm{loc}}\hat{\Sigma}_{\rm{imp}}]\nonumber\\
&&\hspace{-0.2cm}+\tr[\hat{G}_{\rm{loc}}\hat{\Sigma}_{\rm{DC}}]
+\sum_\bT\left(\Phi_{\rm{imp}}-\Phi_{\rm{DC}}\right)\,. 
\label{eq:omega_lda+dmft_2}
\end{eqnarray}
In this expression, $\Omega_{\rm{DFT}}$ is the usual density-functional theory 
expression of the free-energy (calculated at the self-consistent LDA+DMFT charge 
density, however). Taking the zero-temperature limit of this expression leads to 
the following expression~\cite{ama06,sav04} of the total energy at $T$=0: 
\begin{eqnarray}
\hspace{-0.4cm}E_{\rm{LDA+DMFT}}=&&\hspace{-0.2cm}E_{\rm{DFT}}+
\langle\hat{H}_U\rangle-E_{\rm{DC}}\nonumber\\
&&\hspace{-3cm}+\sum_{\bk\alpha\alpha'}H^{\rm{KS}}_{\alpha\alpha'}(\bk) 
\left[\langle\hat{c}^\dagger_{\bk\alpha}
\hat{c}^{\hfill}_{\bk\alpha'}\rangle_{\rm{DMFT}}
-\langle\hat{c}^\dagger_{\bk\alpha}
\hat{c}^{\hfill}_{\bk\alpha'}\rangle_{\rm{KS}}\right]
\label{eq:energy_lda+dmft_2}
\end{eqnarray}
Note that $\alpha,\alpha'$ cover the full electronic Hilbert space.  Finally, it
should be understood that in deriving the LDA+DMFT equations from this 
free-energy functionals, we have implicitly assumed that the orbitals $\chi_m$ 
defining the correlated subset $\mathcal{C}$ were kept fixed. In practice
however, one may want also to optimze these orbitals, e.g. by minimizing the 
free-energy $\Omega$. Furthermore, we emphasize that in the implementation of 
the charge self-consistency described in Appendix~\ref{csc}, it was explicitly 
assumed that these orbitals were recalculated as Wannier functions associated 
with the new KS potential. This is done at each stage of the iteration over the 
charge density. If instead the local orbitals are frozen (e.g. from the LDA 
potential), then they are no longer unitarily related to the new set of KS 
orbitals corresponding to the new potential, and the formulas derived in 
Appendix~\ref{csc} have to be appropriately reconsidered.

\section{Technical details\label{bsappendix}}
In the following we provide further details on the band-structure
calculations performed in this work. Concerning the computations using
the MBPP code, due to the significant ionic character of the treated
compounds, the semicore states of Ba, Sr and V were treated as valence.
Hence the pseudopotentials were constructed for
Sr(($4s4p4d$)/Ba($5s5p5d$), V($3s3p3d$) and
O(($2s2p2d$)/S(($3s3p3d$). The partial-core correction was used in all
constructions. In the crystal calculations, localized functions were
introduced
for all valence states. Thereby, the localized functions are atomic pseudo
wavefunctions, either multiplied with a cut-off function ($sd$) or minus a
spherical Bessel function ($p$). The cut-off radius for the V($3d$) local
functions was chosen as 2.0~a.u.. The plane-wave cutoff energy
$E_{\rm{pw}}^{\rm{(cut)}}$ was 24 Ryd for \srvo3 and 20 Ryd for \bavs3,
and the Perdew-Wang exchange-correlation functional~\cite{per92} was used
in all calculations.

In the FLAPW calculation for \srvo3, the atomic-sphere radii for Sr, V,
and O were chosen to be 2.5, 2.0, and 1.55 a.u., respectively and the FLAPW
basis size was set to include all plane waves up to
$E_{\rm{pw}}^{\rm{(cut)}}$=20.25
Ryd . The local orbitals sector of the basis was used to include the high
lying V($3s$), V($3p$), Sr($4s$), Sr($4p$) and O($2s$) core states as valence
states in
the calculations. Additional V($3d$) local orbitals were used to relax the
linearization of the transition metal $d$-bands. The Hedin-Lundquist
exchange-correlation functional~\cite{hed71} was used in the calculation.

For the self-consistent LMTO calculation of the effective LDA potential
for \srvo3 we have used the prescription described in
Ref.~[\onlinecite{pav05}]. The radii of potential spheres were chosen to be
3.46~a.u., 2.3~a.u. and 2.04~a.u. for Sr, V and O atoms, respectively.
Twelve empty spheres were introduced above the octahedron edges with the
radii 1.01~a.u. The LMTO basis set used in the calculation were
$s(p)d(f)$, $spd$, $(s)p(d)$ and $s(p)(d)$ for Sr, V, O and
empty spheres, respectively. $(l)$ means that the $l$-partial waves were
downfolded within TB-LMTO-ASA. The von Barth-Hedin exchange-correlation
functional~\cite{bar72} was used in the calculation, which is very similar to 
the Hedin-Lundquist parametrization. These elder $E_{\rm{xc}}$
representations utilize the same interpolation formula.

The construction of the MLWFs from the disentangling procedure for \bavs3
was described in section \ref{bavs3wannsec}. In the case of \srvo3, no energy
windows have to be defined, since the \t2g-like bands form an isolated
group of bands.
The NMTO-WFs for \srvo3 were obtained by imposing that for a \t2g orbital, the 
\t2g character on any other V site must vanish. We used $N$=2 and chose the
energies for which this quadratic MTO set is complete as: $\epsilon_0$=-1.29~eV,
$\epsilon_1$=0~eV and $\epsilon_2$=1.61~eV with respect to Fermi level. Finally,
the NMTO set was symmetrically orthonormalized.

%
%

\acknowledgments
We are most grateful to N.~Marzari and to S.~Botti, C.~Els\"asser, G.~Kotliar, 
B.~Meyer, E.~Pavarini, I.~Souza, N.~Vast and J.~Yates for helpful discussions at 
various stages of this work. A.G. would like to thank F.~Mila for hospitality 
and support during an extended visit to EPFL.
Funding for this work has been provided by CNRS, \'Ecole Polytechnique and the
E.U. ``Psi-k $f$-electron'' Network under contract HPRN-CT-2002-00295.
Computations were performed at IDRIS Orsay under project No. 061393.

\bibliographystyle{apsrev}
\bibliography{bibag,bibextra,pubag}

\begin{thebibliography}{123}
\expandafter\ifx\csname natexlab\endcsname\relax\def\natexlab#1{#1}\fi
\expandafter\ifx\csname bibnamefont\endcsname\relax
  \def\bibnamefont#1{#1}\fi
\expandafter\ifx\csname bibfnamefont\endcsname\relax
  \def\bibfnamefont#1{#1}\fi
\expandafter\ifx\csname citenamefont\endcsname\relax
  \def\citenamefont#1{#1}\fi
\expandafter\ifx\csname url\endcsname\relax
  \def\url#1{\texttt{#1}}\fi
\expandafter\ifx\csname urlprefix\endcsname\relax\def\urlprefix{URL }\fi
\providecommand{\bibinfo}[2]{#2}
\providecommand{\eprint}[2][]{\url{#2}}

\bibitem[{\citenamefont{Kohn}(1999)}]{koh99}
\bibinfo{author}{\bibfnamefont{W.}~\bibnamefont{Kohn}}, \bibinfo{journal}{Rev.
  Mod. Phys.} \textbf{\bibinfo{volume}{71}}, \bibinfo{pages}{1253}
  (\bibinfo{year}{1999}).

\bibitem[{\citenamefont{Jones and Gunnarsson}(1989)}]{jon89}
\bibinfo{author}{\bibfnamefont{R.~O.} \bibnamefont{Jones}} \bibnamefont{and}
  \bibinfo{author}{\bibfnamefont{O.}~\bibnamefont{Gunnarsson}},
  \bibinfo{journal}{Rev. Mod. Phys.} \textbf{\bibinfo{volume}{61}},
  \bibinfo{pages}{689} (\bibinfo{year}{1989}).

\bibitem[{\citenamefont{Kohn and Sham}(1965)}]{koh65}
\bibinfo{author}{\bibfnamefont{W.}~\bibnamefont{Kohn}} \bibnamefont{and}
  \bibinfo{author}{\bibfnamefont{L.~J.} \bibnamefont{Sham}},
  \bibinfo{journal}{Phys. Rev.} \textbf{\bibinfo{volume}{140}},
  \bibinfo{pages}{A1133} (\bibinfo{year}{1965}).

\bibitem[{\citenamefont{Perdew and Schmidt}(2001)}]{per01}
\bibinfo{author}{\bibfnamefont{J.~P.} \bibnamefont{Perdew}} \bibnamefont{and}
  \bibinfo{author}{\bibfnamefont{K.}~\bibnamefont{Schmidt}},
  \emph{\bibinfo{title}{in Density Functional Theory and Its Application to
  Materials}} (\bibinfo{publisher}{AIP, Melville, NY}, \bibinfo{year}{2001}).

\bibitem[{\citenamefont{Krieger and Iafrate}(1992)}]{kri92}
\bibinfo{author}{\bibfnamefont{J.~B.} \bibnamefont{Krieger}} \bibnamefont{and}
  \bibinfo{author}{\bibfnamefont{G.~J.} \bibnamefont{Iafrate}},
  \bibinfo{journal}{Phys. Rev. A} \textbf{\bibinfo{volume}{46}},
  \bibinfo{pages}{5453} (\bibinfo{year}{1992}).

\bibitem[{\citenamefont{Georges et~al.}(1996)\citenamefont{Georges, Kotliar,
  Krauth, and Rozenberg}}]{geo96}
\bibinfo{author}{\bibfnamefont{A.}~\bibnamefont{Georges}},
  \bibinfo{author}{\bibfnamefont{G.}~\bibnamefont{Kotliar}},
  \bibinfo{author}{\bibfnamefont{W.}~\bibnamefont{Krauth}}, \bibnamefont{and}
  \bibinfo{author}{\bibfnamefont{M.~J.} \bibnamefont{Rozenberg}},
  \bibinfo{journal}{Rev. Mod. Phys.} \textbf{\bibinfo{volume}{68}},
  \bibinfo{pages}{13} (\bibinfo{year}{1996}).

\bibitem[{\citenamefont{{Georges}}(2004)}]{georges_strong}
\bibinfo{author}{\bibfnamefont{A.}~\bibnamefont{{Georges}}}, in
  \emph{\bibinfo{booktitle}{Lectures on the physics of highly correlated
  electron systems VIII}}, edited by
  \bibinfo{editor}{\bibfnamefont{A.}~\bibnamefont{Avella}} \bibnamefont{and}
  \bibinfo{editor}{\bibfnamefont{F.}~\bibnamefont{Mancini}}
  (\bibinfo{publisher}{American Institute of Physics}, \bibinfo{year}{2004}),
  \bibinfo{note}{cond-mat/0403123}.

\bibitem[{\citenamefont{Kotliar and
  Vollhardt}(2004)}]{kotliar_dmft_physicstoday}
\bibinfo{author}{\bibfnamefont{G.}~\bibnamefont{Kotliar}} \bibnamefont{and}
  \bibinfo{author}{\bibfnamefont{D.}~\bibnamefont{Vollhardt}},
  \bibinfo{journal}{Physics Today} \textbf{\bibinfo{volume}{March 2004}},
  \bibinfo{pages}{53} (\bibinfo{year}{2004}).

\bibitem[{\citenamefont{Kotliar et~al.}(2006)\citenamefont{Kotliar, Savrasov,
  Haule, Oudovenko, Parcollet, and Marianetti}}]{kotliar_review}
\bibinfo{author}{\bibfnamefont{G.}~\bibnamefont{Kotliar}},
  \bibinfo{author}{\bibfnamefont{S.~Y.} \bibnamefont{Savrasov}},
  \bibinfo{author}{\bibfnamefont{K.}~\bibnamefont{Haule}},
  \bibinfo{author}{\bibfnamefont{V.~S.} \bibnamefont{Oudovenko}},
  \bibinfo{author}{\bibfnamefont{O.}~\bibnamefont{Parcollet}},
  \bibnamefont{and} \bibinfo{author}{\bibfnamefont{C.~A.}
  \bibnamefont{Marianetti}}, \bibinfo{journal}{cond-mat/0511085}
  (\bibinfo{year}{2006}).

\bibitem[{\citenamefont{Held et~al.}(2002)\citenamefont{Held, Nekrasov, Keller,
  Eyert, N.~Bl\"{u}mer, Scalettar, Pruschke, Anisimov, and Vollhardt}}]{hel02}
\bibinfo{author}{\bibfnamefont{K.}~\bibnamefont{Held}},
  \bibinfo{author}{\bibfnamefont{I.~A.} \bibnamefont{Nekrasov}},
  \bibinfo{author}{\bibfnamefont{G.}~\bibnamefont{Keller}},
  \bibinfo{author}{\bibfnamefont{V.}~\bibnamefont{Eyert}},
  \bibinfo{author}{\bibfnamefont{A.~K.~M.} \bibnamefont{N.~Bl\"{u}mer}},
  \bibinfo{author}{\bibfnamefont{R.~T.} \bibnamefont{Scalettar}},
  \bibinfo{author}{\bibfnamefont{T.}~\bibnamefont{Pruschke}},
  \bibinfo{author}{\bibfnamefont{V.~I.} \bibnamefont{Anisimov}},
  \bibnamefont{and}
  \bibinfo{author}{\bibfnamefont{D.}~\bibnamefont{Vollhardt}},
  \emph{\bibinfo{title}{in Quantum Simulations of Complex Many-Body Systems:
  From Theory to Algorithms (NIC Series Vol. 10)}} (\bibinfo{year}{2002}).

\bibitem[{\citenamefont{Biermann}(2006)}]{bie06}
\bibinfo{author}{\bibfnamefont{S.}~\bibnamefont{Biermann}},
  \emph{\bibinfo{title}{in Encyclopedia of Materials: Science and Technology
  (online update)}} (\bibinfo{publisher}{Elsevier Ltd}, \bibinfo{year}{2006}).

\bibitem[{\citenamefont{Anisimov et~al.}(1997)\citenamefont{Anisimov,
  Poteryaev, Korotin, Anokhin, and Kotliar}}]{ani97}
\bibinfo{author}{\bibfnamefont{V.~I.} \bibnamefont{Anisimov}},
  \bibinfo{author}{\bibfnamefont{A.~I.} \bibnamefont{Poteryaev}},
  \bibinfo{author}{\bibfnamefont{M.~A.} \bibnamefont{Korotin}},
  \bibinfo{author}{\bibfnamefont{A.~O.} \bibnamefont{Anokhin}},
  \bibnamefont{and} \bibinfo{author}{\bibfnamefont{G.}~\bibnamefont{Kotliar}},
  \bibinfo{journal}{J. Phys.: Condens. Matter} \textbf{\bibinfo{volume}{9}},
  \bibinfo{pages}{7359} (\bibinfo{year}{1997}).

\bibitem[{\citenamefont{{Lichtenstein} and
  {Katsnelson}}(1998)}]{lichtenstein_lda+dmft_1998}
\bibinfo{author}{\bibfnamefont{A.~I.} \bibnamefont{{Lichtenstein}}}
  \bibnamefont{and} \bibinfo{author}{\bibfnamefont{M.~I.}
  \bibnamefont{{Katsnelson}}}, \bibinfo{journal}{Phys. Rev. B}
  \textbf{\bibinfo{volume}{57}}, \bibinfo{pages}{6884} (\bibinfo{year}{1998}).

\bibitem[{\citenamefont{Biermann
  et~al.}(2003{\natexlab{a}})\citenamefont{Biermann, Aryasetiawan, and
  Georges}}]{bie03}
\bibinfo{author}{\bibfnamefont{S.}~\bibnamefont{Biermann}},
  \bibinfo{author}{\bibfnamefont{F.}~\bibnamefont{Aryasetiawan}},
  \bibnamefont{and} \bibinfo{author}{\bibfnamefont{A.}~\bibnamefont{Georges}},
  \bibinfo{journal}{Phys. Rev. Lett.} \textbf{\bibinfo{volume}{90}},
  \bibinfo{pages}{086402} (\bibinfo{year}{2003}{\natexlab{a}}).

\bibitem[{\citenamefont{Biermann
  et~al.}(2003{\natexlab{b}})\citenamefont{Biermann, Aryasetiawan, and
  Georges}}]{bie03_2}
\bibinfo{author}{\bibfnamefont{S.}~\bibnamefont{Biermann}},
  \bibinfo{author}{\bibfnamefont{F.}~\bibnamefont{Aryasetiawan}},
  \bibnamefont{and} \bibinfo{author}{\bibfnamefont{A.}~\bibnamefont{Georges}},
  \bibinfo{journal}{cond-mat/0401653}  (\bibinfo{year}{2003}{\natexlab{b}}).

\bibitem[{\citenamefont{Kotliar and Savrasov}(2001)}]{kot01}
\bibinfo{author}{\bibfnamefont{G.}~\bibnamefont{Kotliar}} \bibnamefont{and}
  \bibinfo{author}{\bibfnamefont{S.~Y.} \bibnamefont{Savrasov}},
  \emph{\bibinfo{title}{in New Theoretical Approaches to Strongly Correlated
  Systems}} (\bibinfo{publisher}{Kluwer Academic Publishers, preprint
  cond-mat/0208241}, \bibinfo{year}{2001}), p. \bibinfo{pages}{259}.

\bibitem[{\citenamefont{Wannier}(1937)}]{wan37}
\bibinfo{author}{\bibfnamefont{G.~H.} \bibnamefont{Wannier}},
  \bibinfo{journal}{Phys. Rev.} \textbf{\bibinfo{volume}{52}},
  \bibinfo{pages}{191} (\bibinfo{year}{1937}).

\bibitem[{\citenamefont{Pavarini et~al.}(2004)\citenamefont{Pavarini, Biermann,
  Poteryaev, Lichtenstein, Georges, and Andersen}}]{pav04}
\bibinfo{author}{\bibfnamefont{E.}~\bibnamefont{Pavarini}},
  \bibinfo{author}{\bibfnamefont{S.}~\bibnamefont{Biermann}},
  \bibinfo{author}{\bibfnamefont{A.}~\bibnamefont{Poteryaev}},
  \bibinfo{author}{\bibfnamefont{A.~I.} \bibnamefont{Lichtenstein}},
  \bibinfo{author}{\bibfnamefont{A.}~\bibnamefont{Georges}}, \bibnamefont{and}
  \bibinfo{author}{\bibfnamefont{O.~K.} \bibnamefont{Andersen}},
  \bibinfo{journal}{Phys. Rev. Lett.} \textbf{\bibinfo{volume}{92}},
  \bibinfo{pages}{176403} (\bibinfo{year}{2004}).

\bibitem[{\citenamefont{Pavarini et~al.}(2005)\citenamefont{Pavarini, Yamasaki,
  Nuss, and Andersen}}]{pav05}
\bibinfo{author}{\bibfnamefont{E.}~\bibnamefont{Pavarini}},
  \bibinfo{author}{\bibfnamefont{A.}~\bibnamefont{Yamasaki}},
  \bibinfo{author}{\bibfnamefont{J.}~\bibnamefont{Nuss}}, \bibnamefont{and}
  \bibinfo{author}{\bibfnamefont{O.~K.} \bibnamefont{Andersen}},
  \bibinfo{journal}{New J. Phys.} \textbf{\bibinfo{volume}{7}},
  \bibinfo{pages}{188} (\bibinfo{year}{2005}).

\bibitem[{\citenamefont{Biermann et~al.}(2005)\citenamefont{Biermann,
  Poteryaev, Lichtenstein, and Georges}}]{bie05}
\bibinfo{author}{\bibfnamefont{S.}~\bibnamefont{Biermann}},
  \bibinfo{author}{\bibfnamefont{A.}~\bibnamefont{Poteryaev}},
  \bibinfo{author}{\bibfnamefont{A.~I.} \bibnamefont{Lichtenstein}},
  \bibnamefont{and} \bibinfo{author}{\bibfnamefont{A.}~\bibnamefont{Georges}},
  \bibinfo{journal}{Phys. Rev. Lett.} \textbf{\bibinfo{volume}{94}},
  \bibinfo{pages}{026404} (\bibinfo{year}{2005}).

\bibitem[{\citenamefont{Poteryaev et~al.}(2004)\citenamefont{Poteryaev,
  Lichtenstein, and Kotliar}}]{pot04}
\bibinfo{author}{\bibfnamefont{A.~I.} \bibnamefont{Poteryaev}},
  \bibinfo{author}{\bibfnamefont{A.~I.} \bibnamefont{Lichtenstein}},
  \bibnamefont{and} \bibinfo{author}{\bibfnamefont{G.}~\bibnamefont{Kotliar}},
  \bibinfo{journal}{Phys. Rev. Lett.} \textbf{\bibinfo{volume}{93}},
  \bibinfo{pages}{086401} (\bibinfo{year}{2004}).

\bibitem[{\citenamefont{Solovyev}(2006)}]{sol06}
\bibinfo{author}{\bibfnamefont{I.~V.} \bibnamefont{Solovyev}},
  \bibinfo{journal}{Phys. Rev. B} \textbf{\bibinfo{volume}{73}},
  \bibinfo{pages}{155117} (\bibinfo{year}{2006}).

\bibitem[{\citenamefont{Anisimov et~al.}(2005)\citenamefont{Anisimov, Kondakov,
  Kozhevnikov, Nekrasov, Pchelkina, Allen, Mo, Kim, Metcalf, Suga
  et~al.}}]{ani05}
\bibinfo{author}{\bibfnamefont{V.~I.} \bibnamefont{Anisimov}},
  \bibinfo{author}{\bibfnamefont{D.~E.} \bibnamefont{Kondakov}},
  \bibinfo{author}{\bibfnamefont{A.~V.} \bibnamefont{Kozhevnikov}},
  \bibinfo{author}{\bibfnamefont{I.~A.} \bibnamefont{Nekrasov}},
  \bibinfo{author}{\bibfnamefont{Z.~V.} \bibnamefont{Pchelkina}},
  \bibinfo{author}{\bibfnamefont{J.~W.} \bibnamefont{Allen}},
  \bibinfo{author}{\bibfnamefont{S.-K.} \bibnamefont{Mo}},
  \bibinfo{author}{\bibfnamefont{H.-D.} \bibnamefont{Kim}},
  \bibinfo{author}{\bibfnamefont{P.}~\bibnamefont{Metcalf}},
  \bibinfo{author}{\bibfnamefont{S.}~\bibnamefont{Suga}}, \bibnamefont{et~al.},
  \bibinfo{journal}{Phys. Rev. B} \textbf{\bibinfo{volume}{71}},
  \bibinfo{pages}{125119} (\bibinfo{year}{2005}).

\bibitem[{\citenamefont{Gavrichkov et~al.}(2005)\citenamefont{Gavrichkov,
  Korshunov, Ovchinnikov, Nekrasov, Pchelkina, and Anisimov}}]{gavri05}
\bibinfo{author}{\bibfnamefont{V.~A.} \bibnamefont{Gavrichkov}},
  \bibinfo{author}{\bibfnamefont{M.~M.} \bibnamefont{Korshunov}},
  \bibinfo{author}{\bibfnamefont{S.~G.} \bibnamefont{Ovchinnikov}},
  \bibinfo{author}{\bibfnamefont{I.~A.} \bibnamefont{Nekrasov}},
  \bibinfo{author}{\bibfnamefont{Z.~V.} \bibnamefont{Pchelkina}},
  \bibnamefont{and} \bibinfo{author}{\bibfnamefont{V.~I.}
  \bibnamefont{Anisimov}}, \bibinfo{journal}{Phys. Rev. B}
  \textbf{\bibinfo{volume}{72}}, \bibinfo{pages}{165104}
  (\bibinfo{year}{2005}).

\bibitem[{\citenamefont{Anisimov et~al.}(2006)\citenamefont{Anisimov,
  Kozhevnikov, Korotin, Lukoyanov, and Khafizullin}}]{ani06}
\bibinfo{author}{\bibfnamefont{V.~I.} \bibnamefont{Anisimov}},
  \bibinfo{author}{\bibfnamefont{A.~V.} \bibnamefont{Kozhevnikov}},
  \bibinfo{author}{\bibfnamefont{M.~A.} \bibnamefont{Korotin}},
  \bibinfo{author}{\bibfnamefont{A.~V.} \bibnamefont{Lukoyanov}},
  \bibnamefont{and} \bibinfo{author}{\bibfnamefont{D.~A.}
  \bibnamefont{Khafizullin}}, \bibinfo{journal}{cond-mat/0602204}
  (\bibinfo{year}{2006}).

\bibitem[{\citenamefont{Ku et~al.}(2002)\citenamefont{Ku, Rosner, Pickett, and
  Scalettar}}]{ku02}
\bibinfo{author}{\bibfnamefont{W.}~\bibnamefont{Ku}},
  \bibinfo{author}{\bibfnamefont{H.}~\bibnamefont{Rosner}},
  \bibinfo{author}{\bibfnamefont{W.~E.} \bibnamefont{Pickett}},
  \bibnamefont{and} \bibinfo{author}{\bibfnamefont{R.~T.}
  \bibnamefont{Scalettar}}, \bibinfo{journal}{Phys. Rev. Lett.}
  \textbf{\bibinfo{volume}{89}}, \bibinfo{pages}{167204}
  (\bibinfo{year}{2002}).

\bibitem[{\citenamefont{Schnell et~al.}(2003)\citenamefont{Schnell, Czycholl,
  and Albers}}]{schn03}
\bibinfo{author}{\bibfnamefont{I.}~\bibnamefont{Schnell}},
  \bibinfo{author}{\bibfnamefont{G.}~\bibnamefont{Czycholl}}, \bibnamefont{and}
  \bibinfo{author}{\bibfnamefont{R.~C.} \bibnamefont{Albers}},
  \bibinfo{journal}{Phys. Rev. B} \textbf{\bibinfo{volume}{68}},
  \bibinfo{pages}{245102} (\bibinfo{year}{2003}).

\bibitem[{\citenamefont{{Andersen}}(1975)}]{andersen_lmto_1975_prb}
\bibinfo{author}{\bibfnamefont{O.~K.} \bibnamefont{{Andersen}}},
  \bibinfo{journal}{Phys. Rev. B} \textbf{\bibinfo{volume}{12}},
  \bibinfo{pages}{3060} (\bibinfo{year}{1975}).

\bibitem[{\citenamefont{Marzari and Vanderbilt}(1997)}]{mar97}
\bibinfo{author}{\bibfnamefont{N.}~\bibnamefont{Marzari}} \bibnamefont{and}
  \bibinfo{author}{\bibfnamefont{D.}~\bibnamefont{Vanderbilt}},
  \bibinfo{journal}{Phys. Rev. B} \textbf{\bibinfo{volume}{56}},
  \bibinfo{pages}{12847} (\bibinfo{year}{1997}).

\bibitem[{\citenamefont{Souza et~al.}(2001)\citenamefont{Souza, Marzari, and
  Vanderbilt}}]{sou01}
\bibinfo{author}{\bibfnamefont{I.}~\bibnamefont{Souza}},
  \bibinfo{author}{\bibfnamefont{N.}~\bibnamefont{Marzari}}, \bibnamefont{and}
  \bibinfo{author}{\bibfnamefont{D.}~\bibnamefont{Vanderbilt}},
  \bibinfo{journal}{Phys. Rev. B} \textbf{\bibinfo{volume}{65}},
  \bibinfo{pages}{035109} (\bibinfo{year}{2001}).

\bibitem[{\citenamefont{Andersen and Saha-Dasgupta}(2000)}]{and00}
\bibinfo{author}{\bibfnamefont{O.~K.} \bibnamefont{Andersen}} \bibnamefont{and}
  \bibinfo{author}{\bibfnamefont{T.}~\bibnamefont{Saha-Dasgupta}},
  \bibinfo{journal}{Phys. Rev. B} \textbf{\bibinfo{volume}{62}},
  \bibinfo{pages}{16219} (\bibinfo{year}{2000}).

\bibitem[{\citenamefont{Andersen et~al.}(2000)\citenamefont{Andersen,
  Saha-Dasgupta, Tank, Arcangeli, Jepsen, and Krier}}]{and00-2}
\bibinfo{author}{\bibfnamefont{O.~K.} \bibnamefont{Andersen}},
  \bibinfo{author}{\bibfnamefont{T.}~\bibnamefont{Saha-Dasgupta}},
  \bibinfo{author}{\bibfnamefont{R.~W.} \bibnamefont{Tank}},
  \bibinfo{author}{\bibfnamefont{C.}~\bibnamefont{Arcangeli}},
  \bibinfo{author}{\bibfnamefont{O.}~\bibnamefont{Jepsen}}, \bibnamefont{and}
  \bibinfo{author}{\bibfnamefont{G.}~\bibnamefont{Krier}},
  \emph{\bibinfo{title}{in Electronic Structure and Physical Properties of
  Solids. The uses of the LMTO method. (Lecture notes in Physics vol. 535)}}
  (\bibinfo{publisher}{Springer, Berlin/Heidelberg}, \bibinfo{year}{2000}).

\bibitem[{\citenamefont{Zurek et~al.}(2005)\citenamefont{Zurek, Jepsen, and
  Andersen}}]{zur05}
\bibinfo{author}{\bibfnamefont{E.}~\bibnamefont{Zurek}},
  \bibinfo{author}{\bibfnamefont{O.}~\bibnamefont{Jepsen}}, \bibnamefont{and}
  \bibinfo{author}{\bibfnamefont{O.~K.} \bibnamefont{Andersen}},
  \bibinfo{journal}{ChemPhysChem} \textbf{\bibinfo{volume}{6}},
  \bibinfo{pages}{1934} (\bibinfo{year}{2005}).

\bibitem[{\citenamefont{Nekrasov et~al.}(2006)\citenamefont{Nekrasov, Held,
  Keller, Kondakov, Pruschke, Kollar, Andersen, Anisimov, and
  Vollhardt}}]{nek06}
\bibinfo{author}{\bibfnamefont{I.~A.} \bibnamefont{Nekrasov}},
  \bibinfo{author}{\bibfnamefont{K.}~\bibnamefont{Held}},
  \bibinfo{author}{\bibfnamefont{G.}~\bibnamefont{Keller}},
  \bibinfo{author}{\bibfnamefont{D.~E.} \bibnamefont{Kondakov}},
  \bibinfo{author}{\bibfnamefont{T.}~\bibnamefont{Pruschke}},
  \bibinfo{author}{\bibfnamefont{M.}~\bibnamefont{Kollar}},
  \bibinfo{author}{\bibfnamefont{O.~K.} \bibnamefont{Andersen}},
  \bibinfo{author}{\bibfnamefont{V.~I.} \bibnamefont{Anisimov}},
  \bibnamefont{and}
  \bibinfo{author}{\bibfnamefont{D.}~\bibnamefont{Vollhardt}},
  \bibinfo{journal}{Phys. Rev. B} \textbf{\bibinfo{volume}{73}},
  \bibinfo{pages}{155112} (\bibinfo{year}{2006}).

\bibitem[{\citenamefont{Yamasaki
  et~al.}(2006{\natexlab{a}})\citenamefont{Yamasaki, Feldbacher, Yang,
  Andersen, and Held}}]{yam06}
\bibinfo{author}{\bibfnamefont{A.}~\bibnamefont{Yamasaki}},
  \bibinfo{author}{\bibfnamefont{M.}~\bibnamefont{Feldbacher}},
  \bibinfo{author}{\bibfnamefont{Y.-F.} \bibnamefont{Yang}},
  \bibinfo{author}{\bibfnamefont{O.~K.} \bibnamefont{Andersen}},
  \bibnamefont{and} \bibinfo{author}{\bibfnamefont{K.}~\bibnamefont{Held}},
  \bibinfo{journal}{Phys. Rev. Lett.} \textbf{\bibinfo{volume}{96}},
  \bibinfo{pages}{166401} (\bibinfo{year}{2006}{\natexlab{a}}).

\bibitem[{\citenamefont{Saha-Dasgupta et~al.}(2005)\citenamefont{Saha-Dasgupta,
  Lichtenstein, and Valenti}}]{tan05}
\bibinfo{author}{\bibfnamefont{T.}~\bibnamefont{Saha-Dasgupta}},
  \bibinfo{author}{\bibfnamefont{A.}~\bibnamefont{Lichtenstein}},
  \bibnamefont{and} \bibinfo{author}{\bibfnamefont{R.}~\bibnamefont{Valenti}},
  \bibinfo{journal}{Phys. Rev. B} \textbf{\bibinfo{volume}{71}},
  \bibinfo{pages}{153108} (\bibinfo{year}{2005}).

\bibitem[{\citenamefont{Sekiyama et~al.}(2004)\citenamefont{Sekiyama, Fujiwara,
  Imada, Suga, Eisaki, Uchida, Takegahara, Harima, Saitoh, Nekrasov
  et~al.}}]{sek04}
\bibinfo{author}{\bibfnamefont{A.}~\bibnamefont{Sekiyama}},
  \bibinfo{author}{\bibfnamefont{H.}~\bibnamefont{Fujiwara}},
  \bibinfo{author}{\bibfnamefont{S.}~\bibnamefont{Imada}},
  \bibinfo{author}{\bibfnamefont{S.}~\bibnamefont{Suga}},
  \bibinfo{author}{\bibfnamefont{H.}~\bibnamefont{Eisaki}},
  \bibinfo{author}{\bibfnamefont{S.~I.} \bibnamefont{Uchida}},
  \bibinfo{author}{\bibfnamefont{K.}~\bibnamefont{Takegahara}},
  \bibinfo{author}{\bibfnamefont{H.}~\bibnamefont{Harima}},
  \bibinfo{author}{\bibfnamefont{Y.}~\bibnamefont{Saitoh}},
  \bibinfo{author}{\bibfnamefont{I.~A.} \bibnamefont{Nekrasov}},
  \bibnamefont{et~al.}, \bibinfo{journal}{Phys. Rev. Lett.}
  \textbf{\bibinfo{volume}{93}}, \bibinfo{pages}{156402}
  (\bibinfo{year}{2004}).

\bibitem[{\citenamefont{Liebsch}(2003)}]{lie03}
\bibinfo{author}{\bibfnamefont{A.}~\bibnamefont{Liebsch}},
  \bibinfo{journal}{Phys. Rev. Lett.} \textbf{\bibinfo{volume}{90}},
  \bibinfo{pages}{096401} (\bibinfo{year}{2003}).

\bibitem[{\citenamefont{Nekrasov et~al.}(2005)\citenamefont{Nekrasov, Keller,
  Kondakov, , Kozhevnikov, Pruschke, Held, Vollhardt, and Anisimov}}]{nek05}
\bibinfo{author}{\bibfnamefont{I.~A.} \bibnamefont{Nekrasov}},
  \bibinfo{author}{\bibfnamefont{G.}~\bibnamefont{Keller}},
  \bibinfo{author}{\bibfnamefont{D.~E.} \bibnamefont{Kondakov}}, ,
  \bibinfo{author}{\bibfnamefont{A.~V.} \bibnamefont{Kozhevnikov}},
  \bibinfo{author}{\bibfnamefont{T.}~\bibnamefont{Pruschke}},
  \bibinfo{author}{\bibfnamefont{K.}~\bibnamefont{Held}},
  \bibinfo{author}{\bibfnamefont{D.}~\bibnamefont{Vollhardt}},
  \bibnamefont{and} \bibinfo{author}{\bibfnamefont{V.~I.}
  \bibnamefont{Anisimov}}, \bibinfo{journal}{Phys. Rev. B}
  \textbf{\bibinfo{volume}{72}}, \bibinfo{pages}{155106}
  (\bibinfo{year}{2005}).

\bibitem[{\citenamefont{Biermann et~al.}(2004)\citenamefont{Biermann,
  Dallmeyer, Carbone, Eberhardt, Pampuch, Rader, Katsnelson, and
  Lichtenstein}}]{bie04}
\bibinfo{author}{\bibfnamefont{S.}~\bibnamefont{Biermann}},
  \bibinfo{author}{\bibfnamefont{A.}~\bibnamefont{Dallmeyer}},
  \bibinfo{author}{\bibfnamefont{C.}~\bibnamefont{Carbone}},
  \bibinfo{author}{\bibfnamefont{W.}~\bibnamefont{Eberhardt}},
  \bibinfo{author}{\bibfnamefont{C.}~\bibnamefont{Pampuch}},
  \bibinfo{author}{\bibfnamefont{O.}~\bibnamefont{Rader}},
  \bibinfo{author}{\bibfnamefont{M.~I.} \bibnamefont{Katsnelson}},
  \bibnamefont{and} \bibinfo{author}{\bibfnamefont{A.~I.}
  \bibnamefont{Lichtenstein}}, \bibinfo{journal}{JETP Letters}
  \textbf{\bibinfo{volume}{80}}, \bibinfo{pages}{612} (\bibinfo{year}{2004}).

\bibitem[{\citenamefont{{Georges} and {Kotliar}}(1992)}]{georges_kotliar_dmft}
\bibinfo{author}{\bibfnamefont{A.}~\bibnamefont{{Georges}}} \bibnamefont{and}
  \bibinfo{author}{\bibfnamefont{G.}~\bibnamefont{{Kotliar}}},
  \bibinfo{journal}{Phys. Rev. B} \textbf{\bibinfo{volume}{45}},
  \bibinfo{pages}{6479} (\bibinfo{year}{1992}).

\bibitem[{\citenamefont{{Anisimov} et~al.}(1991)\citenamefont{{Anisimov},
  {Zaanen}, and {Andersen}}}]{anisimov_lda+u_1991_prb}
\bibinfo{author}{\bibfnamefont{V.~I.} \bibnamefont{{Anisimov}}},
  \bibinfo{author}{\bibfnamefont{J.}~\bibnamefont{{Zaanen}}}, \bibnamefont{and}
  \bibinfo{author}{\bibfnamefont{O.~K.} \bibnamefont{{Andersen}}},
  \bibinfo{journal}{Phys. Rev. B} \textbf{\bibinfo{volume}{44}},
  \bibinfo{pages}{943} (\bibinfo{year}{1991}).

\bibitem[{\citenamefont{Savrasov and Kotliar}(2004)}]{sav04}
\bibinfo{author}{\bibfnamefont{S.~Y.} \bibnamefont{Savrasov}} \bibnamefont{and}
  \bibinfo{author}{\bibfnamefont{G.}~\bibnamefont{Kotliar}},
  \bibinfo{journal}{Phys. Rev. B} \textbf{\bibinfo{volume}{69}},
  \bibinfo{pages}{245101} (\bibinfo{year}{2004}).

\bibitem[{\citenamefont{Czy$\dot{\text{z}}$yk and Sawatzky}(1994)}]{czy94}
\bibinfo{author}{\bibfnamefont{M.~T.} \bibnamefont{Czy$\dot{\text{z}}$yk}}
  \bibnamefont{and} \bibinfo{author}{\bibfnamefont{G.~A.}
  \bibnamefont{Sawatzky}}, \bibinfo{journal}{Phys. Rev. B}
  \textbf{\bibinfo{volume}{49}}, \bibinfo{pages}{14211} (\bibinfo{year}{1994}).

\bibitem[{\citenamefont{Petukhov et~al.}(2003)\citenamefont{Petukhov, Mazin,
  Chioncel, and Lichtenstein}}]{pet03}
\bibinfo{author}{\bibfnamefont{A.~G.} \bibnamefont{Petukhov}},
  \bibinfo{author}{\bibfnamefont{I.~I.} \bibnamefont{Mazin}},
  \bibinfo{author}{\bibfnamefont{L.}~\bibnamefont{Chioncel}}, \bibnamefont{and}
  \bibinfo{author}{\bibfnamefont{A.~I.} \bibnamefont{Lichtenstein}},
  \bibinfo{journal}{Phys. Rev. B} \textbf{\bibinfo{volume}{67}},
  \bibinfo{pages}{153106} (\bibinfo{year}{2003}).

\bibitem[{\citenamefont{Anisimov et~al.}(1993)\citenamefont{Anisimov, Solovyev,
  Korotin, Czy$\dot{\text{z}}$yk, and Sawatzky}}]{ani93}
\bibinfo{author}{\bibfnamefont{V.~I.} \bibnamefont{Anisimov}},
  \bibinfo{author}{\bibfnamefont{I.~V.} \bibnamefont{Solovyev}},
  \bibinfo{author}{\bibfnamefont{M.~A.} \bibnamefont{Korotin}},
  \bibinfo{author}{\bibfnamefont{M.~T.} \bibnamefont{Czy$\dot{\text{z}}$yk}},
  \bibnamefont{and} \bibinfo{author}{\bibfnamefont{G.~A.}
  \bibnamefont{Sawatzky}}, \bibinfo{journal}{Phys. Rev. B}
  \textbf{\bibinfo{volume}{48}}, \bibinfo{pages}{16929} (\bibinfo{year}{1993}).

\bibitem[{\citenamefont{Kohn}(1959)}]{koh59}
\bibinfo{author}{\bibfnamefont{W.}~\bibnamefont{Kohn}}, \bibinfo{journal}{Phys.
  Rev.} \textbf{\bibinfo{volume}{115}}, \bibinfo{pages}{809}
  (\bibinfo{year}{1959}).

\bibitem[{\citenamefont{Boys}(1960)}]{boy60}
\bibinfo{author}{\bibfnamefont{S.~F.} \bibnamefont{Boys}},
  \bibinfo{journal}{Rev. Mod. Phys.} \textbf{\bibinfo{volume}{32}},
  \bibinfo{pages}{296} (\bibinfo{year}{1960}).

\bibitem[{\citenamefont{Blount}(1962)}]{blo62}
\bibinfo{author}{\bibfnamefont{E.~I.} \bibnamefont{Blount}},
  \bibinfo{journal}{Solid State Phys.} \textbf{\bibinfo{volume}{13}},
  \bibinfo{pages}{305} (\bibinfo{year}{1962}).

\bibitem[{\citenamefont{des Cloizeaux}(1963)}]{clo63}
\bibinfo{author}{\bibfnamefont{J.}~\bibnamefont{des Cloizeaux}},
  \bibinfo{journal}{Phys. Rev.} \textbf{\bibinfo{volume}{129}},
  \bibinfo{pages}{554} (\bibinfo{year}{1963}).

\bibitem[{\citenamefont{Zaanen et~al.}(1985)\citenamefont{Zaanen, Sawatzky, and
  Allen}}]{zsa_1985_prl}
\bibinfo{author}{\bibfnamefont{J.}~\bibnamefont{Zaanen}},
  \bibinfo{author}{\bibfnamefont{G.~A.} \bibnamefont{Sawatzky}},
  \bibnamefont{and} \bibinfo{author}{\bibfnamefont{J.~W.} \bibnamefont{Allen}},
  \bibinfo{journal}{Phys. Rev. Lett.} \textbf{\bibinfo{volume}{55}},
  \bibinfo{pages}{418} (\bibinfo{year}{1985}).

\bibitem[{\citenamefont{S\'{a}nchez-Portal
  et~al.}(1995)\citenamefont{S\'{a}nchez-Portal, Artacho, and Soler}}]{san95}
\bibinfo{author}{\bibfnamefont{D.}~\bibnamefont{S\'{a}nchez-Portal}},
  \bibinfo{author}{\bibfnamefont{E.}~\bibnamefont{Artacho}}, \bibnamefont{and}
  \bibinfo{author}{\bibfnamefont{J.}~\bibnamefont{Soler}},
  \bibinfo{journal}{Solid State Comm.} \textbf{\bibinfo{volume}{95}},
  \bibinfo{pages}{685} (\bibinfo{year}{1995}).

\bibitem[{\citenamefont{Andersen and Jepsen}(1984)}]{and84}
\bibinfo{author}{\bibfnamefont{O.~K.} \bibnamefont{Andersen}} \bibnamefont{and}
  \bibinfo{author}{\bibfnamefont{O.}~\bibnamefont{Jepsen}},
  \bibinfo{journal}{Phys. Rev. Lett.} \textbf{\bibinfo{volume}{53}},
  \bibinfo{pages}{2571} (\bibinfo{year}{1984}).

\bibitem[{\citenamefont{L\"owdin}(1951)}]{low51}
\bibinfo{author}{\bibfnamefont{P.~O.} \bibnamefont{L\"owdin}},
  \bibinfo{journal}{J. Chem. Phys.} \textbf{\bibinfo{volume}{19}},
  \bibinfo{pages}{1396} (\bibinfo{year}{1951}).

\bibitem[{\citenamefont{Louie et~al.}(1979)\citenamefont{Louie, Ho, and
  Cohen}}]{lou79}
\bibinfo{author}{\bibfnamefont{S.~G.} \bibnamefont{Louie}},
  \bibinfo{author}{\bibfnamefont{K.~M.} \bibnamefont{Ho}}, \bibnamefont{and}
  \bibinfo{author}{\bibfnamefont{M.~L.} \bibnamefont{Cohen}},
  \bibinfo{journal}{Phys. Rev. B} \textbf{\bibinfo{volume}{19}},
  \bibinfo{pages}{1774} (\bibinfo{year}{1979}).

\bibitem[{\citenamefont{Harrison}(1960)}]{har60}
\bibinfo{author}{\bibfnamefont{W.}~\bibnamefont{Harrison}},
  \emph{\bibinfo{title}{Pseudopotentials in the Theory of Metals}}
  (\bibinfo{publisher}{Benjamin}, \bibinfo{address}{New York},
  \bibinfo{year}{1960}).

\bibitem[{\citenamefont{Heine}(1970)}]{hei70}
\bibinfo{author}{\bibfnamefont{V.}~\bibnamefont{Heine}},
  \emph{\bibinfo{title}{in Solid State Physics Vol. 24}}
  (\bibinfo{publisher}{Academic, New York}, \bibinfo{year}{1970}),
  chap.~\bibinfo{chapter}{1}.

\bibitem[{\citenamefont{Meyer et~al.}(unpublished)\citenamefont{Meyer,
  Els\"{a}sser, Lechermann, and F\"{a}hnle}}]{mbpp_code}
\bibinfo{author}{\bibfnamefont{B.}~\bibnamefont{Meyer}},
  \bibinfo{author}{\bibfnamefont{C.}~\bibnamefont{Els\"{a}sser}},
  \bibinfo{author}{\bibfnamefont{F.}~\bibnamefont{Lechermann}},
  \bibnamefont{and}
  \bibinfo{author}{\bibfnamefont{M.}~\bibnamefont{F\"{a}hnle}},
  \emph{\bibinfo{title}{FORTRAN 90 Program for Mixed-Basis-Pseudopotential
  Calculations for Crystals}}, \bibinfo{organization}{Max-Planck-Institut
  f\"{u}r Metallforschung, Stuttgart} (\bibinfo{year}{unpublished}).

\bibitem[{\citenamefont{Vanderbilt}(1985)}]{van85}
\bibinfo{author}{\bibfnamefont{D.}~\bibnamefont{Vanderbilt}},
  \bibinfo{journal}{Phys. Rev. B} \textbf{\bibinfo{volume}{32}},
  \bibinfo{pages}{8412} (\bibinfo{year}{1985}).

\bibitem[{\citenamefont{Posternak et~al.}(2002)\citenamefont{Posternak,
  Baldereschi, Massidda, and Marzari}}]{pos02}
\bibinfo{author}{\bibfnamefont{M.}~\bibnamefont{Posternak}},
  \bibinfo{author}{\bibfnamefont{A.}~\bibnamefont{Baldereschi}},
  \bibinfo{author}{\bibfnamefont{S.}~\bibnamefont{Massidda}}, \bibnamefont{and}
  \bibinfo{author}{\bibfnamefont{N.}~\bibnamefont{Marzari}},
  \bibinfo{journal}{Phys. Rev. B} \textbf{\bibinfo{volume}{65}},
  \bibinfo{pages}{184422} (\bibinfo{year}{2002}).

\bibitem[{\citenamefont{Jansen and Freeman}(1984)}]{jan84}
\bibinfo{author}{\bibfnamefont{H.~J.~F.} \bibnamefont{Jansen}}
  \bibnamefont{and} \bibinfo{author}{\bibfnamefont{A.~J.}
  \bibnamefont{Freeman}}, \bibinfo{journal}{Phys. Rev. B}
  \textbf{\bibinfo{volume}{30}}, \bibinfo{pages}{561} (\bibinfo{year}{1984}).

\bibitem[{\citenamefont{Massidda et~al.}(1993)\citenamefont{Massidda,
  Posternak, and Baldereschi}}]{mas93}
\bibinfo{author}{\bibfnamefont{S.}~\bibnamefont{Massidda}},
  \bibinfo{author}{\bibfnamefont{M.}~\bibnamefont{Posternak}},
  \bibnamefont{and}
  \bibinfo{author}{\bibfnamefont{A.}~\bibnamefont{Baldereschi}},
  \bibinfo{journal}{Phys. Rev. B} \textbf{\bibinfo{volume}{48}},
  \bibinfo{pages}{5058} (\bibinfo{year}{1993}).

\bibitem[{\citenamefont{Singh}(1991)}]{sin91}
\bibinfo{author}{\bibfnamefont{D.}~\bibnamefont{Singh}},
  \bibinfo{journal}{Phys. Rev. B} \textbf{\bibinfo{volume}{43}},
  \bibinfo{pages}{6388} (\bibinfo{year}{1991}).

\bibitem[{mlw(See: http://www.wannier.org/)}]{mlwf-code}
\emph{\bibinfo{title}{Maximally-localized Wannier Functions code}}
  (\bibinfo{year}{See: http://www.wannier.org/}).

\bibitem[{lmt(See: http://www.fkf.mpg.de/andersen/)}]{lmto}
\emph{\bibinfo{title}{The Stuttgart TB-LMTO-ASA code, version 4.7}}
  (\bibinfo{year}{See: http://www.fkf.mpg.de/andersen/}).

\bibitem[{\citenamefont{Castellani et~al.}(1978)\citenamefont{Castellani,
  Natoli, and Ranninger}}]{cas78}
\bibinfo{author}{\bibfnamefont{C.}~\bibnamefont{Castellani}},
  \bibinfo{author}{\bibfnamefont{C.~R.} \bibnamefont{Natoli}},
  \bibnamefont{and}
  \bibinfo{author}{\bibfnamefont{J.}~\bibnamefont{Ranninger}},
  \bibinfo{journal}{Phys. Rev. B} \textbf{\bibinfo{volume}{18}},
  \bibinfo{pages}{4945} (\bibinfo{year}{1978}).

\bibitem[{\citenamefont{Fr\'{e}sard and Kotliar}(1997)}]{fre97}
\bibinfo{author}{\bibfnamefont{R.}~\bibnamefont{Fr\'{e}sard}} \bibnamefont{and}
  \bibinfo{author}{\bibfnamefont{G.}~\bibnamefont{Kotliar}},
  \bibinfo{journal}{Phys. Rev. B} \textbf{\bibinfo{volume}{56}},
  \bibinfo{pages}{12909} (\bibinfo{year}{1997}).

\bibitem[{\citenamefont{Hirsch and Fye}(1986)}]{hirsch_fye}
\bibinfo{author}{\bibfnamefont{J.~E.} \bibnamefont{Hirsch}} \bibnamefont{and}
  \bibinfo{author}{\bibfnamefont{R.~M.} \bibnamefont{Fye}},
  \bibinfo{journal}{Phys. Rev. Lett.} \textbf{\bibinfo{volume}{25}},
  \bibinfo{pages}{2521} (\bibinfo{year}{1986}).

\bibitem[{\citenamefont{Onoda et~al.}(1991)\citenamefont{Onoda, Ohta, and
  Nagasawa}}]{ono91}
\bibinfo{author}{\bibfnamefont{M.}~\bibnamefont{Onoda}},
  \bibinfo{author}{\bibfnamefont{H.}~\bibnamefont{Ohta}}, \bibnamefont{and}
  \bibinfo{author}{\bibfnamefont{H.}~\bibnamefont{Nagasawa}},
  \bibinfo{journal}{Solid State Comm.} \textbf{\bibinfo{volume}{79}},
  \bibinfo{pages}{281} (\bibinfo{year}{1991}).

\bibitem[{\citenamefont{Imada et~al.}(1998)\citenamefont{Imada, Fujimori, and
  Tokura}}]{imada_mit_review}
\bibinfo{author}{\bibfnamefont{M.}~\bibnamefont{Imada}},
  \bibinfo{author}{\bibfnamefont{A.}~\bibnamefont{Fujimori}}, \bibnamefont{and}
  \bibinfo{author}{\bibfnamefont{Y.}~\bibnamefont{Tokura}},
  \bibinfo{journal}{Rev. Mod. Phys.} \textbf{\bibinfo{volume}{70}},
  \bibinfo{pages}{1039} (\bibinfo{year}{1998}).

\bibitem[{\citenamefont{Fujimori et~al.}(1992)\citenamefont{Fujimori, Hase,
  Namatame, Fujishima, Tokura, Eisaki, Uchida, Takegahara, and
  de~Groot}}]{fujimori_pes_oxides}
\bibinfo{author}{\bibfnamefont{A.}~\bibnamefont{Fujimori}},
  \bibinfo{author}{\bibfnamefont{I.}~\bibnamefont{Hase}},
  \bibinfo{author}{\bibfnamefont{H.}~\bibnamefont{Namatame}},
  \bibinfo{author}{\bibfnamefont{Y.}~\bibnamefont{Fujishima}},
  \bibinfo{author}{\bibfnamefont{Y.}~\bibnamefont{Tokura}},
  \bibinfo{author}{\bibfnamefont{H.}~\bibnamefont{Eisaki}},
  \bibinfo{author}{\bibfnamefont{S.}~\bibnamefont{Uchida}},
  \bibinfo{author}{\bibfnamefont{K.}~\bibnamefont{Takegahara}},
  \bibnamefont{and} \bibinfo{author}{\bibfnamefont{F.~M.~F.}
  \bibnamefont{de~Groot}}, \bibinfo{journal}{Phys. Rev. Lett.}
  \textbf{\bibinfo{volume}{69}}, \bibinfo{pages}{1796} (\bibinfo{year}{1992}).

\bibitem[{\citenamefont{Maiti et~al.}(2001)\citenamefont{Maiti, Sarma,
  Rozenberg, Inoue, Makino, Goto, Pedio, and Cimino}}]{maiti_2001}
\bibinfo{author}{\bibfnamefont{K.}~\bibnamefont{Maiti}},
  \bibinfo{author}{\bibfnamefont{D.~D.} \bibnamefont{Sarma}},
  \bibinfo{author}{\bibfnamefont{M.}~\bibnamefont{Rozenberg}},
  \bibinfo{author}{\bibfnamefont{I.}~\bibnamefont{Inoue}},
  \bibinfo{author}{\bibfnamefont{H.}~\bibnamefont{Makino}},
  \bibinfo{author}{\bibfnamefont{O.}~\bibnamefont{Goto}},
  \bibinfo{author}{\bibfnamefont{M.}~\bibnamefont{Pedio}}, \bibnamefont{and}
  \bibinfo{author}{\bibfnamefont{R.}~\bibnamefont{Cimino}},
  \bibinfo{journal}{Europhys. Lett.} \textbf{\bibinfo{volume}{55}},
  \bibinfo{pages}{246} (\bibinfo{year}{2001}).

\bibitem[{\citenamefont{Maiti}(1997)}]{maiti_phd}
\bibinfo{author}{\bibfnamefont{K.}~\bibnamefont{Maiti}}, Ph.D. thesis,
  \bibinfo{school}{IISC, Bangalore} (\bibinfo{year}{1997}).

\bibitem[{\citenamefont{{Inoue} et~al.}(1995)\citenamefont{{Inoue}, {Hase},
  {Aiura}, {Fujimori}, {Haruyama}, {Maruyama}, and
  {Nishihara}}}]{inoue_casrvo3_1995_prl}
\bibinfo{author}{\bibfnamefont{I.~H.} \bibnamefont{{Inoue}}},
  \bibinfo{author}{\bibfnamefont{I.}~\bibnamefont{{Hase}}},
  \bibinfo{author}{\bibfnamefont{Y.}~\bibnamefont{{Aiura}}},
  \bibinfo{author}{\bibfnamefont{A.}~\bibnamefont{{Fujimori}}},
  \bibinfo{author}{\bibfnamefont{Y.}~\bibnamefont{{Haruyama}}},
  \bibinfo{author}{\bibfnamefont{T.}~\bibnamefont{{Maruyama}}},
  \bibnamefont{and}
  \bibinfo{author}{\bibfnamefont{Y.}~\bibnamefont{{Nishihara}}},
  \bibinfo{journal}{Phys. Rev. Lett.} \textbf{\bibinfo{volume}{74}},
  \bibinfo{pages}{2539} (\bibinfo{year}{1995}).

\bibitem[{\citenamefont{Yoshida et~al.}(2005)\citenamefont{Yoshida, Tanaka,
  Yagi, Ino, Eisaki, Fujimori, and Shen}}]{yos05}
\bibinfo{author}{\bibfnamefont{T.}~\bibnamefont{Yoshida}},
  \bibinfo{author}{\bibfnamefont{K.}~\bibnamefont{Tanaka}},
  \bibinfo{author}{\bibfnamefont{H.}~\bibnamefont{Yagi}},
  \bibinfo{author}{\bibfnamefont{A.}~\bibnamefont{Ino}},
  \bibinfo{author}{\bibfnamefont{H.}~\bibnamefont{Eisaki}},
  \bibinfo{author}{\bibfnamefont{A.}~\bibnamefont{Fujimori}}, \bibnamefont{and}
  \bibinfo{author}{\bibfnamefont{Z.-X.} \bibnamefont{Shen}},
  \bibinfo{journal}{Phys. Rev. Lett.} \textbf{\bibinfo{volume}{95}},
  \bibinfo{pages}{146404} (\bibinfo{year}{2005}).

\bibitem[{\citenamefont{Wadati et~al.}(2006)\citenamefont{Wadati, Yoshida,
  Chikamatsu, Kumigashira, Oshima, Eisaki, Shen, Mizokawa, and
  Fujimori}}]{wad06}
\bibinfo{author}{\bibfnamefont{H.}~\bibnamefont{Wadati}},
  \bibinfo{author}{\bibfnamefont{T.}~\bibnamefont{Yoshida}},
  \bibinfo{author}{\bibfnamefont{A.}~\bibnamefont{Chikamatsu}},
  \bibinfo{author}{\bibfnamefont{H.}~\bibnamefont{Kumigashira}},
  \bibinfo{author}{\bibfnamefont{M.}~\bibnamefont{Oshima}},
  \bibinfo{author}{\bibfnamefont{H.}~\bibnamefont{Eisaki}},
  \bibinfo{author}{\bibfnamefont{Z.~X.} \bibnamefont{Shen}},
  \bibinfo{author}{\bibfnamefont{T.}~\bibnamefont{Mizokawa}}, \bibnamefont{and}
  \bibinfo{author}{\bibfnamefont{A.}~\bibnamefont{Fujimori}},
  \bibinfo{journal}{cond-mat/0603642}  (\bibinfo{year}{2006}).

\bibitem[{\citenamefont{Eguchi et~al.}(2006)\citenamefont{Eguchi, Kiss, Tsuda,
  Shimojima, Mizokami, Yokoya, Chainani, Shin, Inoue, Togashi et~al.}}]{egu06}
\bibinfo{author}{\bibfnamefont{R.}~\bibnamefont{Eguchi}},
  \bibinfo{author}{\bibfnamefont{T.}~\bibnamefont{Kiss}},
  \bibinfo{author}{\bibfnamefont{S.}~\bibnamefont{Tsuda}},
  \bibinfo{author}{\bibfnamefont{T.}~\bibnamefont{Shimojima}},
  \bibinfo{author}{\bibfnamefont{T.}~\bibnamefont{Mizokami}},
  \bibinfo{author}{\bibfnamefont{T.}~\bibnamefont{Yokoya}},
  \bibinfo{author}{\bibfnamefont{A.}~\bibnamefont{Chainani}},
  \bibinfo{author}{\bibfnamefont{S.}~\bibnamefont{Shin}},
  \bibinfo{author}{\bibfnamefont{I.~H.} \bibnamefont{Inoue}},
  \bibinfo{author}{\bibfnamefont{T.}~\bibnamefont{Togashi}},
  \bibnamefont{et~al.}, \bibinfo{journal}{Phys. Rev. Lett.}
  \textbf{\bibinfo{volume}{96}}, \bibinfo{pages}{076402}
  (\bibinfo{year}{2006}).

\bibitem[{\citenamefont{Chamberland and Danielson}(1971)}]{cha71}
\bibinfo{author}{\bibfnamefont{B.}~\bibnamefont{Chamberland}} \bibnamefont{and}
  \bibinfo{author}{\bibfnamefont{P.}~\bibnamefont{Danielson}},
  \bibinfo{journal}{Solid State Comm.} \textbf{\bibinfo{volume}{3}},
  \bibinfo{pages}{243} (\bibinfo{year}{1971}).

\bibitem[{\citenamefont{M\"uller-Hartmann}(1989)}]{mue89}
\bibinfo{author}{\bibfnamefont{E.}~\bibnamefont{M\"uller-Hartmann}},
  \bibinfo{journal}{Z. Phys. B} \textbf{\bibinfo{volume}{74}},
  \bibinfo{pages}{507} (\bibinfo{year}{1989}).

\bibitem[{\citenamefont{Maiti et~al.}(2005)\citenamefont{Maiti, Manju, Ray,
  Mahadevan, and I.~H.~Inoue}}]{mai05}
\bibinfo{author}{\bibfnamefont{K.}~\bibnamefont{Maiti}},
  \bibinfo{author}{\bibfnamefont{U.}~\bibnamefont{Manju}},
  \bibinfo{author}{\bibfnamefont{S.}~\bibnamefont{Ray}},
  \bibinfo{author}{\bibfnamefont{P.}~\bibnamefont{Mahadevan}},
  \bibnamefont{and} \bibinfo{author}{\bibfnamefont{D.~D.~S.}
  \bibnamefont{I.~H.~Inoue}, \bibfnamefont{C.~Carbone}},
  \bibinfo{journal}{cond-mat/0509643}  (\bibinfo{year}{2005}).

\bibitem[{\citenamefont{Morikawa et~al.}(1995)\citenamefont{Morikawa, Mizokawa,
  Kobayashi, Fujimori, Eisaki, Uchida, Iga, and Nishihara}}]{morikawa95}
\bibinfo{author}{\bibfnamefont{K.}~\bibnamefont{Morikawa}},
  \bibinfo{author}{\bibfnamefont{T.}~\bibnamefont{Mizokawa}},
  \bibinfo{author}{\bibfnamefont{K.}~\bibnamefont{Kobayashi}},
  \bibinfo{author}{\bibfnamefont{A.}~\bibnamefont{Fujimori}},
  \bibinfo{author}{\bibfnamefont{H.}~\bibnamefont{Eisaki}},
  \bibinfo{author}{\bibfnamefont{S.}~\bibnamefont{Uchida}},
  \bibinfo{author}{\bibfnamefont{F.}~\bibnamefont{Iga}}, \bibnamefont{and}
  \bibinfo{author}{\bibfnamefont{Y.}~\bibnamefont{Nishihara}},
  \bibinfo{journal}{Phys. Rev. B} \textbf{\bibinfo{volume}{52}},
  \bibinfo{pages}{13711} (\bibinfo{year}{1995}).

\bibitem[{\citenamefont{Dederichs et~al.}(1984)\citenamefont{Dederichs,
  Bl{\"u}gel, Zeller, and Akai}}]{ded84}
\bibinfo{author}{\bibfnamefont{P.~H.} \bibnamefont{Dederichs}},
  \bibinfo{author}{\bibfnamefont{S.}~\bibnamefont{Bl{\"u}gel}},
  \bibinfo{author}{\bibfnamefont{R.}~\bibnamefont{Zeller}}, \bibnamefont{and}
  \bibinfo{author}{\bibfnamefont{H.}~\bibnamefont{Akai}},
  \bibinfo{journal}{Phys. Rev. Lett.} \textbf{\bibinfo{volume}{53}},
  \bibinfo{pages}{2512} (\bibinfo{year}{1984}).

\bibitem[{\citenamefont{McMahan et~al.}(1988)\citenamefont{McMahan, Martin, and
  Satpathy}}]{mcm88}
\bibinfo{author}{\bibfnamefont{A.~K.} \bibnamefont{McMahan}},
  \bibinfo{author}{\bibfnamefont{R.~M.} \bibnamefont{Martin}},
  \bibnamefont{and} \bibinfo{author}{\bibfnamefont{S.}~\bibnamefont{Satpathy}},
  \bibinfo{journal}{Phys. Rev. B} \textbf{\bibinfo{volume}{38}},
  \bibinfo{pages}{6650} (\bibinfo{year}{1988}).

\bibitem[{\citenamefont{Gunnarsson et~al.}(1989)\citenamefont{Gunnarsson,
  Andersen, Jepsen, and Zaanen}}]{gun89}
\bibinfo{author}{\bibfnamefont{O.}~\bibnamefont{Gunnarsson}},
  \bibinfo{author}{\bibfnamefont{O.~K.} \bibnamefont{Andersen}},
  \bibinfo{author}{\bibfnamefont{O.}~\bibnamefont{Jepsen}}, \bibnamefont{and}
  \bibinfo{author}{\bibfnamefont{J.}~\bibnamefont{Zaanen}},
  \bibinfo{journal}{Phys. Rev. B} \textbf{\bibinfo{volume}{39}},
  \bibinfo{pages}{1708} (\bibinfo{year}{1989}).

\bibitem[{\citenamefont{Anisimov and Gunnarsson}(1991)}]{ani91}
\bibinfo{author}{\bibfnamefont{V.~I.} \bibnamefont{Anisimov}} \bibnamefont{and}
  \bibinfo{author}{\bibfnamefont{O.}~\bibnamefont{Gunnarsson}},
  \bibinfo{journal}{Phys. Rev. B} \textbf{\bibinfo{volume}{43}},
  \bibinfo{pages}{7570} (\bibinfo{year}{1991}).

\bibitem[{\citenamefont{Cococcioni and de~Gironcoli}(2005)}]{coc05}
\bibinfo{author}{\bibfnamefont{M.}~\bibnamefont{Cococcioni}} \bibnamefont{and}
  \bibinfo{author}{\bibfnamefont{S.}~\bibnamefont{de~Gironcoli}},
  \bibinfo{journal}{Phys. Rev. B} \textbf{\bibinfo{volume}{71}},
  \bibinfo{pages}{035105} (\bibinfo{year}{2005}).

\bibitem[{\citenamefont{Aryasetiawan et~al.}(2004)\citenamefont{Aryasetiawan,
  Imada, Georges, Kotliar, Biermann, and Lichtenstein}}]{ary04}
\bibinfo{author}{\bibfnamefont{F.}~\bibnamefont{Aryasetiawan}},
  \bibinfo{author}{\bibfnamefont{M.}~\bibnamefont{Imada}},
  \bibinfo{author}{\bibfnamefont{A.}~\bibnamefont{Georges}},
  \bibinfo{author}{\bibfnamefont{G.}~\bibnamefont{Kotliar}},
  \bibinfo{author}{\bibfnamefont{S.}~\bibnamefont{Biermann}}, \bibnamefont{and}
  \bibinfo{author}{\bibfnamefont{A.~I.} \bibnamefont{Lichtenstein}},
  \bibinfo{journal}{Phys. Rev. B} \textbf{\bibinfo{volume}{70}},
  \bibinfo{pages}{195104} (\bibinfo{year}{2004}).

\bibitem[{\citenamefont{Solovyev and Imada}(2005)}]{sol05}
\bibinfo{author}{\bibfnamefont{I.~V.} \bibnamefont{Solovyev}} \bibnamefont{and}
  \bibinfo{author}{\bibfnamefont{M.}~\bibnamefont{Imada}},
  \bibinfo{journal}{Phys. Rev. B} \textbf{\bibinfo{volume}{71}},
  \bibinfo{pages}{045103} (\bibinfo{year}{2005}).

\bibitem[{\citenamefont{Aryasetiawan et~al.}(2006)\citenamefont{Aryasetiawan,
  Karlsson, Jepsen, and Schonberger}}]{ary06}
\bibinfo{author}{\bibfnamefont{F.}~\bibnamefont{Aryasetiawan}},
  \bibinfo{author}{\bibfnamefont{K.}~\bibnamefont{Karlsson}},
  \bibinfo{author}{\bibfnamefont{O.}~\bibnamefont{Jepsen}}, \bibnamefont{and}
  \bibinfo{author}{\bibfnamefont{U.}~\bibnamefont{Schonberger}},
  \bibinfo{journal}{cond-mat/0603138}  (\bibinfo{year}{2006}).

\bibitem[{\citenamefont{Inoue et~al.}(1998)\citenamefont{Inoue, Goto, Makino,
  Hussey, and Ishikawa}}]{inoue98}
\bibinfo{author}{\bibfnamefont{I.~H.} \bibnamefont{Inoue}},
  \bibinfo{author}{\bibfnamefont{O.}~\bibnamefont{Goto}},
  \bibinfo{author}{\bibfnamefont{H.}~\bibnamefont{Makino}},
  \bibinfo{author}{\bibfnamefont{N.~E.} \bibnamefont{Hussey}},
  \bibnamefont{and} \bibinfo{author}{\bibfnamefont{M.}~\bibnamefont{Ishikawa}},
  \bibinfo{journal}{Phys. Rev. B} \textbf{\bibinfo{volume}{58}},
  \bibinfo{pages}{4372} (\bibinfo{year}{1998}).

\bibitem[{\citenamefont{{Booth} et~al.}(1999)\citenamefont{{Booth}, {Figueroa},
  {Lawrence}, {Hundley}, and {Thompson}}}]{boo99}
\bibinfo{author}{\bibfnamefont{C.~H.} \bibnamefont{{Booth}}},
  \bibinfo{author}{\bibfnamefont{E.}~\bibnamefont{{Figueroa}}},
  \bibinfo{author}{\bibfnamefont{J.~M.} \bibnamefont{{Lawrence}}},
  \bibinfo{author}{\bibfnamefont{M.~F.} \bibnamefont{{Hundley}}},
  \bibnamefont{and} \bibinfo{author}{\bibfnamefont{J.~D.}
  \bibnamefont{{Thompson}}}, \bibinfo{journal}{Phys. Rev. B}
  \textbf{\bibinfo{volume}{60}}, \bibinfo{pages}{14852} (\bibinfo{year}{1999}).

\bibitem[{\citenamefont{{Whangbo} et~al.}(2003)\citenamefont{{Whangbo}, {Koo},
  {Dai}, and {Villesuzanne}}}]{whangbo_bavs3_puzzling_jsschem_2003}
\bibinfo{author}{\bibfnamefont{M.-H.} \bibnamefont{{Whangbo}}},
  \bibinfo{author}{\bibfnamefont{H.-J.} \bibnamefont{{Koo}}},
  \bibinfo{author}{\bibfnamefont{D.}~\bibnamefont{{Dai}}}, \bibnamefont{and}
  \bibinfo{author}{\bibfnamefont{A.}~\bibnamefont{{Villesuzanne}}},
  \bibinfo{journal}{J. Solid State Chem.} \textbf{\bibinfo{volume}{175}},
  \bibinfo{pages}{384} (\bibinfo{year}{2003}).

\bibitem[{\citenamefont{Lechermann
  et~al.}(2005{\natexlab{a}})\citenamefont{Lechermann, Biermann, and
  Georges}}]{lec05}
\bibinfo{author}{\bibfnamefont{F.}~\bibnamefont{Lechermann}},
  \bibinfo{author}{\bibfnamefont{S.}~\bibnamefont{Biermann}}, \bibnamefont{and}
  \bibinfo{author}{\bibfnamefont{A.}~\bibnamefont{Georges}},
  \bibinfo{journal}{Phys. Rev. Lett.} \textbf{\bibinfo{volume}{94}},
  \bibinfo{pages}{166402} (\bibinfo{year}{2005}{\natexlab{a}}).

\bibitem[{\citenamefont{Lechermann
  et~al.}(2005{\natexlab{b}})\citenamefont{Lechermann, Biermann, and
  Georges}}]{lecproc}
\bibinfo{author}{\bibfnamefont{F.}~\bibnamefont{Lechermann}},
  \bibinfo{author}{\bibfnamefont{S.}~\bibnamefont{Biermann}}, \bibnamefont{and}
  \bibinfo{author}{\bibfnamefont{A.}~\bibnamefont{Georges}},
  \bibinfo{journal}{Progress of Theoretical Physics Supplement}
  \textbf{\bibinfo{volume}{160}}, \bibinfo{pages}{233}
  (\bibinfo{year}{2005}{\natexlab{b}}).

\bibitem[{\citenamefont{{Graf} et~al.}(1995)\citenamefont{{Graf}, {Mandrus},
  {Lawrence}, {Thompson}, {Canfield}, {Cheong}, and
  {Rupp}}}]{graf_bavs3_pressure_prb_1995}
\bibinfo{author}{\bibfnamefont{T.}~\bibnamefont{{Graf}}},
  \bibinfo{author}{\bibfnamefont{D.}~\bibnamefont{{Mandrus}}},
  \bibinfo{author}{\bibfnamefont{J.~M.} \bibnamefont{{Lawrence}}},
  \bibinfo{author}{\bibfnamefont{J.~D.} \bibnamefont{{Thompson}}},
  \bibinfo{author}{\bibfnamefont{P.~C.} \bibnamefont{{Canfield}}},
  \bibinfo{author}{\bibfnamefont{S.-W.} \bibnamefont{{Cheong}}},
  \bibnamefont{and} \bibinfo{author}{\bibfnamefont{L.~W.}
  \bibnamefont{{Rupp}}}, \bibinfo{journal}{Phys. Rev. B}
  \textbf{\bibinfo{volume}{51}}, \bibinfo{pages}{2037} (\bibinfo{year}{1995}).

\bibitem[{\citenamefont{Mih{\'a}ly et~al.}(2000)\citenamefont{Mih{\'a}ly,
  K{\'e}zsm{\'a}rki, Z{\'a}mborszky, Miljak, Penc, Fazekas, Berger, and
  Forr{\'o}}}]{mih00}
\bibinfo{author}{\bibfnamefont{G.}~\bibnamefont{Mih{\'a}ly}},
  \bibinfo{author}{\bibfnamefont{I.}~\bibnamefont{K{\'e}zsm{\'a}rki}},
  \bibinfo{author}{\bibfnamefont{F.}~\bibnamefont{Z{\'a}mborszky}},
  \bibinfo{author}{\bibfnamefont{M.}~\bibnamefont{Miljak}},
  \bibinfo{author}{\bibfnamefont{K.}~\bibnamefont{Penc}},
  \bibinfo{author}{\bibfnamefont{P.}~\bibnamefont{Fazekas}},
  \bibinfo{author}{\bibfnamefont{H.}~\bibnamefont{Berger}}, \bibnamefont{and}
  \bibinfo{author}{\bibfnamefont{L.}~\bibnamefont{Forr{\'o}}},
  \bibinfo{journal}{Phys. Rev. B} \textbf{\bibinfo{volume}{61}}
  (\bibinfo{year}{2000}).

\bibitem[{\citenamefont{{Inami} et~al.}(2002)\citenamefont{{Inami}, {Ohwada},
  {Kimura}, {Watanabe}, {Noda}, {Nakamura}, {Yamasaki}, {Shiga}, {Ikeda}, and
  {Murakami}}}]{inami_bavs3_prb_2002}
\bibinfo{author}{\bibfnamefont{T.}~\bibnamefont{{Inami}}},
  \bibinfo{author}{\bibfnamefont{K.}~\bibnamefont{{Ohwada}}},
  \bibinfo{author}{\bibfnamefont{H.}~\bibnamefont{{Kimura}}},
  \bibinfo{author}{\bibfnamefont{M.}~\bibnamefont{{Watanabe}}},
  \bibinfo{author}{\bibfnamefont{Y.}~\bibnamefont{{Noda}}},
  \bibinfo{author}{\bibfnamefont{H.}~\bibnamefont{{Nakamura}}},
  \bibinfo{author}{\bibfnamefont{T.}~\bibnamefont{{Yamasaki}}},
  \bibinfo{author}{\bibfnamefont{M.}~\bibnamefont{{Shiga}}},
  \bibinfo{author}{\bibfnamefont{N.}~\bibnamefont{{Ikeda}}}, \bibnamefont{and}
  \bibinfo{author}{\bibfnamefont{Y.}~\bibnamefont{{Murakami}}},
  \bibinfo{journal}{Phys. Rev. B} \textbf{\bibinfo{volume}{66}},
  \bibinfo{pages}{073108} (\bibinfo{year}{2002}).

\bibitem[{\citenamefont{{Fagot} et~al.}(2003)\citenamefont{{Fagot},
  {Foury-Leylekian}, {Ravy}, {Pouget}, and {Berger}}}]{fagot_bavs3_prl_2003}
\bibinfo{author}{\bibfnamefont{S.}~\bibnamefont{{Fagot}}},
  \bibinfo{author}{\bibfnamefont{P.}~\bibnamefont{{Foury-Leylekian}}},
  \bibinfo{author}{\bibfnamefont{S.}~\bibnamefont{{Ravy}}},
  \bibinfo{author}{\bibfnamefont{J.}~\bibnamefont{{Pouget}}}, \bibnamefont{and}
  \bibinfo{author}{\bibfnamefont{H.}~\bibnamefont{{Berger}}},
  \bibinfo{journal}{Phys. Rev. Lett.} \textbf{\bibinfo{volume}{90}},
  \bibinfo{pages}{196401} (\bibinfo{year}{2003}).

\bibitem[{\citenamefont{Fagot et~al.}(2005)\citenamefont{Fagot,
  Foury-Leylekian, Ravy, Pouget, Anne, Popov, Lobanov, and Greenblatt}}]{fag05}
\bibinfo{author}{\bibfnamefont{S.}~\bibnamefont{Fagot}},
  \bibinfo{author}{\bibfnamefont{P.}~\bibnamefont{Foury-Leylekian}},
  \bibinfo{author}{\bibfnamefont{S.}~\bibnamefont{Ravy}},
  \bibinfo{author}{\bibfnamefont{J.-P.} \bibnamefont{Pouget}},
  \bibinfo{author}{\bibfnamefont{M.}~\bibnamefont{Anne}},
  \bibinfo{author}{\bibfnamefont{G.}~\bibnamefont{Popov}},
  \bibinfo{author}{\bibfnamefont{M.~V.} \bibnamefont{Lobanov}},
  \bibnamefont{and}
  \bibinfo{author}{\bibfnamefont{M.}~\bibnamefont{Greenblatt}},
  \bibinfo{journal}{Solid State Sciences} \textbf{\bibinfo{volume}{7}},
  \bibinfo{pages}{718} (\bibinfo{year}{2005}).

\bibitem[{\citenamefont{Nakamura et~al.}(2000)\citenamefont{Nakamura, Yamasaki,
  Giri, Imai, Shiga, Kojima, Nishi, and Metoki}}]{Nak00}
\bibinfo{author}{\bibfnamefont{H.}~\bibnamefont{Nakamura}},
  \bibinfo{author}{\bibfnamefont{T.}~\bibnamefont{Yamasaki}},
  \bibinfo{author}{\bibfnamefont{S.}~\bibnamefont{Giri}},
  \bibinfo{author}{\bibfnamefont{H.}~\bibnamefont{Imai}},
  \bibinfo{author}{\bibfnamefont{M.}~\bibnamefont{Shiga}},
  \bibinfo{author}{\bibfnamefont{K.}~\bibnamefont{Kojima}},
  \bibinfo{author}{\bibfnamefont{M.}~\bibnamefont{Nishi}}, \bibnamefont{and}
  \bibinfo{author}{\bibfnamefont{K.~K.~N.} \bibnamefont{Metoki}},
  \bibinfo{journal}{J. Phys. Soc. Jpn.} \textbf{\bibinfo{volume}{69}},
  \bibinfo{pages}{2763} (\bibinfo{year}{2000}).

\bibitem[{\citenamefont{{Ghedira} et~al.}(1986)\citenamefont{{Ghedira}, {Anne},
  {Chenavas}, {Marezio}, and {Sayetat}}}]{ghedira_bavs3_neutrons_jpc_1986}
\bibinfo{author}{\bibfnamefont{M.}~\bibnamefont{{Ghedira}}},
  \bibinfo{author}{\bibfnamefont{M.}~\bibnamefont{{Anne}}},
  \bibinfo{author}{\bibfnamefont{J.}~\bibnamefont{{Chenavas}}},
  \bibinfo{author}{\bibfnamefont{M.}~\bibnamefont{{Marezio}}},
  \bibnamefont{and}
  \bibinfo{author}{\bibfnamefont{F.}~\bibnamefont{{Sayetat}}},
  \bibinfo{journal}{Journal of Physics C Solid State Physics}
  \textbf{\bibinfo{volume}{19}}, \bibinfo{pages}{6489} (\bibinfo{year}{1986}).

\bibitem[{\citenamefont{Mattheiss}(1995)}]{mattheiss_bavs3_1995}
\bibinfo{author}{\bibfnamefont{L.}~\bibnamefont{Mattheiss}},
  \bibinfo{journal}{Solid State Commun.} \textbf{\bibinfo{volume}{93}},
  \bibinfo{pages}{791} (\bibinfo{year}{1995}).

\bibitem[{\citenamefont{Jepsen and Andersen}(1995)}]{jep95}
\bibinfo{author}{\bibfnamefont{O.}~\bibnamefont{Jepsen}} \bibnamefont{and}
  \bibinfo{author}{\bibfnamefont{O.~K.} \bibnamefont{Andersen}},
  \bibinfo{journal}{Zeitschrift f{\"u}r Physik B}
  \textbf{\bibinfo{volume}{97}}, \bibinfo{pages}{35} (\bibinfo{year}{1995}).

\bibitem[{\citenamefont{Saha-Dasgupta et~al.}(to be
  published)\citenamefont{Saha-Dasgupta, Andersen, Nuss, Poteryaev,
  Lichtenstein, and Georges}}]{tanusri-tbp}
\bibinfo{author}{\bibfnamefont{T.}~\bibnamefont{Saha-Dasgupta}},
  \bibinfo{author}{\bibfnamefont{O.~K.} \bibnamefont{Andersen}},
  \bibinfo{author}{\bibfnamefont{J.}~\bibnamefont{Nuss}},
  \bibinfo{author}{\bibfnamefont{A.}~\bibnamefont{Poteryaev}},
  \bibinfo{author}{\bibfnamefont{A.~I.} \bibnamefont{Lichtenstein}},
  \bibnamefont{and} \bibinfo{author}{\bibfnamefont{A.}~\bibnamefont{Georges}}
  (\bibinfo{year}{to be published}).

\bibitem[{\citenamefont{Yamasaki
  et~al.}(2006{\natexlab{b}})\citenamefont{Yamasaki, Chioncel, Lichtenstein,
  and Andersen}}]{yam06_2}
\bibinfo{author}{\bibfnamefont{A.}~\bibnamefont{Yamasaki}},
  \bibinfo{author}{\bibfnamefont{L.}~\bibnamefont{Chioncel}},
  \bibinfo{author}{\bibfnamefont{A.~I.} \bibnamefont{Lichtenstein}},
  \bibnamefont{and} \bibinfo{author}{\bibfnamefont{O.~K.}
  \bibnamefont{Andersen}}, \bibinfo{journal}{cond-mat/0603305}
  (\bibinfo{year}{2006}{\natexlab{b}}).

\bibitem[{\citenamefont{Andersen et~al.}(to be
  published)\citenamefont{Andersen, Mazin, Jepsen, and
  Johannes}}]{and-coo2-tbp}
\bibinfo{author}{\bibfnamefont{O.~K.} \bibnamefont{Andersen}},
  \bibinfo{author}{\bibfnamefont{I.}~\bibnamefont{Mazin}},
  \bibinfo{author}{\bibfnamefont{O.}~\bibnamefont{Jepsen}}, \bibnamefont{and}
  \bibinfo{author}{\bibfnamefont{M.}~\bibnamefont{Johannes}} (\bibinfo{year}{to
  be published}).

\bibitem[{\citenamefont{Lechermann}(to be published)}]{lecto}
\bibinfo{author}{\bibfnamefont{F.}~\bibnamefont{Lechermann}} (\bibinfo{year}{to
  be published}).

\bibitem[{\citenamefont{Luttinger}(1960)}]{lut60}
\bibinfo{author}{\bibfnamefont{J.~M.} \bibnamefont{Luttinger}},
  \bibinfo{journal}{Phys. Rev.} \textbf{\bibinfo{volume}{119}},
  \bibinfo{pages}{1153} (\bibinfo{year}{1960}).

\bibitem[{\citenamefont{Mitrovic et~al.}(2005)\citenamefont{Mitrovic, Fazekas,
  S$\o$ndergaard, Ariosa, Bari\v{s}i\'{c}, Berger, Clo{\"e}tta, Forr{\'o},
  H{\"o}chst, Kup\v{c}i\'{c} et~al.}}]{mit05}
\bibinfo{author}{\bibfnamefont{S.}~\bibnamefont{Mitrovic}},
  \bibinfo{author}{\bibfnamefont{P.}~\bibnamefont{Fazekas}},
  \bibinfo{author}{\bibfnamefont{C.}~\bibnamefont{S$\o$ndergaard}},
  \bibinfo{author}{\bibfnamefont{D.}~\bibnamefont{Ariosa}},
  \bibinfo{author}{\bibfnamefont{N.}~\bibnamefont{Bari\v{s}i\'{c}}},
  \bibinfo{author}{\bibfnamefont{H.}~\bibnamefont{Berger}},
  \bibinfo{author}{\bibfnamefont{D.}~\bibnamefont{Clo{\"e}tta}},
  \bibinfo{author}{\bibfnamefont{L.}~\bibnamefont{Forr{\'o}}},
  \bibinfo{author}{\bibfnamefont{H.}~\bibnamefont{H{\"o}chst}},
  \bibinfo{author}{\bibfnamefont{I.}~\bibnamefont{Kup\v{c}i\'{c}}},
  \bibnamefont{et~al.}, \bibinfo{journal}{cond-mat/0502144}
  (\bibinfo{year}{2005}).

\bibitem[{\citenamefont{Mo et~al.}(2005)\citenamefont{Mo, Wang, Allen, and
  et~al.}}]{mo05}
\bibinfo{author}{\bibfnamefont{S.-K.} \bibnamefont{Mo}},
  \bibinfo{author}{\bibfnamefont{F.}~\bibnamefont{Wang}},
  \bibinfo{author}{\bibfnamefont{J.~W.} \bibnamefont{Allen}}, \bibnamefont{and}
  \bibinfo{author}{\bibnamefont{et~al.}}, \bibinfo{journal}{private
  communication}  (\bibinfo{year}{2005}).

\bibitem[{\citenamefont{Anisimov et~al.}(2002)\citenamefont{Anisimov, Nekrasov,
  Kondakov, Rice, and Sigrist}}]{ani02}
\bibinfo{author}{\bibfnamefont{V.~I.} \bibnamefont{Anisimov}},
  \bibinfo{author}{\bibfnamefont{I.}~\bibnamefont{Nekrasov}},
  \bibinfo{author}{\bibfnamefont{D.}~\bibnamefont{Kondakov}},
  \bibinfo{author}{\bibfnamefont{T.}~\bibnamefont{Rice}}, \bibnamefont{and}
  \bibinfo{author}{\bibfnamefont{M.}~\bibnamefont{Sigrist}},
  \bibinfo{journal}{Eur. Phys. J. B} \textbf{\bibinfo{volume}{25}},
  \bibinfo{pages}{191} (\bibinfo{year}{2002}).

\bibitem[{\citenamefont{Liebsch}(2005)}]{lie05}
\bibinfo{author}{\bibfnamefont{A.}~\bibnamefont{Liebsch}},
  \bibinfo{journal}{Phys. Rev. Lett.} \textbf{\bibinfo{volume}{95}},
  \bibinfo{pages}{116402} (\bibinfo{year}{2005}).

\bibitem[{\citenamefont{K{\'e}zsm{\'a}rki
  et~al.}(2006)\citenamefont{K{\'e}zsm{\'a}rki, Mih{\'a}ly, Ga{\'a}l,
  Bari\v{s}i\'{c}, Akrap, Berger, Forr{\'o}, Homes, and Mih{\'a}ly}}]{kez06}
\bibinfo{author}{\bibfnamefont{I.}~\bibnamefont{K{\'e}zsm{\'a}rki}},
  \bibinfo{author}{\bibfnamefont{G.}~\bibnamefont{Mih{\'a}ly}},
  \bibinfo{author}{\bibfnamefont{R.}~\bibnamefont{Ga{\'a}l}},
  \bibinfo{author}{\bibfnamefont{N.}~\bibnamefont{Bari\v{s}i\'{c}}},
  \bibinfo{author}{\bibfnamefont{A.}~\bibnamefont{Akrap}},
  \bibinfo{author}{\bibfnamefont{H.}~\bibnamefont{Berger}},
  \bibinfo{author}{\bibfnamefont{L.}~\bibnamefont{Forr{\'o}}},
  \bibinfo{author}{\bibfnamefont{C.~C.} \bibnamefont{Homes}}, \bibnamefont{and}
  \bibinfo{author}{\bibfnamefont{L.}~\bibnamefont{Mih{\'a}ly}},
  \bibinfo{journal}{Phys. Rev. Lett.} \textbf{\bibinfo{volume}{96}},
  \bibinfo{pages}{186402} (\bibinfo{year}{2006}).

\bibitem[{\citenamefont{Fagot et~al.}(2006)\citenamefont{Fagot,
  Foury-Leylekian, Ravy, Pouget, Lorenzo, Joly, Greenblatt, Lobanov, and
  Popov}}]{fag06}
\bibinfo{author}{\bibfnamefont{S.}~\bibnamefont{Fagot}},
  \bibinfo{author}{\bibfnamefont{P.}~\bibnamefont{Foury-Leylekian}},
  \bibinfo{author}{\bibfnamefont{S.}~\bibnamefont{Ravy}},
  \bibinfo{author}{\bibfnamefont{J.-P.} \bibnamefont{Pouget}},
  \bibinfo{author}{\bibfnamefont{E.}~\bibnamefont{Lorenzo}},
  \bibinfo{author}{\bibfnamefont{Y.}~\bibnamefont{Joly}},
  \bibinfo{author}{\bibfnamefont{M.}~\bibnamefont{Greenblatt}},
  \bibinfo{author}{\bibfnamefont{M.~V.} \bibnamefont{Lobanov}},
  \bibnamefont{and} \bibinfo{author}{\bibfnamefont{G.}~\bibnamefont{Popov}},
  \bibinfo{journal}{Phys. Rev. B} \textbf{\bibinfo{volume}{73}},
  \bibinfo{pages}{033102} (\bibinfo{year}{2006}).

\bibitem[{\citenamefont{Paul and Kotliar}(2006)}]{pau06}
\bibinfo{author}{\bibfnamefont{I.}~\bibnamefont{Paul}} \bibnamefont{and}
  \bibinfo{author}{\bibfnamefont{G.}~\bibnamefont{Kotliar}},
  \bibinfo{journal}{cond-mat/0501539}  (\bibinfo{year}{2006}).

\bibitem[{\citenamefont{{Savrasov} et~al.}(2001)\citenamefont{{Savrasov},
  {Kotliar}, and Abrahams}}]{savrasov_kotliar_pu_nature_2001}
\bibinfo{author}{\bibfnamefont{S.~Y.} \bibnamefont{{Savrasov}}},
  \bibinfo{author}{\bibfnamefont{G.}~\bibnamefont{{Kotliar}}},
  \bibnamefont{and} \bibinfo{author}{\bibfnamefont{E.}~\bibnamefont{Abrahams}},
  \bibinfo{journal}{Nature} \textbf{\bibinfo{volume}{410}},
  \bibinfo{pages}{793} (\bibinfo{year}{2001}).

\bibitem[{\citenamefont{Pourovskii et~al.}(to be
  published)\citenamefont{Pourovskii, Amadon, and Georges}}]{pouto}
\bibinfo{author}{\bibfnamefont{L.}~\bibnamefont{Pourovskii}},
  \bibinfo{author}{\bibfnamefont{B.}~\bibnamefont{Amadon}}, \bibnamefont{and}
  \bibinfo{author}{\bibfnamefont{A.}~\bibnamefont{Georges}} (\bibinfo{year}{to
  be published}).

\bibitem[{\citenamefont{Chioncel et~al.}(2003)\citenamefont{Chioncel, Vitos,
  Abrikosov, Koll\'{a}r, Katsnelson, and Lichtenstein}}]{chi03}
\bibinfo{author}{\bibfnamefont{L.}~\bibnamefont{Chioncel}},
  \bibinfo{author}{\bibfnamefont{L.}~\bibnamefont{Vitos}},
  \bibinfo{author}{\bibfnamefont{I.~A.} \bibnamefont{Abrikosov}},
  \bibinfo{author}{\bibfnamefont{J.}~\bibnamefont{Koll\'{a}r}},
  \bibinfo{author}{\bibfnamefont{M.~I.} \bibnamefont{Katsnelson}},
  \bibnamefont{and} \bibinfo{author}{\bibfnamefont{A.~I.}
  \bibnamefont{Lichtenstein}}, \bibinfo{journal}{Phys. Rev. B}
  \textbf{\bibinfo{volume}{67}}, \bibinfo{pages}{235106}
  (\bibinfo{year}{2003}).

\bibitem[{\citenamefont{Min\'{a}r et~al.}(2005)\citenamefont{Min\'{a}r,
  Chioncel, Perlov, Ebert, Katsnelson, and Lichtenstein}}]{min05}
\bibinfo{author}{\bibfnamefont{J.}~\bibnamefont{Min\'{a}r}},
  \bibinfo{author}{\bibfnamefont{L.}~\bibnamefont{Chioncel}},
  \bibinfo{author}{\bibfnamefont{A.}~\bibnamefont{Perlov}},
  \bibinfo{author}{\bibfnamefont{H.}~\bibnamefont{Ebert}},
  \bibinfo{author}{\bibfnamefont{M.~I.} \bibnamefont{Katsnelson}},
  \bibnamefont{and} \bibinfo{author}{\bibfnamefont{A.~I.}
  \bibnamefont{Lichtenstein}}, \bibinfo{journal}{Phys. Rev. B}
  \textbf{\bibinfo{volume}{72}}, \bibinfo{pages}{045125}
  (\bibinfo{year}{2005}).

\bibitem[{\citenamefont{Amadon et~al.}(2006)\citenamefont{Amadon, Biermann,
  Georges, and Aryasetiawan}}]{ama06}
\bibinfo{author}{\bibfnamefont{B.}~\bibnamefont{Amadon}},
  \bibinfo{author}{\bibfnamefont{S.}~\bibnamefont{Biermann}},
  \bibinfo{author}{\bibfnamefont{A.}~\bibnamefont{Georges}}, \bibnamefont{and}
  \bibinfo{author}{\bibfnamefont{F.}~\bibnamefont{Aryasetiawan}},
  \bibinfo{journal}{Phys. Rev. Lett.} \textbf{\bibinfo{volume}{96}},
  \bibinfo{pages}{066402} (\bibinfo{year}{2006}).

\bibitem[{\citenamefont{Perdew and Wang}(1992)}]{per92}
\bibinfo{author}{\bibfnamefont{J.~P.} \bibnamefont{Perdew}} \bibnamefont{and}
  \bibinfo{author}{\bibfnamefont{Y.}~\bibnamefont{Wang}},
  \bibinfo{journal}{Phys. Rev. B} \textbf{\bibinfo{volume}{45}},
  \bibinfo{pages}{13244} (\bibinfo{year}{1992}).

\bibitem[{\citenamefont{Hedin and Lundquist}(1971)}]{hed71}
\bibinfo{author}{\bibfnamefont{L.}~\bibnamefont{Hedin}} \bibnamefont{and}
  \bibinfo{author}{\bibfnamefont{B.~I.} \bibnamefont{Lundquist}},
  \bibinfo{journal}{J. Phys. C} \textbf{\bibinfo{volume}{4}},
  \bibinfo{pages}{2064} (\bibinfo{year}{1971}).

\bibitem[{\citenamefont{von Barth and Hedin}(1972)}]{bar72}
\bibinfo{author}{\bibfnamefont{U.}~\bibnamefont{von Barth}} \bibnamefont{and}
  \bibinfo{author}{\bibfnamefont{L.}~\bibnamefont{Hedin}}, \bibinfo{journal}{J.
  Phys. C} \textbf{\bibinfo{volume}{5}}, \bibinfo{pages}{1692}
  (\bibinfo{year}{1972}).

\end{thebibliography}

\end{document}